\documentclass{aa}
\usepackage{aalongtable}
\usepackage{graphicx}
\usepackage{amssymb}
\usepackage{natbib}
\bibpunct{(}{)}{;}{a}{}{,}
\usepackage{supertabular}
\usepackage{rotating}
\usepackage{setspace}
\usepackage{lscape}
\def\vsini{$V\!\sin i$}
\def\teff{$T_{\rm eff}$}

\def\logg{$\log~g$}

\def\kms{km~s$^{-1}$}

\def\rv{RV}
\def\ttms{$\frac{\tau}{\tau_{MS}}$}

\begin{document}
%
\title{Early-type objects in NGC\,6611 and Eagle Nebula.}

\titlerunning{Early-type objects in M16.}
\author{
C. Martayan \inst{1,2}
\and  M. Floquet \inst{2}
\and  A.M. Hubert  \inst{2}
\and  C. Neiner \inst{2}
\and  Y. Fr\'emat \inst{1}
\and  D. Baade \inst{3}
\and  J. Fabregat  \inst{4}
}
\offprints {C. Martayan}
\mail{Martayan@oma.be}
\institute{Royal Observatory of Belgium, 3 avenue circulaire, 1180 Brussels, Belgium 
\and GEPI, Observatoire de Paris, CNRS, Universit\'e Paris Diderot; 5 place Jules Janssen 92195 Meudon Cedex, France
\and European Organisation for Astronomical Research in the Southern Hemisphere, 
Karl-Schwarzschild-Str. 2, D-85748 Garching b. Muenchen, Germany
\and Observatorio Astron\'omico de Valencia, edifici Instituts d'investigaci\'o, 
Poligon la Coma, 46980 Paterna Valencia, Spain }
\date{Received /Accepted}
\abstract
{}
{An important question about Be stars is whether Be stars are born as Be stars
or whether they become Be stars during their evolution. It is necessary to
observe young clusters to answer this question.}
{To this end, observations of stars in NGC\,6611  and the star-formation region of Eagle Nebula have been carried out with
the ESO-WFI in slitless spectroscopic mode and at the VLT-GIRAFFE (R$\simeq$ 6400--17000). 
The targets for the GIRAFFE observations were pre-selected from the literature and our catalogue of
emission-line stars based on the WFI study. GIRAFFE
observations allowed us to study accurately the population of the early-type stars with and without emission lines. 
For this study, we determined the fundamental parameters of OBA stars thanks to the GIRFIT code. We
also studied the status of  the objects (main sequence or pre-main sequence stars) by using
IR data, membership probabilities, and location in HR diagrams.}
{The nature of the  early-type stars with emission-line stars in NGC\,6611 
 and its surrounding environment is derived. The slitless
observations with the WFI clearly indicate a small number of emission-line stars
in M16. We observed with GIRAFFE 101 OBA stars,
among them 9 are emission-line stars with circumstellar emission in
H$\alpha$.  We found that: W080 could be a new He-strong star, like W601. 
W301 is a possible classical Be star, W503 is a mass-transfer eclipsing binary with an accretion disk, 
and the other ones are possible Herbig Ae/Be stars.
We also found that the rotational velocities of main sequence B stars are 18\% lower than those of pre-main
sequence B stars, in good agreement with theory about the evolution of rotational velocities. 
Combining adaptive optics, IR data, spectroscopy, and radial velocity indications, we found that 27\% of
the B-type stars are binaries.  We also redetermined the age of NGC\,6611 found equal to 1.2--1.8 Myears 
in good agreement with the most recent determinations.}
{}

\maketitle

\section{Introduction}
The origin of the Be phenomenon, i.e. periods of spectral emission due to the
presence of a circumstellar envelope around Be stars, is still debated. 
Rapid rotation seems to be a major key in triggering this phenomenon. To
understand the Be phenomenon, it is important to know at which phase of the
stellar evolution on the main sequence (MS) it appears. 

According to a statistical study of Be stars in clusters, \citet{fabregat2000}
concluded that it may occur in the second half of the MS phase. Taking into
account effects due to fast rotation, \citet{zorec2005} estimated that the
appearance of the Be phenomenon among field early-type stars is probably
mass-dependent and that it may appear at any time during the MS phase.
\citet{marta2006b,marta2007a} showed that the appearance of Be stars is mass-,
and metallicity-dependent. In the  field of Milky Way (MW), they found that less massive Be stars
only appear during the second part of the MS, while massive Be stars
appear mainly during the first part of the MS, and the intermediate-mass Be
stars appear during the whole MS. 

To confirm this,
it is necessary to observe young or very young
open clusters with emission-line stars (ELS).

\addtocounter{figure}{+1}
\begin{figure*}[t]
\begin{tabular}{cc}
    \centering
    \includegraphics[angle=0, width=8cm]{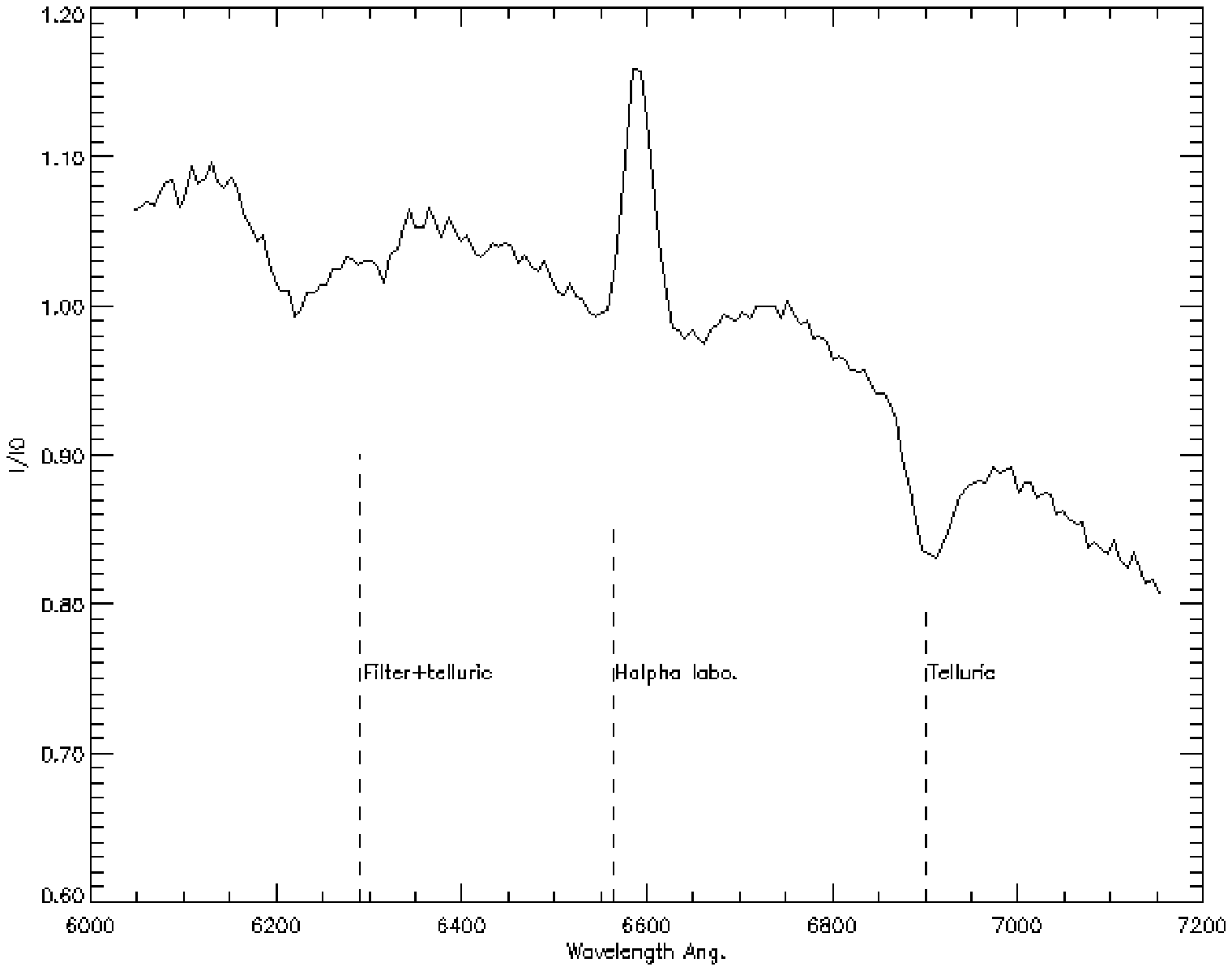} &  \includegraphics[angle=0, width=8cm]{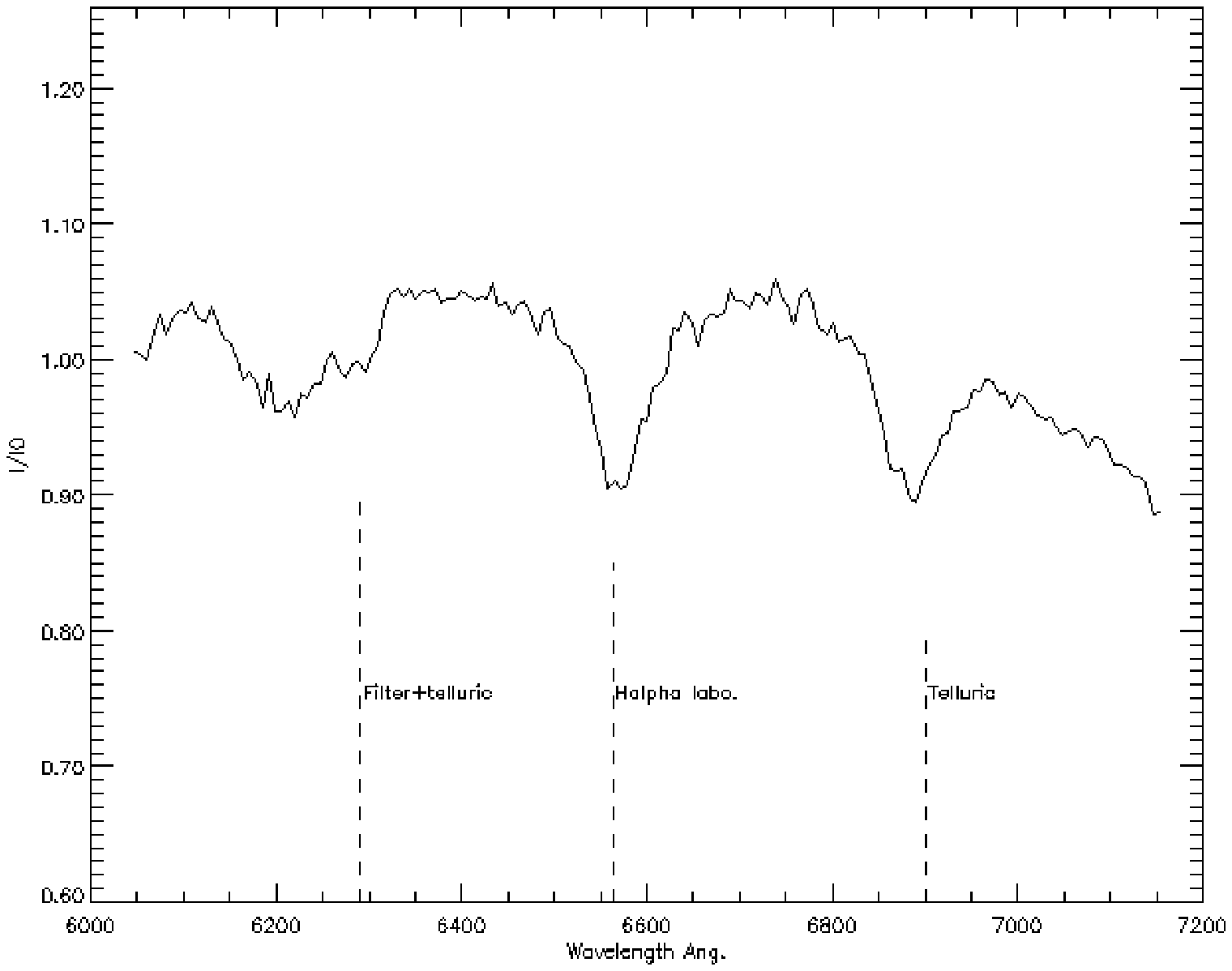} \\
    \includegraphics[angle=0, width=8cm]{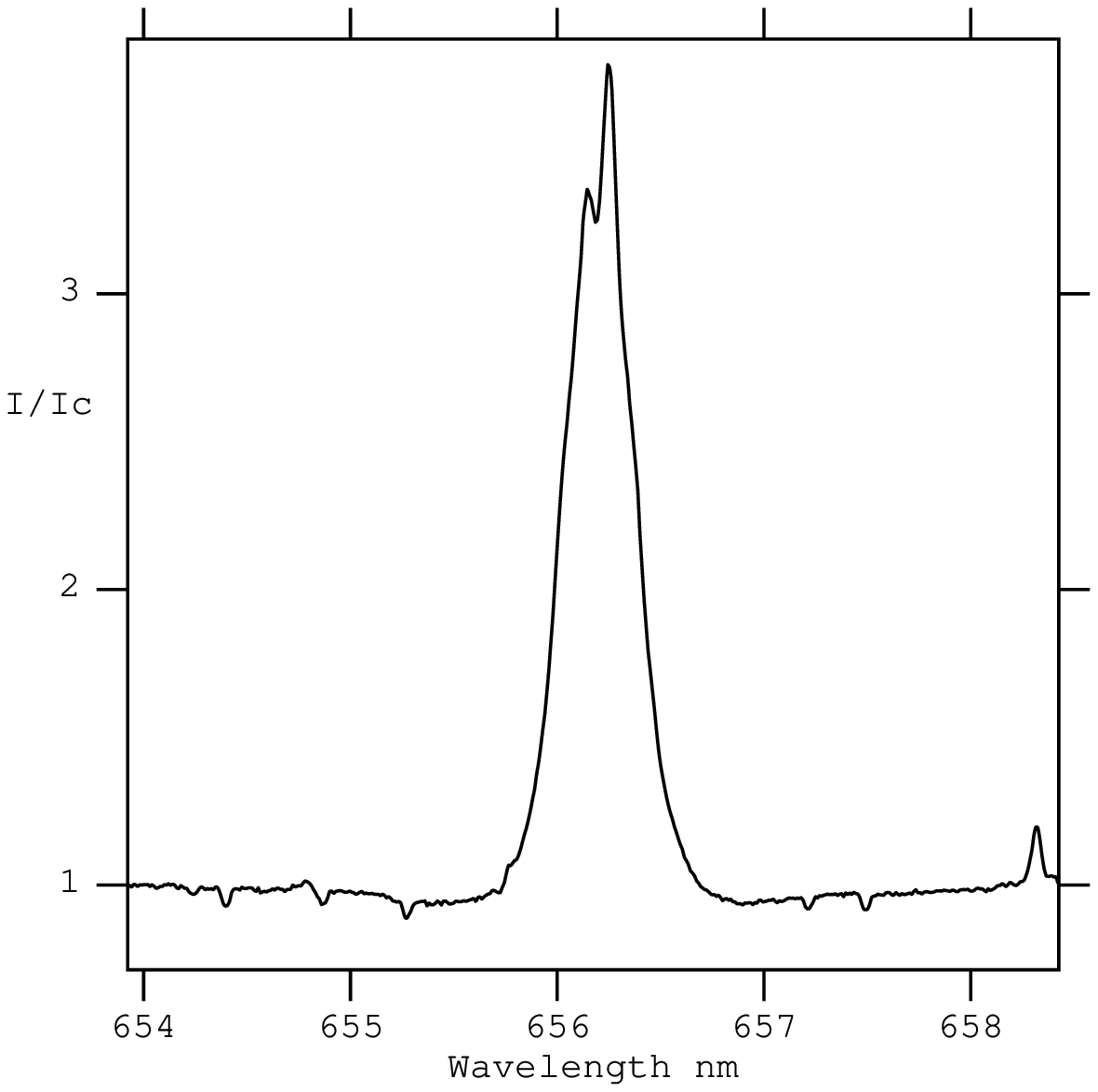} &  \includegraphics[angle=0, width=8cm]{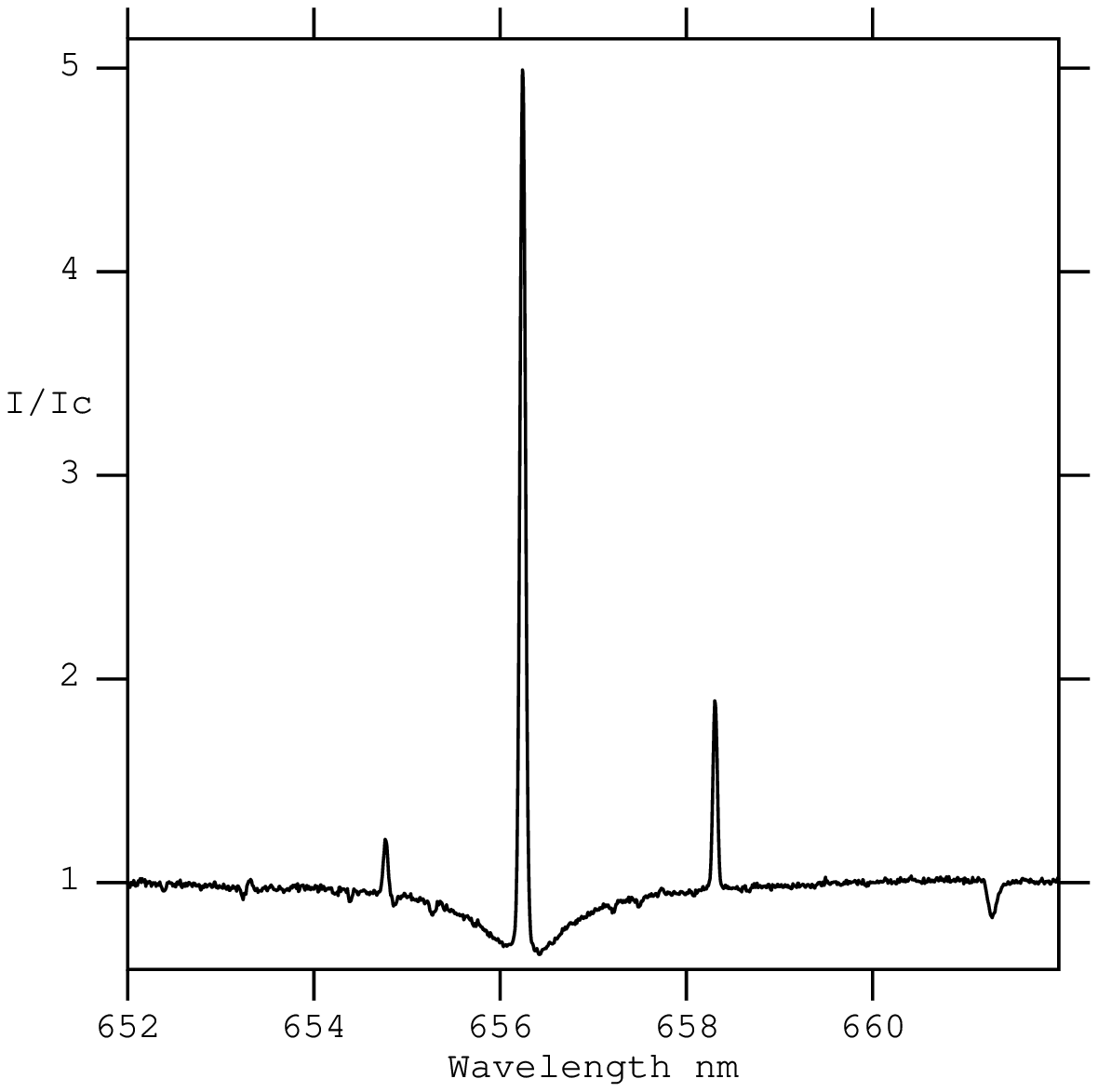} \\
\end{tabular}
    \caption{Left: H$\alpha$ spectra of the true ELS star W483. 
    Top-left: spectrum, scaled to the mean value of its
    continuum, obtained with the WFI in slitless spectroscopic mode. 
    Bottom-left: spectrum,  normalized to its continuum, 
     obtained with the VLT-GIRAFFE. In this last spectrum, the circumstellar and nebular
     H$\alpha$ emissions are visible.     
     Right: H$\alpha$ spectra of the star W371. 
     Top-right: spectrum, scaled to the mean value of its
    continuum, obtained with the WFI in slitless spectroscopic mode.
     Bottom-right: spectrum with nebular emission line, normalized to its continuum, obtained with the VLT-GIRAFFE.
     Only nebular emission is present.}
    \label{specWFI1}
\end{figure*}
%
%
%

\object{NGC\,6611} in \object{M16} is a very young cluster previously known to contain a
large number of ELS \citep{hillenbrand1993,dewinter1997}, which have been included in
the reference database SIMBAD and WEBDA. However, the investigation of the ELS character was made
from low and moderate resolution spectra, which did not allow to distinguish 
intrinsic stellar emission from nebular emission as noted by \citet{hillenbrand1993}. Using
deep objective prism spectroscopy, which is not sensitive to nebular lines, 
\citet{herbig2001} only identified a small number of ELS at the opposite of the other studies. 
This was recently confirmed by \citet{evans2005} who observed the more massive population of NGC 6611 with
high-resolution spectroscopy (FLAMES and FEROS at ESO). 
With the ESO-WFI in slitless spectroscopic mode and the VLT-GIRAFFE, we performed
observations from late O to early A type stars in NGC 6611 and surrounding fields, which are
thought to be still in formation stages \citep[see e.g.][]{indeb07}, in order to analyse the
B star population with and without emission lines.  In the present paper we report on the
detection of new ELS in the NGC 6611 region, we determined the fundamental parameters and studied: 
(i) the evolution of rotational velocities between pre-main sequence phase and main sequence, 
(ii) the age-distributions of objects in M16, 
(iii) the evolutionary status of each B-type star, as well as the nature 
of the emission/absorption stars: pre-main sequence stars or HAeBe (ELS), or main-sequence stars
or classical Be stars (ELS). 


\section{Observations/Reduction}

NGC\,6611 lies in the star-formation region of the complex Eagle nebula. 
The ESO-WFI in its slitless spectroscopic mode and the high-resolution VLT-FLAMES  spectroscopy 
allow us to check, on the case by case basis, if there is emission and if it is of nebular or circumstellar origin.


\subsection{ESO-WFI in spectroscopic mode}
\label{obsWFI}

\begin{table*}[t]
\centering
\caption[]{Comparison of candidate ELS or Be stars found in NGC\,6611 by different authors and in function
of the technique used. \citet[][and references therein]{hillenbrand1993}, \citet{dewinter1997},
\citet{evans2005} used slit and/or fibre spectroscopy, \citet{herbig2001}  used slitless spectroscopy, and in our
study we use slitless and fibre spectroscopy (WFI/GIRAFFE). ``Abs'' is for absorption line in H$\alpha$, ``Em'' is for
emission line in H$\alpha$, ``x'' for no spectrum (WFI or GIRAFFE) or for no exploitable spectrum
(WFI).}
\centering
\small{
\begin{tabular}{l@{\ }l@{\ }l@{\ }l@{\ }l@{\ }l@{\ \ \ \ \ \ \ }l@{\ }l@{\ }l@{\ }l@{\ }l@{\ }l@{\ \ \ \ \ \ \ }l@{\ }l@{\ }l@{\ }l@{\ }l@{\ }l}
\hline
\hline	
Stars & H93 & W97 & H01 & E05 & M07 & Stars & H93 & W97 & H01 & E05 & M07 & Stars & H93 & W97 & H01 & E05 & M07 \\
\hline	
WFI[N6611]017 & & &     &     &Em/Em &     W266 &      & Em  & cont&	 & x/x &	W374 &       & Em  & Abs &     & Abs/x  \\     
W031 &      &     &     &     & x/Em &     W267 &      & Em  & Abs & Abs & x/Abs  &	W388 &       & Em  & Abs &     & Abs/Abs \\    
W080 &      &     &     &     & x/Em &     W273 &      & Em  & Abs &	 & x/Abs  &	W389 & Em   &	   & Abs &     & Abs/Abs \\    
W112 & Em   &     & Abs &     & x/x  &     W280 & Em   &     &     & Abs & x/x &	W396 &       & Em  & Abs &     & Abs/x \\      
W138 & Em   &     & Abs &     & Abs/x &    W281 & Em   &     & Abs &	 & Abs/Abs &	W400 &       & Em  & Abs & Abs & Abs/Abs \\    
W181 & Em   &     &     &     & Abs/x &    W297 & abs  & Em  & Abs & Abs & Abs/x &	W455 &       & Em  & Abs & Abs & Abs/Abs \\    
W198 & Em   &     &     &     &  Abs/x &    W299 & Em   & Em  & Abs &	 & x/Abs &	W469 & Em   &Abs & Abs & Abs & Abs/Abs \\      
W202 &      & Em  & Abs &     & Abs/Abs &  W300 & Em   & Em  & Abs &	 & Abs/Abs &	W472 & Em   &	   & Abs & Abs & Abs/Abs \\    
W203 & Em   &     & Abs &     & Abs/Abs &  W301 &      &     &     & Em  & Abs/Em &	W483 &       & Em  &	 & Em  & Em/Em \\      
W205 &      &     & Abs & Em  & Abs/x &    W306 & Em   & Em  & Abs &	 & Abs/Abs &	W484 &       & Em  & Abs & Abs & x/Abs \\      
W207 & Em   & Em  & Abs &     & Abs/x &   W307 & Em   &      & Abs &	  & Abs/Abs&	W494 &       & Em  & Em  &     & Em/x \\      
W210 & Em   &     &     & Abs & Abs/x  &   W310 & Em   &      &     &	  & Abs/x  & 	W496 & Em   &     & Abs &     & x/x \\        
W213 &      & Em  & Abs &     & Abs/x &    W311 & Em   &      & Abs & Abs& Abs/x & 	W500 &       &     &	 & Em  & Em/Em \\      
W221 & Em   &     & Abs &     & Abs/x  &   W313 &	& Em  & Abs& Abs & Abs/Abs &	W503 & Em   &	   & Em  &  Em & Em/Em \\     
W227 & Em   &     & Abs & Abs & Abs/x &    W322 & Em   &      & Abs &	  & Abs/x & 	W504 &       & Em  & Abs& Abs & Abs/Abs  \\    
W232 &      &  Em &     &     & Abs/x   &  W323 & Em   &      & Abs &	  & Abs/Abs &   W525 &       & Em  &	 &     & x/x \\        
W235 & Em   &     & Em  & Em  & Em/Em  &  W336 &	& Em  & Abs & Abs & Abs/Abs &	W536 &       & Em  &	 & Abs & Abs/Abs \\    
W243 & Em   &     & Abs &     & Abs/Abs &  W339 &	& Em  & Abs &	  & x/x  &	W601 & Abs  &	   &	 &     & Abs/Em \\     
W245 &      & Em  & Em  &     & Em/x  &   W351 & Em   &      & Abs & Abs & Abs/Abs  &   W611 &       & Em  &	 &     & x/x \\        
W251 & Em   &     & Abs &     & x/Abs &    W371 & Em   &      & Abs &	  & Abs/Abs&	W617 &       & Em  &	 &     & Abs?/x \\     
W262 & Em   &     & Abs &     & x/x &           &&&&&						     &&&&& \\
\hline
\multicolumn{18}{l}{H93 = \citet{hillenbrand1993}; W97 = \citet{dewinter1997}; H01 = \citet{herbig2001}; }\\
\multicolumn{18}{l}{E05 = \citet{evans2005}; M07 = this study (slitless/slit)}\\
\end{tabular}
\label{nbreEm}
}
\end{table*}

The Wide Field Imager (WFI) is attached to the 2.2m MPG telescope at La Silla.
Observations\footnote{ESO run 69.D-0275A}  with WFI were obtained on September 26, 2002. 
The WFI has a field of view of 33\arcmin~$\times$~34\arcmin. We used the R50 grism and
the Rc162 filter centered on H$\alpha$ with a bandpass of 200 nm and a resolution lower
than 1000 \citep[see][]{WFI}.  The two images were obtained with respectively 120 and 450s 
exposure times, which allow to detect the continuum of faint sources (V$\sim$19). 
However, we only need brightest stars (up to V$\sim$15), which represent  the OB-type population in the 
formation region; their spectra have sufficient signal to noise ratio to detect reliably H$\alpha$ emission.
Note that the first image (120s exposed) is used to study the brightest stars, which are saturated 
in the second more exposed image of our data.

This special slitless mode is sensitive neither to the
nebular lines nor to the weak circumstellar emission. However, it is easy to detect ELS
with sufficient circumstellar emission in H$\alpha$. The image of the field of NGC\,6611
($\alpha$(2000)~=~18h18min42s, $\delta$(2000)~=~-13$^{\circ}$46\arcmin57.6\arcsec) in
slitless spectroscopic mode is shown in Fig.~\ref{WFIN6611}. 

The image reduction was performed with IRAF\footnote{IRAF is distributed by the
National Optical Astronomy Observatories, which is operated by the Association of
Universities for Research in Astronomy (AURA), Inc., under cooperative agreement with the
National Science Foundation.} tasks and the MSCRED package. The extraction of spectra was
performed using the SExtractor code \citep{bertin1996} with special adapted
convolution masks  (provided by E. Bertin). About 15000 spectra have been
extracted. Spectra of certain stars in the field are not available because they
fall in the interCCDs spaces. Also note that certain spectra are contaminated by cosmic
rays, by superimposed zeroth orders of other spectra or/and by other spectra. 

The analysis of extracted spectra was performed individually with our IDL-based
code (lecspec4). Examples of slitless spectra for one ELS with a circumstellar emission
line at H$\alpha$ (W483) and for one non-ELS (W371) are shown in Fig.~\ref{specWFI1} left and
right respectively. A rough wavelength calibration, based on the theoretical
dispersion law of the grism and Rc filter, is provided for the spectra (see
Fig.~\ref{specWFI1}. Astrometric calibration was performed with the
ASTROM package \citep{wallace2003} and refined with the UCAC2 catalogue. 

We explored the NGC 6611 field towards the 19th V magnitude; several faint stars, like 
\object{[OSP2002] BRC 30 4} from \citet{ogura02} (who used the same instrumentation than \citet{herbig2001}), 
exhibit H$\alpha$ emission. These targets are not B-type objects, but TTauri stars as explained 
by \citet{ogura02} and \citet{herbig2001}. 
However, we concentrate our study only on the stars with V magnitude up to 15 for our purpose. 

A short list of 7 H$\alpha$ ELS was finally obtained with WFI and are indicated in Table~\ref{nbreEm}; 
6 of them are in common with those recently confirmed/discovered 
by \citet{herbig2001} and \citet{evans2005}, 1 ELS is a new detection. 
Note that ELS with weak emission (like W601 or W205) cannot be detected with such an instrumentation.

Similar observations were obtained for the open cluster \object{Westerlund 1}
in a field centered on ($\alpha$(2000)~=~16h46min47.9s, $\delta$(2000)~=~-45$^{\circ}$48\arcmin59\arcsec).
The exposure times for the 2 images are 120 and 900s. 
We found several ELS previously known as Wolf-Rayet or Be stars. 
However, the spectra extracted are difficult to exploit due
to the very faint luminosity of the stars.
We mention here the existence of these observations to whom it may interest.


\subsection{VLT-GIRAFFE}

This work also makes use of spectra obtained with the multifibre spectrograph
VLT-FLAMES \citep{pasquini02} in Medusa mode (131 fibres) at medium and high resolutions.
Observations with GIRAFFE\footnote{ESO run 73.D-0133A} were carried out on April 14-15,
2004 in the young cluster NGC\,6611 and in its surrounding field, as part of the Guaranteed
Time Observation programmes of the Paris Observatory (P.I.: F. Hammer).  The observed
field (25\arcmin~in diameter) is centered at $\alpha$(2000)~=~18h18min50s and
$\delta$(2000)~=~$-13^{\circ}$49\arcmin30\arcsec. 

The HR15N setup (647--679 nm, R=17000) was used to identify H$\alpha$ ELS, to study the
H$\alpha$ characteristics, and to determine the interstellar reddening of each target.
The LR3 (450.1--507.8 nm, R=7500) setup was used for the H$\beta$ line characteristics 
and the LR2 (396.4--456.7 nm, R=6400) setup for the determination of fundamental
parameters. Two consecutive spectra were obtained in each setup and summed up. 
The exposure time was 2$\times$1800s, 2$\times$1000s, and 2$\times$900s 
in the HR15N, LR3, LR2 setups, respectively.

The VLT-GIRAFFE targets were pre-selected from our WFI catalogue of ELS and from 
\citet{hillenbrand1993}, \citet{dewinter1997}, and \citet{herbig2001}. OBA stars without
emission were also pre-selected from CDS/SIMBAD. The location of observed stars with
the VLT-GIRAFFE are shown in Fig.~\ref{loc6611}.

Bias, flat-fields and wavelength calibration (ThAr) exposures were obtained for each
stellar exposure and used to reduce the spectra. The data reduction was performed with
the dedicated software GIRBLDRS developed at the Geneva Observatory (see
http://girbldrs.sourceforge.net) and with several tasks of the IRAF package for
extraction, calibration and sky correction of the spectra. The heliocentric correction
(-28 \kms) at the epoch of observations was applied to each stellar spectrum.

We note that almost all the spectra exhibit a narrow emission-line component onto the
core of H line profiles with a Full Width at Half Maximum (FWHM) close to the spectral
resolution. This component is due to the ambient nebulosity, which also produces many
other nebular lines visible in the spectra, such as [SII] and [NII] lines.


\section{Results}

\subsection{Comparison between WFI and VLT-GIRAFFE spectroscopy in NGC\,6611}

In Fig.~\ref{specWFI1}, we compare the spectra obtained in
the WFI slitless spectroscopic mode and the VLT-GIRAFFE high resolution spectroscopy.
In Fig.~\ref{specWFI1}-left, we show the example of the star W483 with a ``true''
circumstellar emission line in H$\alpha$, visible with the two
instrumentations. We also show in Fig.~\ref{specWFI1}-right the star W371, a Be star
according to \citet{hillenbrand1993}. No emission is however detected with the
WFI-spectro for this star. With the VLT-GIRAFFE observations, we found only a
nebular emission (small FWHM) and no circumstellar emission. 

Finally, in Table~\ref{nbreEm} we compare the ELS found in NGC\,6611 
depending on the techniques and studies. The presence of strong nebular
emission could explain the high number of \emph{false} Be stars
previously observed in this cluster with low resolution slit or fibre
spectroscopy by several authors.

\subsection{Spectroscopy of circumstellar ELS}

With the VLT-GIRAFFE, we found 9 \emph{true} ELS: 3 of them (WFI[N6611]017, W031,
W080) are newly detected ones; 6 of them (W235, W301, W483, W500, W503, W601)
pre-selected from previous studies \citep{hillenbrand1993,dewinter1997,herbig2001}
or recently discovered \citep{evans2005,alecian08} are confirmed. 

We measure the equivalent width of the H$\alpha$ emission line (EW$\alpha$)
as a first clue to determine their nature. Note that several new ELS show
faint EW$\alpha$, which could only be detected with a large telescope such as the
VLT and a high spectral resolution such as the one on the HR15N setting of
GIRAFFE. Results are given in Table~\ref{paramELS}. EW$\alpha$ have been corrected
from the nebular contribution, in the same way as we proceeded for Be stars in the
Magellanic Clouds \citep[see][]{marta06a}. The H$\alpha$ line of the 9 ELS with
circumstellar (CS) emission observed with VLT-GIRAFFE is shown in
Figs~\ref{Hastar017} to \ref{HaW601}. The WFI spectrum is also shown when
available. Unfortunately, the slitless spectrum was saturated for W235, W503 and
W601. The stars WFI[N6611]017, W235, W483, W500, and W503 also show a CS emission
in the H$\beta$ line.

\subsection{Fundamental parameters of OBA and ELS stars}
\label{FPD}

As in \citet{marta2006b,marta2007a} we make use of the GIRFIT least-square
procedure \citep{fremat2006} to derive the fundamental parameters: effective
temperature (\teff), surface gravity (\logg), projected rotational velocity
(\vsini), and radial velocity (RV). This procedure fits the observations with
theoretical spectra interpolated in a grid of stellar fluxes computed with the
SYNSPEC programme and from model atmospheres calculated with TLUSTY \citep[][and
references therein]{hubeny1995} or/and with ATLAS9
\citep{kurucz1993,castelli1997}. The grid of model atmospheres we used to build
the GIRFIT input of stellar fluxes was obtained for the metallicity of the Milky
Way (MW). 
For a more detailed description of the grid of
model atmospheres and of the fitting criteria we adopt in the GIRFIT procedure, we
refer the reader to Sect. 3 of \citet{marta2006b}. Resulting fits and the
corresponding residuals obtained for the CS ELS are depicted  in
Figs~\ref{Hastar017} to \ref{HaW601} (upper panel). In the residuals several
structures appear: \ion{Fe}{ii} emission lines at 4233, 4173-78, and
4385 \AA\ are present in W235 and W500. However, in W483 and W503 false
structures appear,  e.g. false P Cygni lines, probably due to the uncertainty
on \rv\ used in the fit (\rv\ is determined at $\pm$10\kms).

Finally, we determine the spectral classification of each star with a method based
on the fundamental parameters. The calibration we established to estimate these
spectral types is described in \citet{marta2006b}. 

Early-type stars are listed in Table~\ref{tabPFNGC6611}, sorted by ID number from
the literature or Simbad \citep[WXXX, from][]{walker1961} or from our
WFI-catalogue (WFI[N6611]XXX). The fundamental parameters \teff, \logg, \vsini,
and \rv\ obtained by fitting the observed spectra, as well as the spectral
classification deduced from \teff--\logg~plane calibration \citep[CFP determination][]{marta2006b} 
are reported in  this Table. The classification
provided by \citet{evans2005} for some of our targets observed with the same
instrumentation is also given.

\addtocounter{figure}{+1}

\begin{figure}[htpb]
    \centering
    \resizebox{\hsize}{!}{\includegraphics[angle=-90]{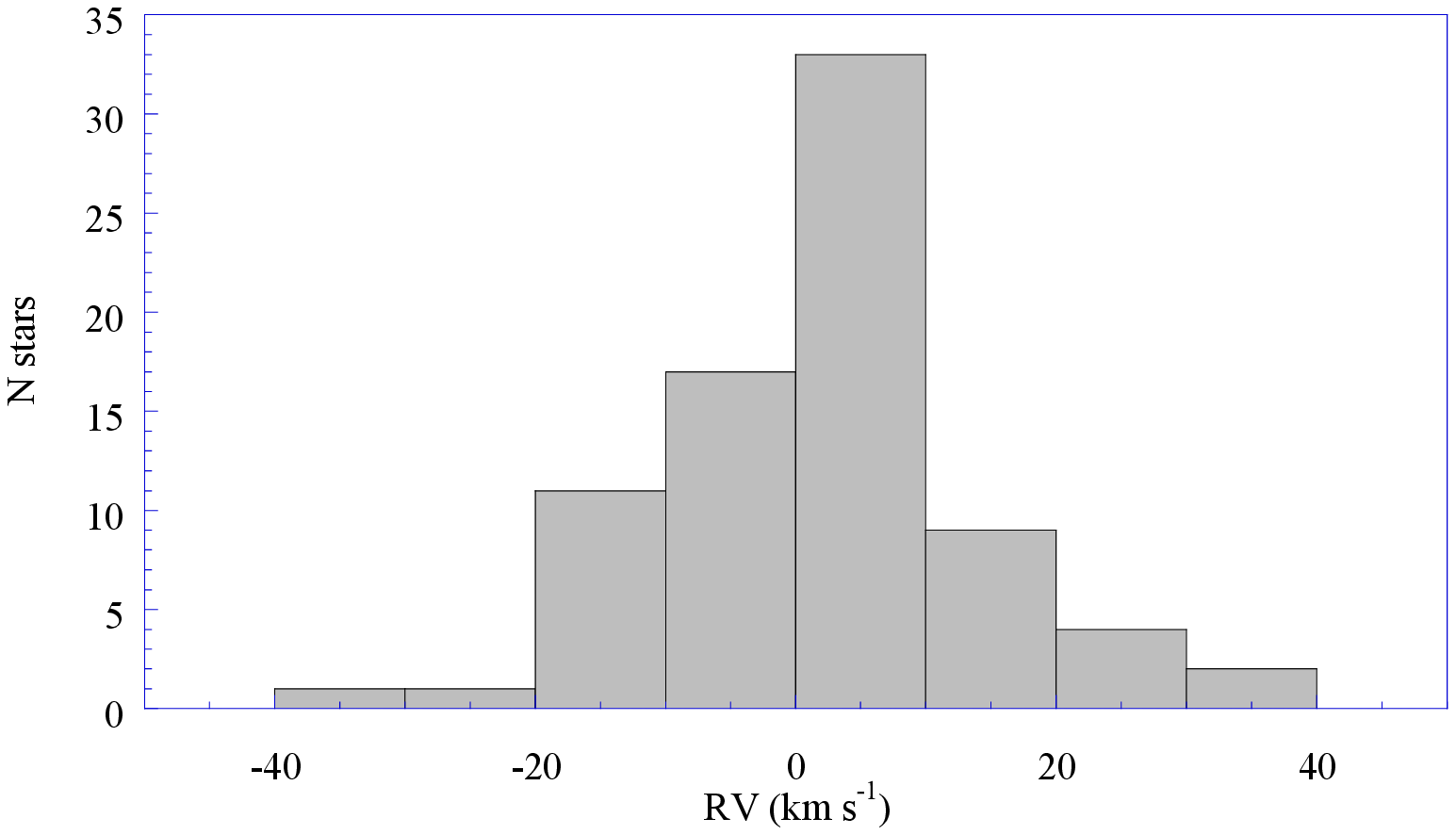}}    
    \caption{Distribution of stellar radial velocities. The typical error-bar is
$\pm$10 \kms.}
    \label{rvstar}
\end{figure}

We compare the fundamental parameters determined by \citet{dufton2006} with ours
for the 14 stars we have in common ((this study - \citeauthor{dufton2006})/this
study $\times$ 100 in percents).  We found on average a difference of 0.1\%
(average: 19946 K) for \teff,  4.1\% (average: 4.1 dex) for \logg, and 9.5\%
(average: 179 \kms) for \vsini. 

We note that the distribution of \rv\ shown in Fig.~\ref{rvstar} is similar to the
one determined by \citet{evans2005} for this cluster.

To determine whether the stars are members or not of the open cluster NGC 6611,
we used the membership probabilities from \citet{ref1} and \citet{ref2}. We considered that 
the star is member of NGC 6611 if the average of the 2 membership probabilities is higher than 50 \% or
the probability is higher than 50 \% if only one of the two probabilities is present. The membership probabilities
are reported in Table~\ref{preuvePMS} and are used in the following sections and figures.

\begin{figure}[h]
    \centering
    \resizebox{\hsize}{!}{\includegraphics[angle=-90]{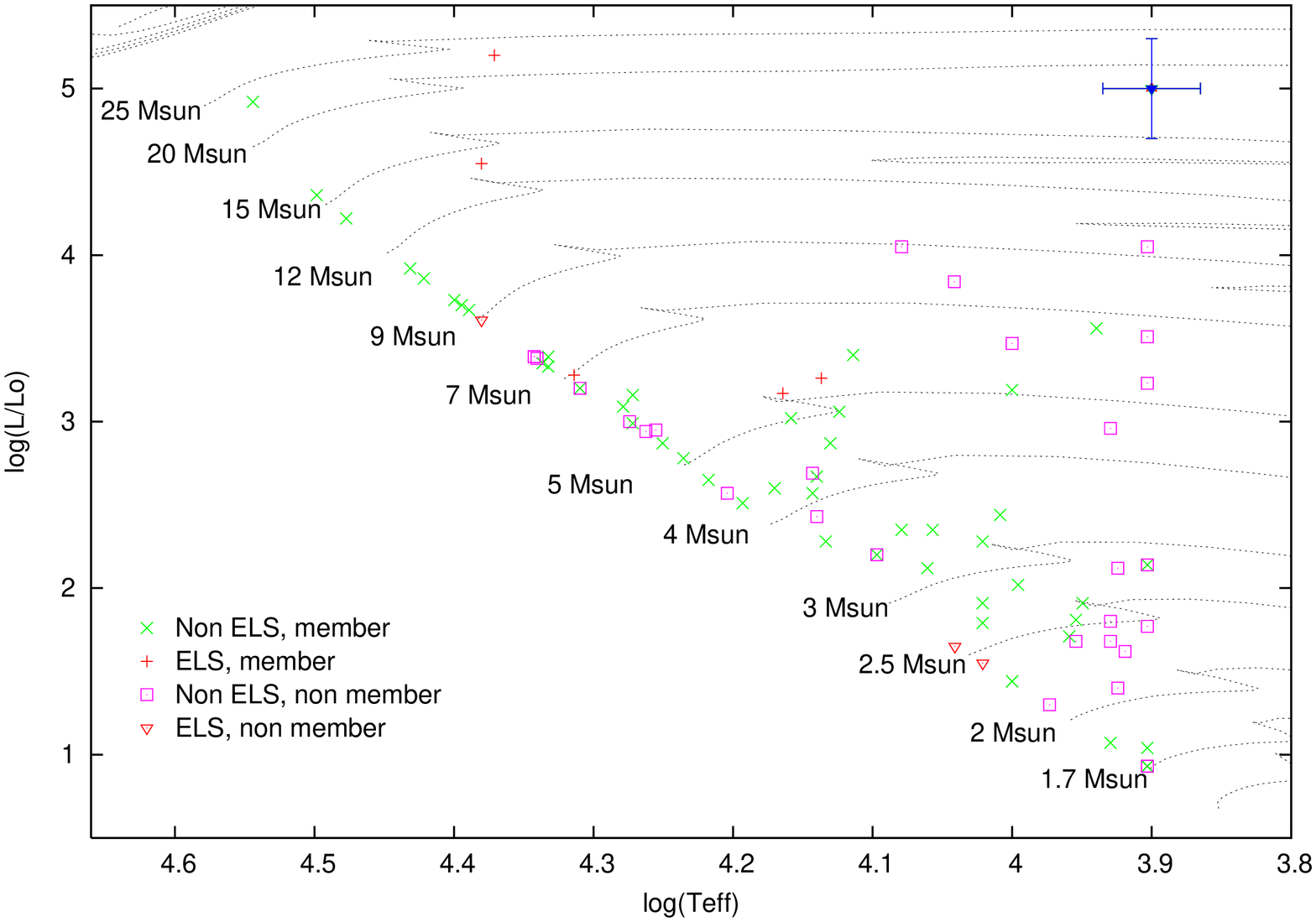}}
    \caption{HR diagram. Red '+' are for the ELS members of NGC\,6611, red triangles are for the ELS  non-members
    of the open cluster. Green 'x' are for the non-ELS members of NGC\,6611, and the
    pink squares are for the non-ELS  non-members of the open cluster. The
    tracks for Z=0.020 come from \citet{schaller1992}.
    The star is considered as member of the cluster if the average of membership
    probabilities is higher than 50\% or the membership probability higher than 50 \% 
    in case of only one is present, for more detail about the membership see Sect.~\ref{FPD}.}
    \label{HRNGC6611}
\end{figure}

To derive the luminosity, mass, age, and radius of OBA stars from their
fundamental parameters, we interpolated in the HR-diagram grids
\citep{schaller1992} calculated for MS stars without rotation at the solar
metallicity (Z = 0.020). The obtained luminosity, mass, radius, and age of most
OBA stars of the sample are given in Table~\ref{tabinterpB}. The masses are in
good agreement with the calibration of \citet{huang2006} for B-type stars at the
ZAMS. The position in the HR diagram of each observed star, members of the
cluster and non-members, is shown in Fig.~\ref{HRNGC6611}.

This figure shows that the most massive non-emission B-type stars (5$<$ $M/M_{\odot}$
$\le$25) are located at the ZAMS, while most of the less massive stars (2$\le$
$M/M_{\odot}$ $\le$5) are above the ZAMS. The latter may appear evolved and too
old for this region, but are in fact probably PMS stars, which are going to reach
the ZAMS.

\subsection{Binaries}

We identify binaries in our sample: (i) by cross-correlation with the study
of \citet{duchene2001} of visual binaries detected with adaptive optics; and (ii)
by using the radial velocity discrepancies between our measurements and those of \citet{evans2005}. 
We define two categories, first with RV discrepancies between 10 and 15 \kms, 
second with RV discrepancies higher than 15 \kms.

We confirm the binary nature of W025, W125, W175, W188, W243, W267, W275, W299,
W313, W343, W364, W400, W472, and W536 already detected by \citet{duchene2001} or
\citet{evans2005}. Moreover, we detected 8 other possible binaries: W161, W239,
W409, W444, W469, W473, W503, and W582. Among them, W503 is an ELS and W343 is a
SB2 with 2 B-type components. The indications of binarity from spectroscopy are
reported in Table~\ref{tabPFNGC6611}.

\citet{evans2005} found in their sample a proportion of $\sim$10\% of binary B-type stars, 
based on the \rv\ scatter in their
observations. We find a ratio of binaries equal to 22\% of our entire
sample and 27\% of the B-type objects, which is in good agreement with the
proportion reported by \citet{porter2003}. 
The ratio of binaries among our limited
sample of A-type objects is equal to 8\%, certainly underestimated.

We searched for lightcurves for the binaries as well as the ELS in our sample in
the ASAS database \citep{asas}. Unfortunately when photometry is found, the time-sampling, time coverage, and
the quality of the data do not allow to study reliably the potential periodicity of the variations.
The best lightcurve is the one of W503, which shows 3 eclipses without possibility to find a periodicity.


%
\begin{figure}[!t]
    \centering
    \resizebox{\hsize}{!}{\includegraphics[angle=-90]{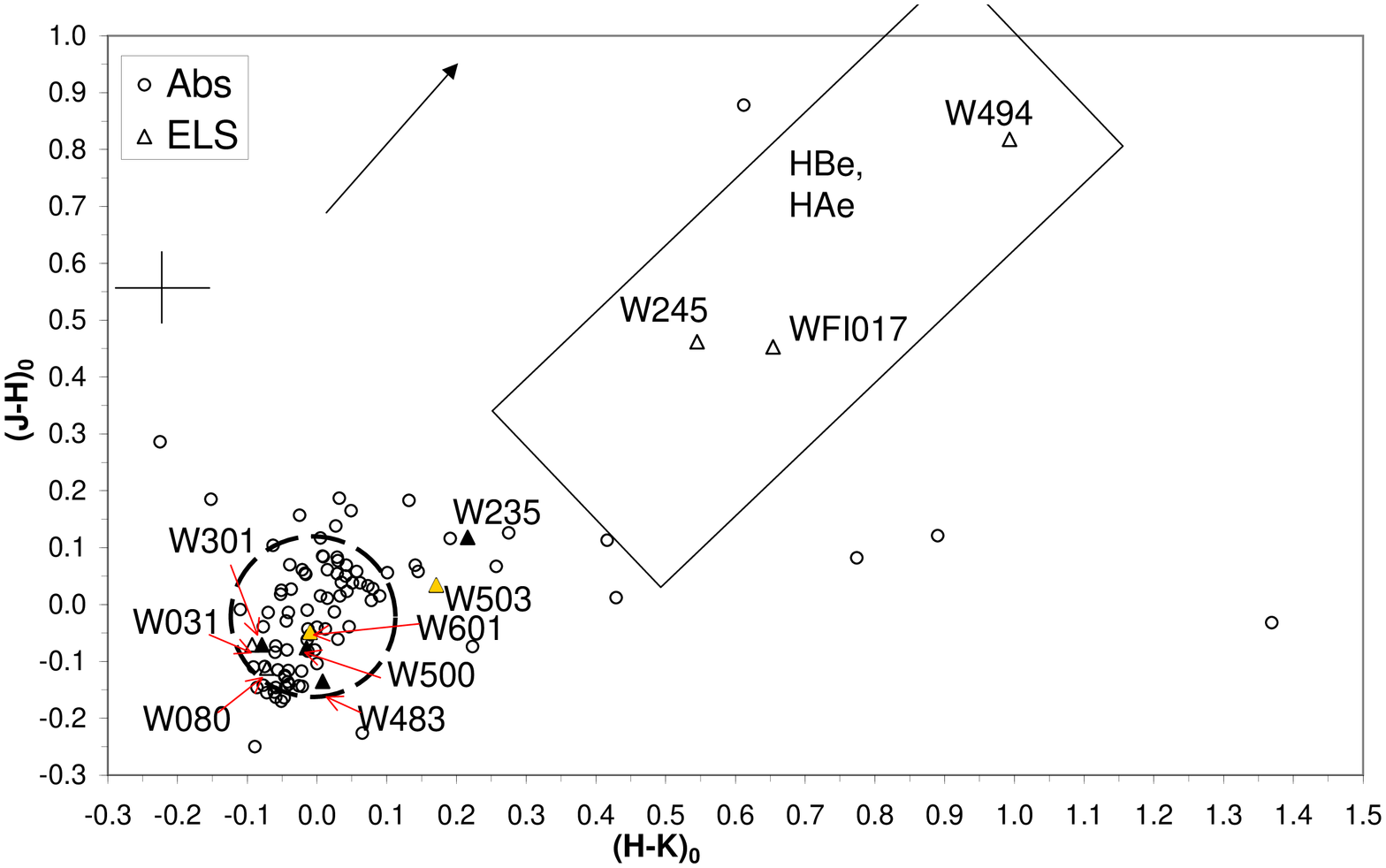}}
    \caption{Colour-coulour diagram based on 2MASS magnitudes corrected from the local
    extinction for the stars observed in NGC\,6611 and in the Eagle Nebula. 
    The large box shows the area of Herbig HAe/Be stars and 
    the small dashed ellipse shows the area of classical Be stars,
    according to \citet{hernandez2005}. Note that several Herbig Ae/Be and PMS
    stars can exist in the area of classical Be stars. ELS are indicated by a
    triangle, filled in black if the star is a member of NGC 6611, filled in yellow 
    if the membership is uncertain, and empty if the star is not a member. 
    WFI017 is for WFI[N6611]017. The open circles are for
    non-ELS (without indication of the membership to NGC 6611). The cross in the  
    top left corner indicates the mean error, and the arrow indicates the reddening vector.
    For more details about the membership see Sect.~\ref{FPD} and Sect.~\ref{EMdiscussion}.} 
    \label{IR2MASSgraphe}
\end{figure}
%

%
%

%
%

\subsection{Infrared study}
\label{IRS}

We search for infrared (IR) excess in each star, which could be attributed to a
circumstellar environment. First, with the \citet{herbig1975} calibration of
interstellar extinction based on the equivalent widths of diffuse interstellar
absorption bands at 4430 \AA\ and 6613 \AA, we  determine the local E(B-V)
for each star with or without emission lines. We also consider the two stars
W245 and W494 reported as ELS by \citet{herbig1975}  and confirmed by our WFI
spectra but for which no GIRAFFE spectra are available. For these 2 stars we use a
mean E(B-V) value.

Then, we search for their J, H, K-band magnitudes in the 2MASS 
catalogue and 3.6, 4.5, 5.8 and 8.0 $\mu$m  magnitudes in the GLIMPSE
archive of SPITZER (release: spring07, 06/12/2007). We determine the colour
indices (J-H)$_{0}$ and (H-K)$_{0}$ (see Table~\ref{IR2MASS}) as well as the
colour indices (3.6-4.5)$_{0}$  and (5.8-8.0)$_{0}$. The 2MASS colour-colour
diagram is shown in Fig.~\ref{IR2MASSgraphe}. We marked off in this graph the area
defining classical Be stars  and Herbig Ae/Be stars with a strong IR
excess, as defined by \citet{hernandez2005}.

The figure~\ref{IR2MASSgraphe} shows that several stars have a strong IR excess in 2MASS and lie in or
close to the area of PMS stars or Herbig Ae/Be. This is the case for the ELS
WFI017, W245, and W494, which fall into the HAeBe area.
Note that Herbig Ae/Be stars could also have weak IR excess and fall in the classical Be star area 
as it could be shown by using data from \citet{vieira03} and \citet{the93} for example. 
Note also that between the area of MS stars and HAeBe, classical Be as well as HAeBe could also lie 
as it could be shown by using data from \citet{vieira03} and \citet{the93} for HAeBe and from 
\citet{currie07} for classical Be stars.
However, the star W235 seems to have an IR excess stronger than those of classical Be stars 
and must be considered as an HBeAe.
The excess for W503 is probably due to the disk of the binary.
For the other ELS (W031, W080, W301, W483, W500, W601), this Fig.~\ref{IR2MASSgraphe}
indicates a possible MS status.
Note that several non-ELS have also a strong IR excess, which indicates a possible PMS nature.
This is the case for W026, W299, W568, and 4 others with a very strong excess: W307, W409, W596, W627, W633.
Note that the strong reddening of W299 and the medium excess of W400 could also be explained by their binary nature.

%

%
\begin{figure}[!t]
    \centering
    \resizebox{\hsize}{!}{\includegraphics[angle=-90]{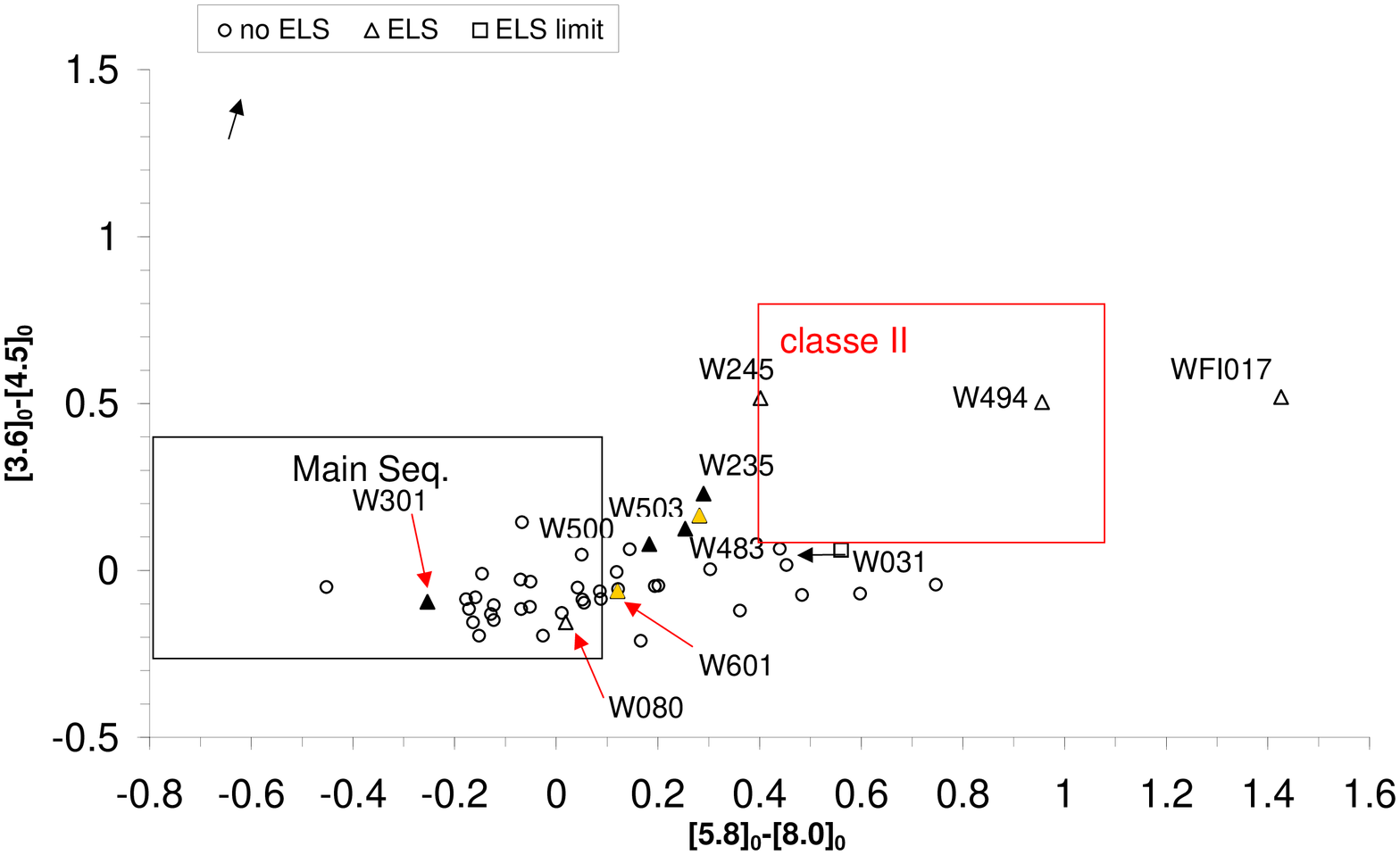}}
    \caption{Colour-coulour diagram from IRAC/SPITZER magnitudes corrected from
    the local extinction for the stars observed in NGC\,6611. The red box is the
    area of class II objects following \citet{allen04}. The black box is the area of MS stars.
    Note that the boxes are not dereddened here.
    ELS are indicated by a triangle, filled in black if the star is a member of NGC 6611, 
    filled in yellow if the membership is uncertain, and empty if the star is not a member.
    Non-ELS are indicated with a circle (without indication of the membership to NGC 6611). 
    W031, which does not have a measured  magnitude at 8.0 $\mu$m but only the 
    limiting magnitude of the GLIMPSE archive is noted with a diamond. Note
    that its location  is a limit and could move to the left. 
    The arrow (in the top left corner) indicates the reddening vector.
    For more details about the membership see Sect.~\ref{FPD} and Sect.~\ref{EMdiscussion}.}
    \label{IRACgraphe}
\end{figure}
%

%
%

We also used the SPITZER colour-colour diagram to investigate the nature of the
ELS stars (see Fig~\ref{IRACgraphe}). In this graph the class II area,
attributed to young stars with an accretion disk (PMS stars) as defined  by
\citet{allen04}, is indicated as well as the MS area.
The star W494 falls in the class II area, W245 is at its limit.
According to the lower limit of its (5.8-8.0)$_{0}$ colour index, 
the ELS star W031 could also be at the limit of the class II area.

The star WFI017, close to the centre of the HAeBe area in the 2MASS colour-colour
diagram falls outside the PMS star domain (class II area) in the SPITZER 
colour-colour diagram, sharing the domain of very young objects (class I area,
\citet{allen04}); this optical source might then be a young class II source
projected onto a locally dense cloud which could explain its abnormal excess in
the (5.8-8.0)$_{0}$ colour index.
According to \citet{indeb07}, WFI17 has an accretion disk.
W235, W503, W500, W483 have a mid-IR excess and fall in the area between MS and 
class II stars. W301 seems to be clearly a MS star, while W080 and W601 are 
at the edge of MS area to intermediate area between MS and class II areas.
The non-ELS W026, W239, W400, W444, W473, W520, W536 have also a strong infrared excess
and are close to the area of class II.

As it could be shown by using data from the literature (see above for the references) for HAeBe and for classical Be stars
CBe as well as HBeAe and PMS stars could fall  in the intermediate area between class II and MS.
See also \citet{sic06} who shew that stars with weak emission could also fall in MS and in the intermediate area between class II and MS.

The indication of infra-red excess is reported for each star in Table~\ref{preuvePMS}.

\begin{figure}[t]
    \centering
    \resizebox{\hsize}{!}{\includegraphics[angle=-90]{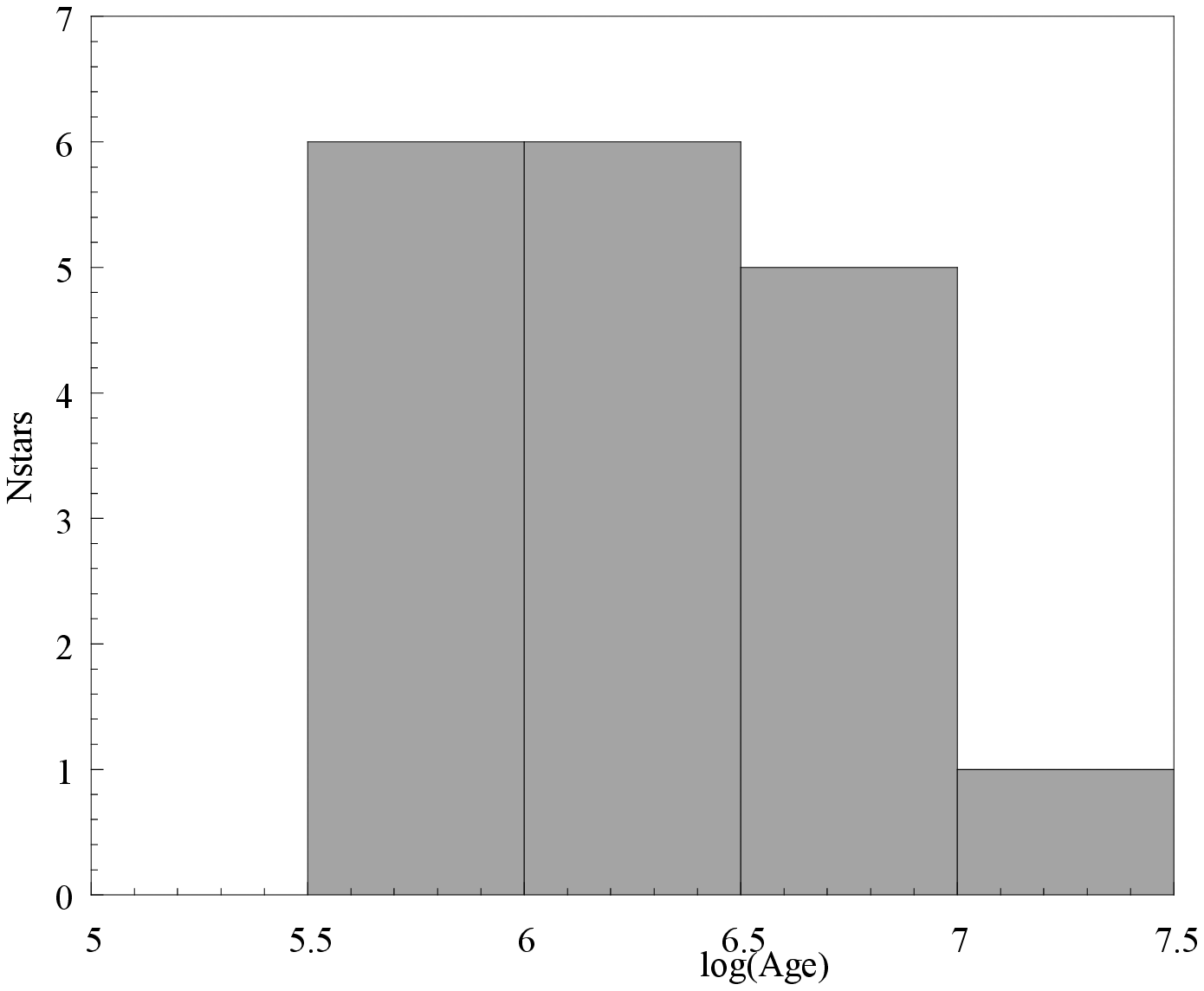}}
    \caption{Distributions of stellar ages for MS members of NGC 6611.}
    \label{distages6611cl}
\end{figure}
\begin{figure}[t]
    \centering
    \resizebox{\hsize}{!}{\includegraphics[angle=-90]{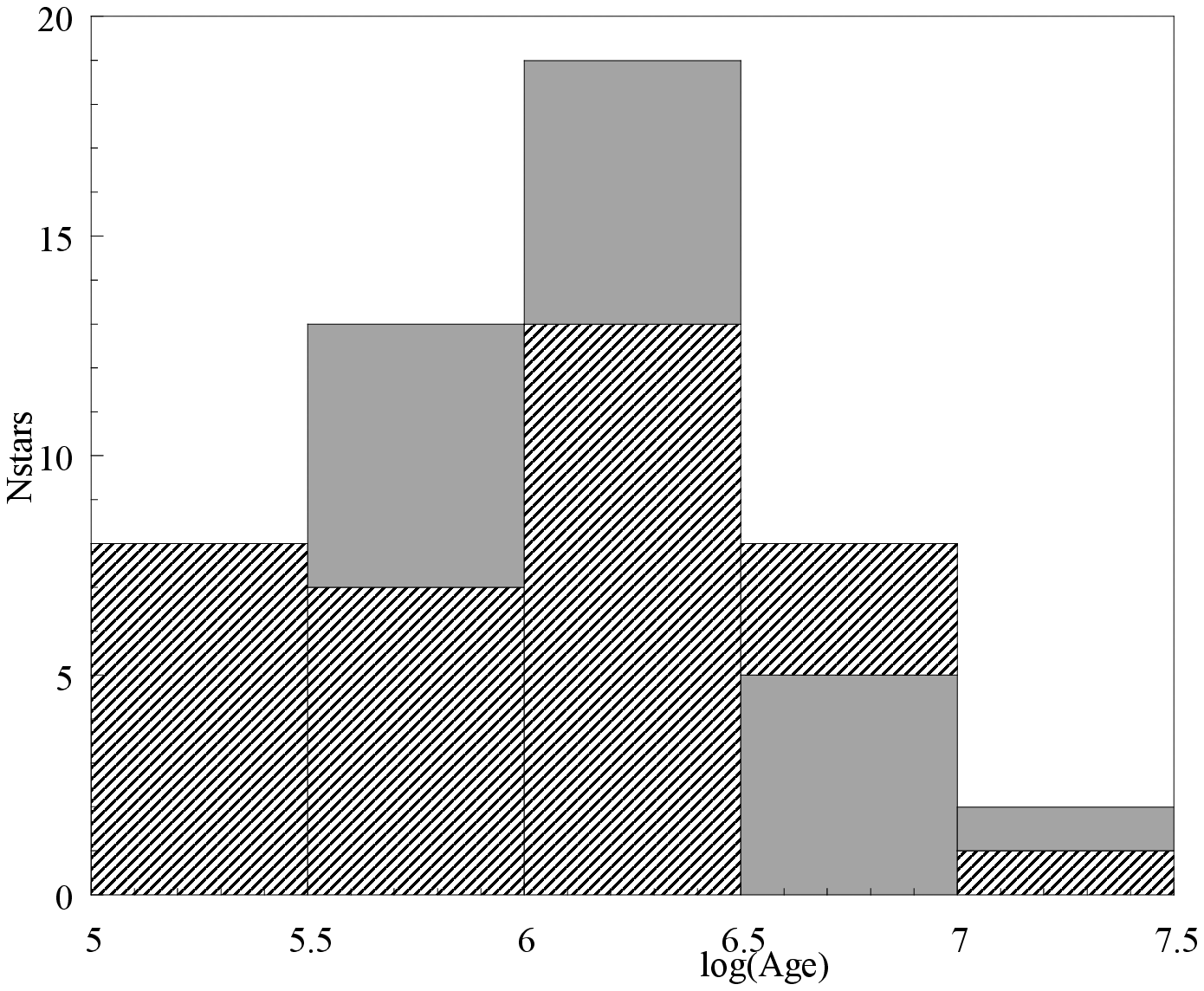}}
    \caption{Distributions of stellar ages for PMS (cross-hatched bars) + MS (filled grey bars) members of NGC 6611.}
    \label{distages6611cl2}
\end{figure}

\subsection{Age of the cluster NGC\,6611}
\label{age}

We obtained the age of NGC\,6611 by using gaussian fitting values of the
age-distributions of the stars and PMS objects belonging to this cluster, as shown in Figs.~\ref{distages6611cl} 
and \ref{distages6611cl2}. For this purpose, we only take into account the stars that have their average membership probability
higher than 50\% (see Table~\ref{preuvePMS} and Sect.~\ref{FPD}). We also determine the age by merging the
PMS objects with MS stars.

We identified the potential PMS and MS stars (see Table~\ref{preuvePMS}) 
by using indications provided by IR data or from the fact that their age derived
from the MS tracks seems too old for the members of the cluster. 
Then, we derive the age of PMS stars by interpolation
in the PMS tracks given by \citet{iben1965} and/or from \citet{palla1993},
while for MS stars, as in \citet{marta2006b,marta2007a}, we determine the age of
each  star by interpolation in HR tracks of MS stars \citep{schaller1992}. The
distribution of stellar ages of MS stars in the cluster is displayed in
Fig.~\ref{distages6611cl} and for the PMS+MS in Fig.~\ref{distages6611cl2}.
The age finally retained for each star of the global
sample is reported in Table~\ref{preuvePMS}. Then, we consider that the age of the
cluster has to be derived only from the MS stars members of the cluster,  i.e.
the more massive stars, which are close to the ZAMS, or from the MS stars with PMS objects merged. 
We find log(age NGC\,6611)=6.25$\pm$0.30 (i.e. 1.78 Myears) from MS stars, and 6.08 $\pm$0.25 (i.e. 1.20 Myears) from MS+PMS stars
in good agreement with the log(age)=6.0--6.5 recently determined by \citet{dufton2006} 
or with log(age)=6.11 (1.3 Myears) by \citet{bonatto06}. Note that the age
determined here is younger than the previous estimates from the photometric
studies (6.882, see WEBDA). 


\addtocounter{table}{+4}

\begin{table}[th!]
\centering
\caption[]{Comparison of the mean rotational velocities for PMS and MS
stars. The mean \vsini, the median \vsini, and the statistical
error are given in cols. 3, 4, and 5, respectively. In the upper part of the
table, we first give the values  for the whole PMS or  B-type PMS
samples,  and then with the binaries removed from the
samples. In the lower part of the table, we give the same statistics as in the
upper part but for the MS stars. Note that only the stars with a reliable status are kept here.} 
\centering
\small{
\begin{tabular}{@{\ }lll@{\ \ }l@{\ \ }l@{\ \ }l@{\ }}
\hline
\hline	
Sub-sample & Number     & \multicolumn{4}{l}{\vsini~in \kms} 		             \\
           & of objects & Aver.   & Med.    & error  & D06 \\
\hline	
All PMS          &  32 &  169 &  155 &  $\pm$20 &  \\
PMS B-type       &  18 &  213 &  201 &  $\pm$25 & 187 \\
PMS B-type - bin &  12 &  238 &  236 &  $\pm$31 &  \\
\hline
All MS           &  26 &  162 &  157 &  $\pm$15 &  \\
MS B-type        &  23 &  165 &  183 &  $\pm$17 & 132 \\
MS B-type - bin  &  18 &  183 &  199 &  $\pm$18 &  \\
\hline
\multicolumn{4}{l}{D06: \citet[][ their Table 6]{dufton2006}}\\
\end{tabular}
\label{statV}
}
\end{table}

\subsection{Rotational velocity statistics}

It is very difficult to compare the rotational velocities of the objects in- or
outside of NGC\,6611 due to the small number of stars in these sub-samples.
Moreover, the membership probability varies highly from a study to an other. 
It is nevertheless interesting to statistically study the rotational velocities of
MS and PMS stars, irrespective of their membership. To increase the number of
objects in the sub-samples, we thus merge the PMS objects in- and outside of
the cluster and we do the same for the MS stars. We remove, in a second step, the
binaries. Note that only the stars with a reliable status (MS or PMS) are kept here 
(see Sect.~\ref{nonels} and Table~\ref{preuvePMS}).
We then compare the mean and median stellar rotational velocities of the sub-samples of PMS and MS
stars as well as of B-type PMS and MS objects. The results are given in
Table~\ref{statV}.

\begin{table}[h]
\centering
\caption[]{Ratios of the mean or median rotational velocities for the  B-type
PMS and MS objects, with or without binaries included in the sample.} 
\centering
\begin{tabular}{ccc}
\hline
\hline	
        & $\frac{MS B}{PMS B}$ & $\frac{MS B - bin}{PMS B - bin}$ \\
\hline	
Average &  0.77 &  0.77 \\
Median  &  0.91 &  0.84 \\
\hline
\end{tabular}
\label{decvel}
\end{table}

\begin{table*}[t]
\centering
\caption[]{ Parameters and indications for true circumstellar ELS.
In col. 11 the presence of infrared excesses based on 2MASS (2M) and on SPITZER (SPI) data are indicated. 
In col. 12, the membership (MP) to the open cluster NGC 6611 is indicated: y (for reliable membership), u (for uncertain membership),
and n (for non-membership).
In the last column, the nature of the ELS is
given as well as the presence of an accretion envelope (class I protostar), 
or whether the star is a MS object or at the ZAMS.}
\tiny{
\centering
\begin{tabular}{l@{\ \ }l@{\ \ }l@{\ \ }l@{\ \ }l@{\ \ }l@{\ \ }l@{\ \ }l@{\ \ }l@{\ \ }l@{\ \ }l@{\ \ }l@{\ \ }l@{\ \ }l}
\hline
\hline	
Stars & EW$\alpha$  & \teff & \logg & \vsini & CFP & $M/M_{\odot}$ & age MS  & age PMS & E[B-V] & IR excess  & MP & age kept & Nature\\
      & (\AA)       & K     & dex   & \kms   &     &               & Myears  & Myears  &        & 2M/SPI &  & Myears & \\

\hline
WFI17$^{0}$   & 15          		& 9600  & 4.2 & 169 & A1V   & 2.3  & 239  &  3     & 0.742 & yes/yes & n & 3    &  HAe, class I\\
W031          & 2           		& 9000 	& 3.7 & 156 & A2IV  & 2.7  & 432  &  1     & 0.898 & no/yes? & n & 1    &  probably HAe\\
W080          & undef$^{1}$ 		& 24000 & 4.3 & 183 & B1V   & 8.9  & 0.5  &  0.1   & 1.866 & no/no   & n & 0.5  &  possible He-strong, ZAMS\\
W235          & $>$60$^{2}$/70$^{3}$ 	& 24000: & 3.5: & 482: & B1IV  & 13.9: & 12: &  0.02  & 0.978 & yes/yes & y & 0.02 & HBe\\
W301          & 2           		& 20600 & 4.2 & 115 & B2V   & 7.0  & 7    &  \_  & 0.905 & no/no   & y & 7    & CBe, MS\\
W483          & 11/11$^{3}$          	& 14600 & 3.6 & 186 & B3IV  & 5.4  & 79   &  0.1   & 0.747 & no/yes  & y & 0.1  &  HBe\\
W500          & 13/13$^{3}$          	& 13700 & 3.4 & 289 & B5IV  & 5.4  & 85   &  0.12  & 0.684 & no/yes & y & 0.12 &  HBe\\
W503          & 11/17$^{3}$          	& 23500: & 3.0: & 236: & B1III & 21.7: & 7: & 0.02  & 0.687 & yes?/yes & u & 0.02 & MS?, bin\\
W601          & undef$^{1,4}$ 		& 22500 & 4.0 &  190 & B1V   &  10.2 & \_  &  0.016 & 0.691 & no/no & u & 0.016 &  He-strong, ZAMS\\
\hline
W245          & \_                      & \_  & \_  & \_  & A0$^{5}$  & \_  & \_  & \_  & \_ &  yes/yes & n & \_  & HAeBe \\
W494          & \_                      & \_  & \_  & \_  & B$^{5}$  & \_  & \_  & \_  & \_  & yes/yes & n & \_  & HAeBe \\
\hline
\multicolumn{14}{l}{ $^{0}$: WFI17 is written for WFI[N6611]017; $^{1}$ : EW undefined, P Cygni profile, very weak emission; $^{2}$: H$\alpha$ saturated;}\\
\multicolumn{14}{l}{ $^{3}$: EW measured in spectra from \citet{evans2005}; $^{4}$: Fundamental parameters from \citet{alecian08}, $^{5}$: from Simbad.}\\
\end{tabular}
\label{paramELS}
}
\end{table*}

The comparison shows that B-type PMS objects rotate with a higher rotational
velocity than MS stars.  The trend of these velocities is in fair agreement 
with the trend of rotational velocities determined by \citet[][Table 6]{dufton2006} 
as shown in Table~\ref{statV}. 
This result could be explained by the fact that the stars at the ZAMS 
undergo a first contraction with a redistribution of their internal angular
momentum, generating a decrease of their rotational velocity. Note that, in the
theoretical diagrams by \citet{meynet2000}, the rotational velocity decreases by
$\sim$20\% during this contraction.  Here we find on average for B-type
stars a decrease of 18\% (i.e. ratio $\simeq$ 82\%) as shown in
Table~\ref{decvel}, in excellent agreement with the theoretical calculation. 
Note that, this result is also important to constrain the models of the stellar evolution with rotation 
from PMS phase to MS.

The fact that we find stellar rotational velocities higher for the PMS candidates
than for the true MS stars supports our conclusion that these objects are
actually PMS stars and not evolved MS stars. Due to stellar winds
\citep[see][]{meynet2000}, the stellar rotational velocities of evolved
stars would be lower than non-evolved stars or PMS stars.



\section{Discussion on the nature of the stars}
\label{EMdiscussion}

\subsection{True circumstellar ELS}
In this section we discuss the nature of each ELS observed in NGC\,6611 and its
surrounding field. We use information reported in previous sections and summarized
in Table~\ref{paramELS}.
Note that the masses of the following stars (W503 excluded) are in the range 
of masses for PMS stars \citep[M$<$15M$\odot$][]{iben1965}.

\begin{itemize}

\item {\bf W245 and W494:}
We confirm the result of \citet{herbig2001} about W245 and W494, 
which are also found as ELS with our WFI observations.
Their infrared excess indicate that they are Herbig Ae/Be stars.
However, W245 and W494 are not member of NGC 6611 but lie in the ambient star-formation region 
of the Eagle Nebula and as WFI017 (see below) are representative to its PMS (A, B)-population. \\

\item {\bf WFI017:}
The emission of WFI[N6611]017 is moderate and the spectrum does not display any
\ion{Fe}{ii} permitted and forbidden emission lines. However, this star shows a
strong IR excess, in particular in the SPITZER colour/colour diagram, like very
young objects. This large IR excess is compatible with the one of HAeBe
stars.  This star is not a member of NGC 6611 and the age for this star derived from 
MS tracks is very old (239 Myears). This old age is not compatible with a 
PMS status.  
In fact, this star is representative of a PMS population 
not member of NGC 6611 itself but member of the surrounding star-formation region of the Eagle Nebula,
as shown by \citet[][private communication]{indeb07}.
Thus the true age of this star was determined with PMS tracks (3 Myears).
In conclusion, we propose this star as a possible HAe star.\\

\item {\bf W031:} 
The H$\alpha$ emission is weak. W031 has a potential IR excess in the SPITZER 
colour/colour diagram, even if the magnitude at 8 $\mu$m is not well defined,
which favours a PMS status. The star is not a member of
NGC 6611, and its RV indicates a field star.
The age of W031 (432 Myears) is very old, and except if the star is a foreground star, 
its age is not compatible with the star-formation region of the Eagle Nebula.
Moreover \citet{indeb07} found two PMS objects very close to W031.
With these indications, it seems that W031 could be more probably a HAe than a CBe, 
and its age is probably 1 Myear.\\

\item {\bf W235:}
The EW of the H$\alpha$ emission line is 70 {\AA}. This value is unusual for 
CBe, but is often seen in Herbig Ae/Be stars.
The 2MASS and SPITZER diagrams also indicate a probable Herbig star.
This star is clearly a member of NGC 6611 and its RV is compatible with 
the NGC 6611 RV distribution of stars. Its age from HR tracks (12 Myears) 
is too old for NGC 6611. 
This star is already known as a HBe star and we thus confirm its nature.  \\

\item {\bf W301:}
The H$\alpha$ emission is weak. The 2MASS and SPITZER diagrams indicate a MS
star. The star is a member of the cluster and its RV is compatible with 
the NGC 6611 RV distribution of stars. Its MS age determined is compatible
with the age of the cluster and its evolutionary status (\ttms=16\%) falls 
in the area of classical Be stars. We thus propose W301 as a classical Be star.\\

\item {\bf W483:}
The H$\alpha$ emission is moderate. The infrared excess found in SPITZER diagram cannot 
settle the status of this star, we have to consider the membership of the star to NGC 6611 
and to compare the ages. 
The star is a member of the cluster and its RV is compatible with 
the NGC 6611 RV distribution of stars. Its MS age determined is too high (79 Myears) as compared
to the age of the cluster. We thus confirm W483 as a HBe star.\\

\item {\bf W500:}
 The emission in H$\alpha$ is moderate. As for W483, the infrared study cannot 
help to conclude on the nature of this star. 
However, the star is a member of the cluster and its RV is compatible with 
the NGC 6611 RV distribution of stars. Its MS age determined is too high (85 Myears) as compared
to the age of the cluster. We thus confirm W500 as a HBe star. \\

\item {\bf W503:}
The H$\alpha$ emission is moderate. The profiles of the H$\alpha$, H$\beta$ and
H$\gamma$ emission lines are asymmetric and show 2 violet (V) and red (R) peaks
with V$>$R, as often observed in Herbig Ae/Be stars. The infrared study indicates
an infrared excess but cannot help to conclude on the nature of this star.
The membership of this star to NGC 6611 is uncertain (94\% and 40 \% 
from \citet{ref1} and \citet{ref2} respectively) and the RV cannot be used
due to the binarity of this system.
For this star, the emission lines profiles as well as the infrared excess can be interpreted
by a mass-transfer binary, in which the line emission arises from an accretion disk.
The mass of this star was estimated to $\sim$22M$\odot$, too high for HBe/Ae, 
but due to the binarity of this star, the mass determined is not reliable.
Note that this star is a binary as many Herbig Ae/Be stars.
However, we propose W503 as a possible MS star.\\

\item {\bf W601:}
This star and W080 have common properties: their H$\alpha$ profile is P Cygni-like
and their IR excess is very weak. Their fundamental parameters are very similar.
The membership of W601 to NGC 6611 is uncertain (93\% and 43 \% 
from \citet{ref1} and \citet{ref2} respectively). This star is an He-strong star 
and hosts a magnetic field \citep{alecian08} 
as found in about 8\% of Herbig Ae/Be stars and is proposed as an Herbig Be star close to the ZAMS by \citet{alecian08}.
In fact, if this star is not a member of NGC 6611 itself, as WFI017, W601 is a member 
of the surrounding star-formation region of the Eagle Nebula.\\

\item {\bf W080:} 
As indicated above, its spectral profiles and IR excess are similar to those of W601.
W080 is not a member of NGC 6611 but is a member of the ambient star-formation region of the Eagle Nebula, while its RV is not
different from the average RV of the RV distribution in NGC 6611.
Moreover, W080 seems to be very close to a new young open cluster, 
which undertook to emerge to the ambient nebula (see Fig.~\ref{loc6611}).
Its age (0.5 Myear) is compatible with this star-formation region.
W080 could be either a classical Be star or a Herbig Be star at the ZAMS, whose disk has been blown away
by the stellar wind. Consequently, the H$\alpha$ line may exhibit a P Cygni
profile as the HAeBe star AB Aur and according to \citet{finken84}.
However, we found that the NLTE models do not reproduce correctly 
the HeI lines in the LR2 domain as shown in Fig.~\ref{HaW080}-top.
We also compare in Fig.~\ref{W080W601fitsHe} the HeI 6678 {\AA} line to the NLTE models for W080 and for W601, 
already known as a magnetic He-strong star.
W080 and W601 show similar line profiles and similar fundamental parameters, the NLTE models 
are not able to reproduce their HeI line profiles. As W601 is already known as a magnetic He-strong,
we propose W080 as a potential new He-strong star. Moreover, this star lies in an area where the polarisation 
measured by \citet{bastien04} is large and of interstellar origin, a very good grain alignement 
is needed, probably calling for a correspondingly large local interstellar magnetic
field in the parent molecular cloud, near W080.

\end{itemize}

\subsection{Previously suspected ``Be'' stars}

The other potential Be stars from the literature are seen as non-ELS in our data
but often with strong nebular lines. See the WFI slitless and VLT-GIRAFFE
spectra examples in Fig.~\ref{specWFI1} for the star W371, which exhibits
strong nebular lines and no circumstellar line in H$\alpha$. 
This is the case for the stars listed in Table~\ref{nbreEm}. The stars W138, W202, W203, 
W207, W210, W213, W221, W227, W243, W251, W267, W273, W281, W297,
W299, W300, W306, W307, W311, W313, W322, W323, W336, W351, W371, W374,
W388, W389, W396, W400, W455, W469, W472, W484, W504, W536, and W617, are non-ELS
in our data and we confirm the status of non-Be stars as indicated by \citet{herbig2001,evans2005}. 
W205 is found as an 04Vf by \citet{evans2005} but do not display sufficient emission to be seen in our WFI data.

We infirm the nature of Be stars for the stars W181, W198, W232, W310, W617, which have spectra in absorption in our data.


\subsection{Non-ELS}
\label{nonels}
To determine if the star is a MS star or a PMS star, we used the same criteria as for the ELS.
First, if the star has a strong infrared excess compatible with the area of PMS objects, we consider then the star as a PMS.
Second, if the star is member of NGC 6611 and its MS age too old for the youth of NGC 6611, we consider that the age is not the true
one, and the star must be considered as a PMS object and its age was interpolated in PMS tracks.
Third, if the star has no infrared excess and if its membership to NGC 6611 is uncertain, we considered in general by default 
that the star is a potential MS star (noted as ``MS'') even if its age seems too old for the ambient star-formation region 
(noted as ``MS?''), while the stars with several hundreds of Myears are noted as ``PMS?''.
We also use indications about the nature of the stars from previous studies of \citet{walker1961,ref1,ogura02,indeb07}.
Note that all non-ELS considered here as PMS objects have masses lower than 8$M_{\odot}$, in agreement with the masses expected
for this kind of objects \citep[M$<$15$M_{\odot}$][]{iben1965}.
All the details about each star are given in Table~\ref{preuvePMS}.


\section{Conclusion}
Thanks to our observations with 2 different instrumentations, the ESO-WFI in
slitless spectroscopic mode which is not sensitive to the ambient nebula and the
VLT-GIRAFFE fibre multiobjects high-resolution spectrograph, we were able
to find a small number of true circumstellar ELS (in H$\alpha$) among the brightest population of the very young cluster
NGC 6611 and the star formation region of Eagle nebula. With spectra
obtained at the VLT, we were able to study accurately their nature: Herbig Ae/Be
or classical Be star. We also conducted the same study for the other
non-ELS. Finally, only 11 true ELS with circumstellar or wind emission were
found. The other previous potential Be stars from the literature are actually
stars with a strong nebular emission pollution in H$\alpha$.

We determined the fundamental parameters for 85 stars and gave general
information for several others. Among our sample of B-type stars, we found 27\%
of them as binaries. Concerning rotational velocities, we found that the
B-type MS stars rotate 18\% slower than B-type PMS objects, in good
agreement with published theoretical models at the ZAMS. 
This value could be used to constrain the 
models currently developed for the stellar evolution with rotation 
from the younger (PMS phase) to the older ages (G. Meynet, private communication). 
With IR data, we found that the low-mass
stars are mainly PMS stars without circumstellar emission.

We redetermined the age of NGC\,6611, found equal to 1.2--1.8 Myears, in
good agreement with recent estimates. With clues from spectroscopy, IR, HR
ages, membership probabilities, RV, and evolutionary status, we found that:
there is a MS population and a PMS population in NGC 6611 itself but also in 
the surrounding ambient star-formation region of the Eagle Nebula.
Among the true circumstellar H$\alpha$ ELS, we found that: WFI017, W245, W494, W235, W483, 
W500 are Herbig Ae/Be stars; W301 is a classical Be star, W503 a binary 
with an accretion disk, and W080 is a possible He-strong magnetic star like W601.
This study confirms that the appearance of Be stars is mass- and age-dependent.


\begin{acknowledgements}

C.M. acknowledges Drs E. Bertin, E. Al\'ecian, C. Catala, R. Indebetouw, G. Meynet, E. Puga, and C.
Evans for their useful information for our study. This research has made use of
the Simbad and Vizier databases maintained at CDS, Strasbourg, France,
of NASA's Astrophysics Data System Bibliographic Services, and of
the NASA/ IPAC Infrared Science Archive, which is operated by the Jet Propulsion
Laboratory, California Institute of Technology, under contract with the National
Aeronautics and Space Administration.
C.M. acknowledges funding from the ESA/Belgian Federal Science Policy in the 
framework of the PRODEX program (C90290).
We also acknowledge the anonymous referee for his very useful comments.
\end{acknowledgements}

\bibliographystyle{aa}
\bibliography{article5bib}






\Online
\appendix
\onecolumn
\begin{figure*}[!htpb]
    \centering
    \resizebox{\hsize}{!}{\includegraphics[angle=0]{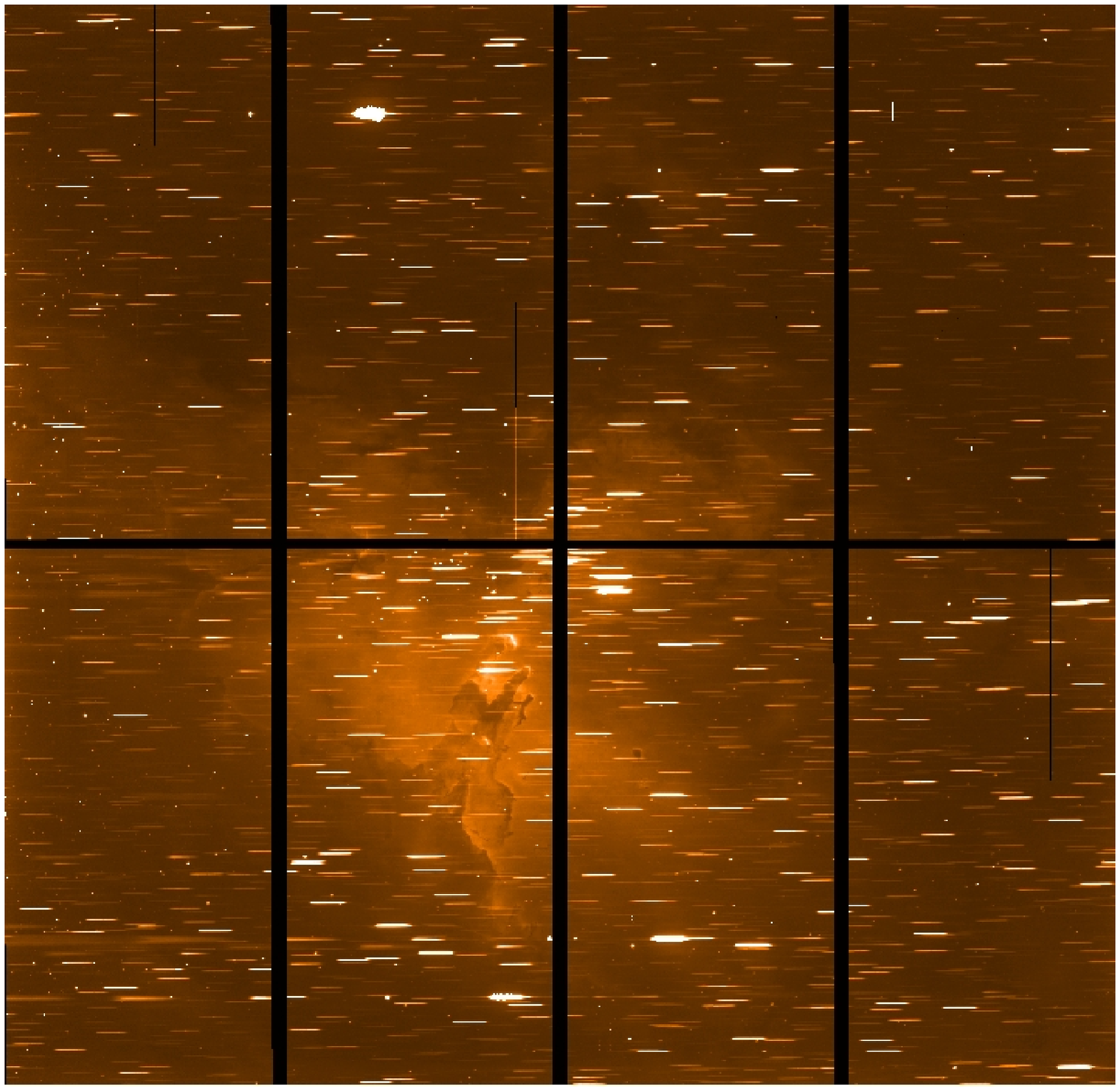}}
    \caption{Field of NGC\,6611 obtained with the ESO-WFI in slitless
    spectroscopic mode with the filter RC (200 nm bandpass) centered on
    H$\alpha$. The center of this image is ($\alpha$(2000)~=~18h18min42s, 
    $\delta$(2000)~=~-13$^{\circ}$46\arcmin57.6\arcsec), while the cluster lies
    at $\alpha$(2000)~=~18h18min48s,
    $\delta$(2000)~=~-13$^{\circ}$48\arcmin24\arcsec. North is at the top, East
    on the left. The stars appear as spectra. The diffuse background is due to
    the H$\alpha$ emission line of the nebulosity. Note the elephant-trunk
    features, which correspond to shocks and often called ``the pillars of
    creation''. Also note the  $0^{th}$ order of the spectra which appear
    as points in the image. The 2 images we obtained are combined in this
    figure.}
    \label{WFIN6611}
\end{figure*}

\newpage

\addtocounter{figure}{+1}

\begin{figure*}[!h]
    \centering
    \resizebox{\hsize}{!}{\includegraphics[angle=0]{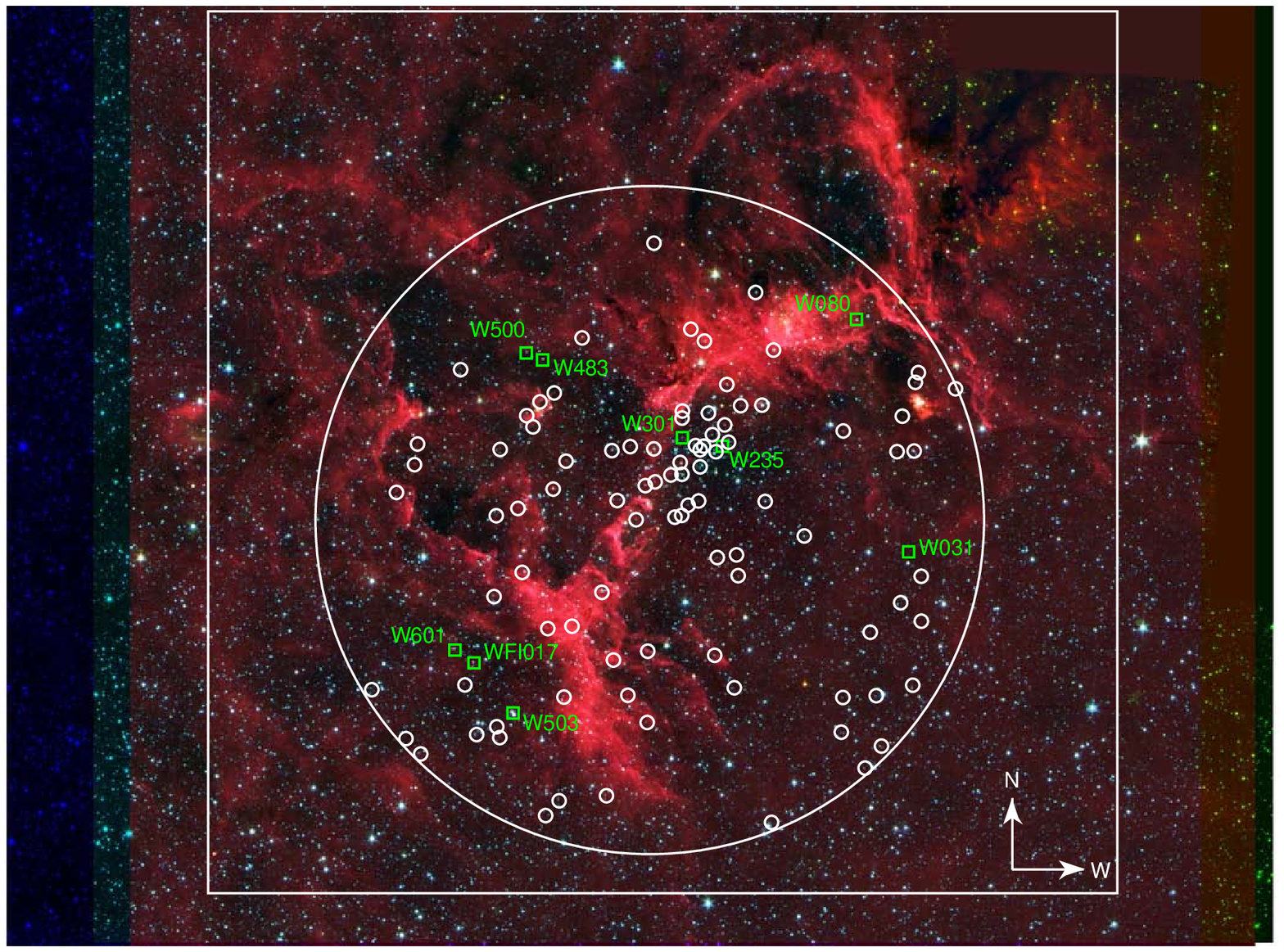}}
    \caption{Location of the stars observed with the VLT-GIRAFFE in the
    field of NGC\,6611. The image is a RGB mosaic with 3 channels of SPITZER (B:
    3.6$\mu$m, G: 4.5$\mu$m, and R: 8$\mu$m). The small white circles are  for
    non-ELS stars, the green boxes are for  ELS stars. The large white
    circle shows the field of GIRAFFE and the large white  box the field of
    the WFI-spectro. WFI017 stands for WFI[N6611]017. 
    }
    \label{loc6611}
\end{figure*}
%
%
%

\addtocounter{table}{+1}

\tiny{
\begin{longtable}{llllllllllllll}
\caption{\label{tabPFNGC6611} Fundamental parameters for stars observed with the
VLT-GIRAFFE in the open cluster NGC\,6611 and its surrounding field. In col.~1,
the name of the object is given: WXXX from \citet{walker1961} or WFIXXX for
WFI[N6611]XXX from our catalogue. The coordinates ($\alpha$(2000),
$\delta$(2000)) are taken from the UCAC2 catalogue and are given in  cols.
2 and 3. The magnitude V and colour index B-V are given in cols. 4 and 5.
The signal to noise (S/N) ratio is given in col.~6. In cols. 7, 8 and 9  \teff\ in K, \logg\ in dex and \vsini\ in \kms\ are given. The errors in
\teff\ range from $\pm$500 to $\pm$1500K for spectrum with high to
low S/N. The errors in \logg\ range from $\pm$0.1 to $\pm$0.2dex for
spectra with a high to low S/N. The errors in \vsini\ range from $\pm$10\kms\ to
$\pm$30\kms\ for spectra with a high to low S/N. ``CFP'' corresponds to the
spectral classification based on the fundamental parameters. Col. ``obs''
indicates with ``Bew'' if the star is detected as a Be star with the VLT-GIRAFFE
spectra and with our WFI-spectro study. If ``no L2'' is written, there is no
spectrum in this setting for the concerning stars (too bright). The 12th column
``lit'' gives indications from the literature and from \citet{evans2005}. The
last column gives the radial velocity (error $\pm$10\kms) corrected from the
heliocentric velocity.}\\
\hline
\hline
Star & $\alpha$ & $\delta$ & V & B-V & S/N &  \teff & \logg & \vsini & CFP &  obs & lit. & RV\\
\hline
\endfirsthead
\caption{continued.}\\
\hline
\hline
Star & $\alpha$ & $\delta$ & V & B-V & S/N &  \teff & \logg & \vsini & CFP &  obs & lit. & RV\\
\hline
\hline		
\endhead
\hline
\endfoot
WFI[N6611]014a & 18 19 06.043 & -14 0 33.40 & 14.55 & 0.68 & 10 & 12000: & 2.6: & 348: & B8III-II: & & & 28:\\
WFI[N6611]019 & 18 19 27.561 & -13 57 40.56 & 14.56 & 0.70 & 15 & 11000 & 2.6 & 253 & B9III-II &  &  &   29 \\
WFI[N6611]125 & 18 18 31.213 & -14 0 48.48 & 13.94 & 0.65 & 15 & 8000 & 1.9  & 100  & A7II &  &  &  28 \\
WFI[N6611]152 & 18 19 29.014 & -13 48 27.81 & 13.71 & 0.71 & 30 & 8300 & 3.8 & 60 & A4IV &  &  &  28\\
W002 & 18 18 02.929 & -13 44 34.77 & 10.56 & 0.35 &  &  &  &  &  & no L2 & \_/B8III & \\
W018 & 18 18 08.181 & -13 51 35.81 & 14.16 & 0.60 & 30 & 8000 & 4.3 & 117 & A7V &  & A0sh/\_ &  26\\
W024 & 18 18 08.647 & -13 43 58.59 & 14.22 & 0.76 & 15 & 8000 & 3.6 & 68 & A7IV-III &  &  &  -21 \\
W025 & 18 18 09.283 & -13 46 54.55 & 12.93 & 0.98 & 80 & 24500 & 4.3 & 68 & B1V& bin$^{1}$ & B0.5V/\_  &  25 \\
W026 & 18 18 09.110 & -13 44 20.48 & 13.46 & 0.60 & 40 & 8900 & 3.7 & 33 & A2IV &   & &  7\\
W035 & 18 18 11.348 & -13 52 35.21 & 14.11 & 0.52 & 60 & 10000 & 4.5 & 275 & A0V &  & &  4\\
W036 & 18 18 11.116 & -13 45 36.35 & 13.40 & 0.61 & 25 & 8000 & 4.2 & 121 & A7V & & \_/A7II &  0\\
W041 & 18 18 11.921 & -13 46 56.05 & 14.15 & 0.85 & 20 & $<$8000 &  & & $<$A7 &  & &  -6\\
W064 & 18 18 16.031 & -13 53 41.58 & 13.74 & 0.53 & 30 & 8000 & 2.3 & 166 & A7II &  & &  -9\\
W090 & 18 18 20.207 & -13 46 09.70 & 11.73 & 0.38 & 175 & 14800 & 4.1 & 262 & B4V &   & B3-5V/B5V &  32\\
W125 & 18 18 26.196 & -13 50 05.49 & 10.01 & 0.47 &  &  &  &  &  & no L2, bin$^{2}$ & B1.5V/B1V+?&  17:\\
W161 & 18 18 30.956 & -13 43 08.23 & 11.29 & 1.05 & 130 & 35000 & 4.0 & 117 & bin$^{3}? $ &  & O8V/O8.5V &  10\\
W166 & 18 18 32.222 & -13 48 48.06 & 10.37 & 0.57 &  &  &  &  &  &  no L2 & O8.5/O9V &  21:\\
W175 & 18 18 32.730 & -13 45 11.88 & 10.09 & 0.84 & &  &  &  &  & no L2, bin$^{1,2}$ & O8.5/O5Vf+,SB2 &  \\
W188 & 18 18 33.719 & -13 40 58.83 & 13.13 & 1.34 & 60 & 31500 & 4.3 & 239 & B0V & bin$^{1}$ & B0V/\_ &  -17\\
W194 & 18 18 36.380 & -13 51 34.70 & 13.90 & 0.53 & 10 &   &   &   & ? &  & &  26 \\
W201 & 18 18 36.973 & -13 55 46.42 & 13.55 & 0.68 & 15 & 10000 & 2.7 & 48 & A0III-II &  &  &  14\\
W202 & 18 18 36.013 & -13 45 13.12 & 14.40 & 0.96 & 55 & 12500 & 4.1 & 310 & B7V & &  A0/\_ &  13\\
W203 & 18 18 36.640 & -13 50 48.02 & 14.19 & 0.47 & 35 & 9100 & 3.9 & 36 & A2IV & & &  -7 \\
W223 & 18 18 37.865 & -13 46 35.15 & 11.20 & 0.59 & 220 & 25100 & 4.4 & 95 & B1V & & B1V/\_ &  5 \\
W228 & 18 18 38.136 & -13 44 25.47 & 13.51 & 0.93 & 70 & 21500 & 4.5 & 86 & B2V &  &  B2V/\_ &  -3\\
W231 & 18 18 38.464 & -13 45 56.22 & 12.71 & 0.75 & 135 & 24800 & 4.4 & 200 & B1V &  & B1V/\_ &  -1\\
W238 & 18 18 39.598 & -13 50 54.00 & 13.37 & 0.67 & 50 & 9000 & 3.8 & 13 & A2IV &  &  &  -4\\
W239 & 18 18 39.993 & -13 54 33.51 & 11.48 & 0.36 & 240 & 20400 & 4.3 & 95 & B2V & bin$^{3}?$  & B1V/B1.5V &  16 \\
W243 & 18 18 39.810 & -13 46 56.50 & 13.80 & 0.63 & 55 & 15600 & 4.4 & 30 & B4V & bin$^{1}$ & &  38\\
W251 & 18 18 40.357 & -13 46 18.08 & 13.34 & 0.69 & 110 & 20400 & 4.4 & 78 & B2V &  & &  6\\
W259 & 18 18 40.965 & -13 45 29.65 & 11.61 & 0.72 & 210 & 30000 & 4.4 & 148 & B0V &  & \_/B0.5V &  14\\
W267 & 18 18 41.692 & -13 46 43.86 & 13.13 & 0.52 & 125 & 14400 & 3.7 & 218 & B5IV & bin$^{1}$  & \_/B3V & 0\\
W269 & 18 18 41.586 & -13 42 48.02 & 13.98 & 0.93 & 55 & 22000 & 4.5 & 198 & B2V &  &  &  -4\\
W273 & 18 18 42.251 & -13 47 30.42 & 14.21 & 0.71 & 85 & 9900 & 3.8 & 245 & A0IV &  &  &  -13\\
W275 & 18 18 42.250 & -13 46 52.10 & 12.12 & 0.46 & 100 & 21900 & 4.5 & 75 & B1V & bin$^{1}$  & &  8\\
W276 & 18 18 42.480 & -13 48 47.02 & 13.74 & 0.67 & 100 & 13600 & 4.2 & 210 & B6V &  &  & -2 \\
W281 & 18 18 42.950 & -13 46 42.80 & 13.80 & 0.59 & 30 & 8700 & 2.4 & 85 & A3III-II &  & &  -1\\
W289 & 18 18 44.087 & -13 48 56.49 & 12.60 & 0.50 & 170 & 18700 & 4.3 & 161 & B2V &  & \_/B3V &  9\\
W292 & 18 18 43.683 & -13 42 21.40 & 14.09 & 0.78 & 30 & 10500 & 4.1 & 86 & B9V &  &  &  6\\
W299 & 18 18 45.071 & -13 49 19.42 & 14.46 & 0.69 & 30 & 8000 & 4.4 & 126 & A6V &  bin$^{1}$  & &  -16\\
W300 & 18 18 45.039 & -13 47 47.17 & 12.69 & 0.52 & 145 & 18800 & 4.3 & 262 & B2V &   & &  6\\
W305 & 18 18 44.970 & -13 45 25.08 & 13.51 & 1.07 & 80 & 27000 & 4.4 & 205 & B1V &  & &  7\\
W306 & 18 18 45.030 & -13 45 41.02 & 12.77 & 0.68 & 100 & 21700 & 4.3 & 245 & B1V &   & &  2\\
W307 & 18 18 45.320 & -13 47 20.60 & 14.18 & 0.97 & 25 & 18700 & 4.1 & 135 & B2V &   & &  2\\
W313 & 18 18 46.132 & -13 49 23.43 & 12.92 & 0.50 & 140 & 13800 & 3.9 & 307 & B6V &  bin$^{1,3}$  & \_/B5III &  3\\
W323 & 18 18 46.743 & -13 47 48.68 & 13.48 & 0.57 & 100 & 13800 & 4.1 & 257 & B4V &   & &  4\\
W336 & 18 18 49.168 & -13 48 04.23 & 13.29 & 0.52 & 110 & 13500 & 3.7 & 154 & B5IV &   & \_/B5III &  7\\
W343 & 18 18 49.373 & -13 46 50.05 & 11.72 & 0.85 & 155 & \_ & \_ & \_ & \_ & SB2, bin$^{1}$ & &  28 \\
W344 & 18 18 50.309 & -13 54 24.32 & 13.63 & 0.87 & 10 & $<$8000 &   &   & $<$A8 & & &  -9\\
W347 & 18 18 49.363 & -13 39 08.33 & 14.38 & 0.88 & 10 & 8000 & 3.3 & 38 & A7III & & & -14\\
W351 & 18 18 50.683 & -13 48 12.72 & 11.26 & 0.45 & 340 & 26400 & 4.3 & 232 & B1V &   & \_/B1V &  2\\
W364 & 18 18 52.098 & -13 49 29.20 & 13.44 & 0.53 & 100 & 11500 & 4.0 & 105 & B7V & bin$^{1}$  & &  5\\
W368 & 18 18 53.372 & -13 56 03.23 & 13.62 & 0.61 & 30 & 8000 & 2.5 & 144 & A7III-II &  & &  -16\\
W371 & 18 18 53.012 & -13 46 45.07 & 13.44 & 0.65 & 120 & 16500 & 4.3 & 213 & B4V &   & &  12\\
W388 & 18 18 55.010 & -13 48 46.10 & 13.70 & 0.58 & 35 & 13900 & 4.0 & 140 & B7V &   & &  10\\
W389 & 18 18 55.590 & -13 54 44.25 & 14.34 & 0.51 & 60 & 8500 & 3.8 & 213 & A5IV &   & &  9\\
W400 & 18 18 55.832 & -13 46 54.05 & 12.87 & 0.60 & 100 & 11400 & 3.8 & 44 & B8IV & bin$^{1,2}$  & \_/B9III &  7\\
W409 & 18 18 57.369 & -13 52 12.21 & 12.84 & 0.40 & 140 & 17200 & 4.3 & 192 & B3V & bin$^{3}?$ & \_/B2.5V &  9\\
W444 & 18 19 00.428 & -13 42 41.02 & 12.74 & 0.81 & 100 & 22000 & 4.3 & 110 & B1V & bin$^{3}$? & \_/B1.5V &  -10\\
W445 & 18 19 01.977 & -13 53 28.29 & 14.16 & 0.48 & 80 & 9400 & 4.3 & 169 & A1V &  & & -6\\
W455 & 18 19 02.889 & -13 47 17.67 & 12.11 & 0.60 & 15 & 8000 & 3.3 & 36 & A6III &   & \_/A5II &  -16\\
W469 & 18 19 04.877 & -13 48 20.44 & 10.69 & 0.40 &  &  &  &  &  & bin$^{2}$ no L2 & \_/B0.5Vn &  \\
W472 & 18 19 04.712 & -13 44 44.54 & 12.97 & 0.50 & 80 & 13000 & 3.2 & 121 & B6III &  bin$^{2}$  & \_/B3V+? & 3\\
W473 & 18 19 05.706 & -13 53 33.58 & 12.57 & 0.30 & 110 & 10000 & 2.9 & 266 & A0III-II & bin$^{3}$?  & \_/A0II &  -33\\
W484 & 18 19 06.903 & -13 45 04.50 & 12.46 & 0.46 & 80 & 12500 & 4.1 & 110 & B7V &  & \_/B8III &  -13\\
W490 & 18 19 07.991 & -13 46 00.71 & 13.07 & 0.44 & 130 & 10500 & 3.7 & 300 & B9IV &  & &  -9\\
W495 & 18 19 08.953 & -13 45 35.58 & 14.32 & 0.54 & 30 & 8500 & 4.4 & 103 & A3V &   & &  -3\\
W496 & 18 19 09.653 & -13 51 27.43 & 14.06 & 0.63 & 25 & 8500 & 3.7 & 102 & A3IV &   & &  -19\\
W504 & 18 19 10.300 & -13 49 03.85 & 12.78 & 0.42 & 145 & 12000 & 3.9 & 49 & B8V &   & \_/B9III &  8\\
W515 & 18 19 13.040 & -13 46 51.37 & 13.40 & 0.77 & 12 & 8000 & 4.5 & 10 & A7V &   & &  20\\
W519 & 18 19 13.643 & -13 49 20.19 & 13.71 & 0.72 & 37 & 8500 & 2.8 & 153 & A3III-II &   & &  -3\\
W520 & 18 19 13.984 & -13 52 21.71 & 11.64 & 0.46 & 186 & 13300 & 3.5 & 296 & B5IV &   & \_/B5IIIn &  -14\\
W536 & 18 19 18.481 & -13 55 40.09 & 11.46 & 0.22 & 196 & 21500 & 4.2 & 26 & B2V &  bin$^{2}$  & \_/B1.5V +? &  27\\
W541 & 18 19 19.127 & -13 43 52.35 & 13.31 & 0.60 & 76 & 19000 & 4.2 & 59 & B2V &   & \_/B1-3V &  10\\
W550 & 18 19 25.759 & -13 46 39.13 & 13.66 & 0.68 & 21 & 8000 & 4.5 & 129 & A8V &  & &  -23\\
W567 & 18 18 16.804 & -13 58 46.45 & 11.99 & 0.36 & 82 & 18000 & 4.2 & 345 & B2V &  & &  6\\
W568 & 18 18 20.474 & -13 57 25.94 & 13.57 & 0.72 & 10 & $<$8000 &  &  & $<$A8 &   & &  22\\
W570 & 18 18 15.094 & -13 56 03.84 & 12.74 & 0.71 & 14 & $<$8000 &  &  & $<$A8 &   & &  28\\
W582 & 18 18 20.234 & -13 56 08.14 & 12.33 & 0.59 & 40 & 8400 & 3.4 & 90 & A3IV &   bin? & &  17\\
W587 & 18 18 56.667 & -13 59 48.79 & 11.97 & 0.41 & 179 & 18300 & 4.3 & 131 & B2V &   & &  1\\
W588 & 18 19 03.960 & -14 00 00.20 & 12.18 & 0.38 & 21 & 8400 & 4.0 & 222 & A3IV &   & & 29\\
W590 & 18 18 50.389 & -13 57 04.36 & 12.29 & 0.31 & 50 & 10200 & 3.5 & 43 & B8IV &   & &  25\\
W591 & 18 19 03.213 & -13 56 07.39 & 11.74 & 0.40 & 146 & 13900 & 3.9 & 92 & B5V &   & &  0\\
W596 & 18 19 25.295 & -13 58 14.97 & 13.03 & 0.68 & 15 & $<$8000 &  &  & $<$A8 &   & &  -18\\
W597 & 18 19 13.106 & -13 57 38.28 & 12.30 & 0.39 & 124 & 17800 & 4.3 & 237 & B2V &  & &  6\\
W607 & 18 19 32.824 & -13 55 50.67 & 12.48 & 0.47 & 65 & 16000 & 4.3 & 69 & B3V &   & &  8\\
W625 & 18 18 08.154 & -13 53 16.63 & 14.07 & 0.55 & 37 & 10500 & 4.0 & 254 & A0V &   & &  12\\
W626 & 18 18 09.471 & -13 55 40.36 & 13.92 & 0.76 & 9 & $<$8000 &  &  & $<$A8 &   & &  29\\
W627 & 18 18 14.285 & -13 57 57.03 & 13.97 & 0.49 & 22 & 10000 & 4.5 & 181 & A0V &   & &  -5\\
W632 & 18 19 13.569 & -13 57 13.85 & 14.36 & 0.54 & 39 & 9000 & 3.9 & 55 & A2V &   & &  -7\\
W633 & 18 19 16.670 & -13 57 31.50 & 13.76 & 0.66 & 8 & $<$8000 &  &  & $<$A8 &   & & 28:\\
W639 & 18 19 26.230 & -13 47 25.50 & 13.08 & 0.61 & 13 & 8000 & 4.3 & 5 & A6V &  & &  28:\\
\hline
WFI[N6611]017 & 18 19 17.083 & -13 54 50.64 & 14.29 & 0.64 & 55 & 9600 & 4.2 & 169 & A1V & Bew {\bf ELS} & &  22\\
W031 & 18 18 10.121 & -13 50 41.15 & 14.40 & 0.40 & 50 & 9000 & 3.7 & 156 & A2IV & {\bf ELS} & &  33\\
W080 & 18 18 18.201 & -13 41 59.25 & 13.82 & 1.64 & 35 & 24000 & 4.3 & 183 & B1V & {\bf ELS} & B2V/\_ &  3\\
W235 & 18 18 38.817 & -13 46 44.28 & 10.98 & 0.82 & 170 & 24000: & 3.5: & 482: & B1IV & {\bf HBe} & HBe/HBe &  12\\
W301 & 18 18 44.980 & -13 46 24.90 & 12.22 & 0.57 & 210 & 20600 & 4.2 & 115 & B2V & {\bf ELS} & \_/B2V&  5\\
W483 & 18 19 06.506 & -13 43 30.47 & 10.99 & 0.41 & 195 & 14600 & 3.6 & 186 & B3IV & Bew {\bf ELS} & \_/B3V &  -13\\
W500 & 18 19 09.019 & -13 43 14.95 & 11.28 & 0.43 & 190 & 13700 & 3.4 & 289 & B5IV & Bew {\bf ELS} & \_/B5e & -16\\
W503 & 18 19 11.068 & -13 56 43.08 & 9.75 & 0.49 & 90 & 23500: & 3.0: & 236: & B1III & Bew$^{3}$ {\bf ELS} & \_/B1:e &  -9\\
W601 & 18 19 20.031 & -13 54 21.67 & 10.68 & 0.36 &  &  22500$^{4}$ & 4.0$^{4}$ & 190$^{4}$ &  & \tiny{no L2,{\bf ELS}} & B1.5V/\_ &  \\
\end{longtable}
}
\begin{flushleft}
bin$^{1}$ the star was detected as a binary by \citet{duchene2001}\\
bin$^{2}$ the difference in \rv\ between this paper and \citet{evans2005} is larger than 15\kms\\
bin$^{3}$ the difference in \rv\ between this paper and \citet{evans2005} is between 10 and 15\kms\\
$^{4}$ fundamental parameters from \citet{alecian08}.
\end{flushleft}


%
\begin{longtable}{ccccc}
\caption{\label{tabinterpB} Associated parameters for stars: mass (col. 2),
luminosity (col. 3), radius (col. 4), and age in Myears (col. 5). The
corresponding errors are: $\pm$0.5 to 1.5 in $M/M_{\odot}$, $\pm$0.3 in
$\log(L/L_{\odot})$, $\pm$0.5 to 1.5 in $R/R_{\odot}$, and $\pm$1 to 25 Myears
for the age, depending on the value of the parameter and the initial S/N ratio
of spectrum. }\\
\hline
\hline
Star & $M/M_{\odot}$ & $\log(L/L_{\odot})$ & $R/R_{\odot}$ & Age (Myears) \\
\hline
\endfirsthead
\caption{continued.}\\
\hline
\hline
Star & $M/M_{\odot}$ & $\log(L/L_{\odot})$ & $R/R_{\odot}$ & Age (Myears) \\
\hline
\hline		
\endhead
\hline
\endfoot
WFI014a & 8.9 & 4.05 & 24.9 & 27 \\
WFI019 & 7.8 & 3.84 & 23.5 & 37 \\
WFI125 & 8.9 & 4.05 & 56.5 & 27 \\
WFI152 & 2.3 & 1.62 & 3.2 & 646 \\
W002 & \_ & \_ & \_ & \_ \\
W018 & 1.7 & 0.93 & 1.5 & 33 \\
W024 & 2.4 & 1.77 & 4.0 & 717 \\
W025 & 9.4 & 3.67 & 3.8 & 1 \\
W026 & 2.7 & 1.91 & 3.8 & 448 \\
W035 & 2.3 & 1.44 & 1.8 & 28 \\
W036 & 1.8 & 1.04 & 1.8 & 510 \\
W041 & \_ & \_ & \_ & \_ \\
W064 & 6.5 & 3.51 & 30.4 & 55 \\
W090 & 4.3 & 2.60 & 3.1 & 62 \\
W125 & \_ & \_ & \_ & \_ \\
W161 & 23.0 & 4.92 & 7.9 & 3 \\
W166 & \_ & \_ & \_ & \_ \\
W175 & \_ & \_ & \_ & \_ \\
W188 & 15.7 & 4.36 & 5.2 & 0.5 \\
W194 & \_ & \_ & \_ & \_ \\
W201 & 6.2 & 3.47 & 18.7 & 64 \\
W202 & 3.4 & 2.20 & 2.7 & 122 \\
W203 & 2.4 & 1.71 & 2.9 & 473 \\
W223 & 9.8 & 3.73 & 3.9 & 0.9 \\
W228 & 7.4 & 3.33 & 3.4 & 1.4 \\
W231 & 9.6 & 3.70 & 3.9 & 1 \\
W238 & 2.5 & 1.81 & 3.3 & 459 \\
W239 & 6.8 & 3.20 & 3.2 & 2.5 \\
W243 & 4.4 & 2.51 & 2.5 & 4.9 \\
W251 & 6.8 & 3.20 & 3.2 & 2.5 \\
W259 & 14.1 & 4.22 & 4.8 & 0.5 \\
W267 & 5.0 & 3.02 & 5.3 & 82 \\
W269 & 7.7 & 3.39 & 3.4 & 1.2 \\
W273 & 2.9 & 2.02 & 3.5 & 333 \\
W275 & 7.7 & 3.38 & 3.4 & 1.3 \\
W276 & 3.7 & 2.28 & 2.5 & 53 \\
W281 & 6.7 & 3.56 & 27 & 52 \\
W289 & 5.9 & 2.99 & 3.0 & 5.5 \\
W292 & 2.7 & 1.79 & 2.4 & 232 \\
W299 & 1.7 & 0.93 & 1.5 & 33 \\
W300 & 6.0 & 3.00 & 3.0 & 5.3 \\
W305 & 11.3 & 3.92 & 4.2 & 0.4 \\
W306 & 7.5 & 3.35 & 3.4 & 1.3 \\
W307 & 6.3 & 3.16 & 3.7 & 25 \\
W313 & 4.3 & 2.67 & 3.8 & 104 \\
W323 & 3.9 & 2.43 & 2.9 & 81 \\
W336 & 4.6 & 2.87 & 5.0 & 106 \\
W343 & \_ & \_ & \_ & \_ \\
W344 & \_ & \_ & \_ & \_ \\
W347 & 2.8 & 2.14 & 6.2 & 442 \\
W351 & 10.8 & 3.86 & 4.1 & 0.6 \\
W364 & 3.2 & 2.12 & 2.9 & 193 \\
W368 & 5.5 & 3.23 & 22.0 & 83 \\
W371 & 4.7 & 2.65 & 2.6 & 3 \\
W388 & 4.1 & 2.57 & 3.4 & 94 \\
W389 & 2.4 & 1.68 & 3.2 & 587 \\
W400 & 3.5 & 2.35 & 3.9 & 211 \\
W409 & 5.1 & 2.78 & 2.8 & 8 \\
W444 & 7.7 & 3.39 & 3.4 & 1.2 \\
W445 & 2.1 & 1.30 & 1.7 & 37 \\
W455 & 2.8 & 2.14 & 6.2 & 442 \\
W469 & \_ & \_ & \_ & \_ \\
W472 & 5.8 & 3.40 & 10.1 & 73 \\
W473 & 5.1 & 3.19 & 13.4 & 92 \\
W484 & 3.4 & 2.20 & 2.7 & 122 \\
W490 & 3.3 & 2.28 & 4.2 & 259 \\
W495 & 1.9 & 1.07 & 1.6 & 38 \\
W496 & 2.5 & 1.80 & 3.7 & 512 \\
W504 & 3.5 & 2.35 & 3.5 & 180 \\
W515 & 1.7 & 0.93 & 1.5 & 33 \\
W519 & 4.5 & 2.96 & 14.2 & 128 \\
W520 & 4.8 & 3.06 & 6.6 & 110 \\
W536 & 7.6 & 3.39 & 3.6 & 6 \\
W541 & 6.2 & 3.09 & 3.3 & 12 \\
W550 & 1.7 & 0.93 & 1.5 & 33 \\
W567 & 5.7 & 2.95 & 3.1 & 16 \\
W568 & \_ & \_ & \_ & \_ \\
W570 & \_ & \_ & \_ & \_ \\
W582 & 2.8 & 2.12 & 5.5 & 457 \\
W587 & 5.7 & 2.94 & 3.0 & 6 \\
W588 & 2.1 & 1.40 & 2.4 & 637 \\
W590 & 3.4 & 2.44 & 5.4 & 286 \\
W591 & 4.3 & 2.69 & 3.9 & 102 \\
W596 & \_ & \_ & \_ & \_ \\
W597 & 5.4 & 2.87 & 2.9 & 7 \\
W607 & 4.5 & 2.57 & 2.6 & 4.1 \\
W625 & 2.8 & 1.91 & 2.8 & 271 \\
W626 & \_ & \_ & \_ & \_ \\
W627 & 2.3 & 1.44 & 1.8 & 28 \\
W632 & 2.4 & 1.68 & 2.9 & 498 \\
W633 & \_ & \_ & \_ & \_ \\
W639 & 1.7 & 0.93 & 1.5 & 33 \\
\hline
WFI017 & 2.3 & 1.47 & 2.0 & 239 \\
W031 & 2.7 & 1.93 & 3.8 & 432 \\
W080 & 8.9 & 3.61 & 3.7 & 0.5 \\
W235 & 13.9: & 4.55: & 11: & 12: \\
W301 & 7.0 & 3.28 & 3.5 & 7 \\
W483 & 5.4 & 3.17 & 6.1 & 79 \\
W500 & 5.4 & 3.26 & 7.8 & 85 \\
W503 & 21.7: & 5.20: & 24.9: & 7: \\
W601 & \_ & \_ & \_ & \_ \\
\end{longtable}

\begin{flushleft}
\begin{landscape}
\tiny{
\begin{table*}[tbph]
\caption{Equivalent widths of interstellar absorption bands at
4430\AA\ (col. 2) and at 6613 \AA~(col. 3) and their corresponding E[B-V] (cols.
4 and 5) from the calibration of \citet{herbig1975}. The mean E(B-V) is
given in col. 6. In cols. 7, 8 and 9 the J, H and K magnitudes from 2MASS are
given. In cols. 10 and 11 the (J-H)$_{0}$ and (H-K)$_{0}$  colours obtained
with the mean E(B-V) and 2MASS magnitudes are given. In cols. 12, 13, 14, 
and 15 the magnitudes at 3.6, 4.5, 5.8 and 8 $\mu$m from SPITZER are given
(archive, release Spring07). In cols. 16 and 17 the (3.6-4.5)$_{0}$ and
(5.8-8)$_{0}$ colours obtained with the mean E(B-V) and SPITZER magnitudes
are given. The corresponding errors are: $\pm$0.0010 to 0.0020 for the EW 
and $\pm$0.005 in the E[B-V]. The mean errors on the 2MASS magnitudes and 2MASS
colour indexes are $\pm$0.060, while for SPITZER magnitudes the mean error is
$\pm$ 0.045. Note that for the calculations one more digit was kept.}
\centering
\begin{tabular}{@{\ }l@{\ \ \ }l@{\ \ \ }l@{\ \ \ }l@{\ \ \ }l@{\ \ \ }l@{\ \ \ }l@{\ \ \ }l@{\ \ \ }l@{\ \ \ }l@{\ \ \ }l@{\ \ \ }l@{\ \ \ }l@{\ \ \ }l@{\ \ \ }l@{\ \ \ }l@{\ \ \ }l@{\ }}
\hline
\hline
Star & EW & EW & E(B-V) & E(B-V) & $<$E(B-V)$>$ & J & H & K & (J-H)$_{0}$ & (H-K)$_{0}$ & 3.6 & 4.5 & 5.8 & 8.0 & (3.6-4.5)$_{0}$ &	(5.8-8)$_{0}$ \\
Star & 4430 & 6613 & 4430 & 6613 &  & & &  &  &  &  &  &  &  &  & \\
\hline
WFI014a & \_ & 0.016 & \_ & 0.638 & 0.638 & 12.986 & 12.713 & 12.573 & 0.061 & 0.015 & 12.517 & 12.463 & \_ & \_ & \_ & \_ \\
WFI019 & \_ & 0.015 & \_ & 0.622 & 0.622 & 12.966 & 12.656 & 12.598 & 0.104 & -0.063 & 12.527 & 12.470 & 11.939 & \_ & \_ & \_ \\
WFI125 & \_ & 0.013 & \_ & 0.541 & 0.541 & 12.520 & 12.158 & 11.920 & 0.183 & 0.132 & 11.902 & \_ & \_ & \_ & \_ & \_ \\
WFI152 & 0.134 & 0.021 & 0.589 & 0.837 & 0.713 & 11.946 & 11.722 & 11.558 & -0.013 & 0.025 & 11.478 & 11.398 & 11.148 & \_ & \_ & \_ \\
W002 & \_ & 0.014 & \_ & 0.557 & 0.557 & 9.643 & 9.538 & 9.472 & -0.080 & -0.043 & 9.439 & 9.441 & 9.452 & 9.361 & -0.062 & 0.086 \\
W018 & 0.091 & 0.017 & 0.400 & 0.703 & 0.552 & 12.704 & 12.405 & 12.106 & 0.116 & 0.191 & 12.259 & 12.086 & \_ & \_ & \_ & \_ \\
W024 & \_ & 0.021 & \_ & 0.842 & 0.841 & 12.500 & 12.235 & 12.141 & -0.014 & -0.070 & 12.071 & 12.019 & 11.875 & \_ & \_ & \_ \\
W025 & 0.307 & 0.022 & 1.352 & 0.886 & 1.119 & 10.195 & 9.766 & 9.490 & 0.058 & 0.057 & 9.485 & 9.451 & 9.452 & 9.390 & -0.087 & 0.052 \\
W026 & 0.256 & 0.020 & 1.128 & 0.801 & 0.965 & 11.996 & 11.664 & 11.047 & 0.012 & 0.429 & 11.614 & 11.493 & 11.696 & 11.234 & 0.016 & 0.453 \\
W035 & 0.149 & 0.023 & 0.656 & 0.919 & 0.787 & 12.796 & 12.350 & 12.348 & 0.185 & -0.152 & 12.003 & 12.047 & 12.200 & \_ & \_ & \_ \\
W036 & 0.125 & 0.015 & 0.550 & 0.594 & 0.572 & 11.877 & 11.672 & 11.555 & 0.015 & 0.005 & 11.558 & 11.340 & 11.126 & \_ & \_ & \_ \\
W041 & 0.185 & 0.015 & 0.816 & 0.594 & 0.705 & 12.270 & 11.951 & 11.805 & 0.085 & 0.008 & 11.699 & 11.734 & 11.564 & \_ & \_ & \_ \\
W064 & 0.111 & 0.014 & 0.488 & 0.557 & 0.522 & 11.997 & 11.774 & 11.631 & 0.050 & 0.041 & 11.484 & 11.380 & 11.453 & 11.398 & 0.047 & 0.050 \\
W090 & 0.141 & 0.017 & 0.622 & 0.695 & 0.659 & 10.808 & 10.706 & 10.599 & -0.117 & -0.022 & 10.609 & 10.589 & 10.614 & 10.566 & -0.052 & 0.042 \\
W125 & \_ & 0.021 & \_ & 0.862 & 0.862 & 8.823 & 8.692 & 8.596 & -0.155 & -0.072 & 8.568 & 8.590 & 8.512 & 8.573 & -0.116 & -0.069 \\
W161 & 0.280 & 0.029 & 1.236 & 1.183 & 1.209 & 8.156 & 7.732 & 7.453 & 0.023 & 0.043 & 7.323 & 7.219 & 7.187 & 7.246 & -0.027 & -0.070 \\
W166 & \_ & 0.029 & \_ & 1.175 & 1.175 & 8.876 & 8.736 & 8.596 & -0.250 & -0.089 & 8.588 & 8.588 & 8.528 & 8.506 & -0.127 & 0.011 \\
W175 & \_ & 0.022 & \_ & 0.898 & 0.898 & 7.590 & 7.223 & 7.006 & 0.069 & 0.042 & 7.149 & 6.907 & 6.799 & 6.858 & 0.145 & -0.067 \\
W188 & 0.408 & 0.041 & 1.800 & 1.675 & 1.737 & 9.534 & 9.074 & 8.776 & -0.116 & -0.041 & 8.624 & 8.631 & 8.561 & 8.571 & -0.195 & -0.026 \\
W194 & 0.070 & 0.023 & 0.309 & 0.927 & 0.618 & 12.400 & 11.909 & 12.013 & 0.286 & -0.225 & 11.930 & 11.922 & 11.716 & \_ & \_ & \_ \\
W201 & \_ & 0.016 & \_ & 0.646 & 0.646 & 11.948 & 11.681 & 11.571 & 0.053 & -0.016 & 11.478 & 11.567 & 11.644 & \_ & \_ & \_ \\
W202 & 0.215 & 0.022 & 0.948 & 0.882 & 0.915 & 11.821 & 11.440 & 11.231 & 0.077 & 0.030 & 11.057 & 11.006 & 11.034 & \_ & \_ & \_ \\
W203 & 0.090 & 0.019 & 0.397 & 0.756 & 0.576 & \_ & \_ & \_ & \_ & \_ & 12.258 & 12.215 & 11.828 & \_ & \_ & \_ \\
W223 & 0.279 & 0.021 & 1.230 & 0.862 & 1.046 & 9.543 & 9.306 & 9.193 & -0.110 & -0.091 & 9.099 & 9.141 & 9.069 & 9.223 & -0.155 & -0.164 \\
W228 & 0.279 & 0.026 & 1.230 & 1.045 & 1.137 & 10.789 & 10.384 & 10.082 & 0.028 & 0.080 & 10.221 & \_ & 9.842 & \_ & \_ & \_ \\
W231 & 0.266 & 0.023 & 1.172 & 0.947 & 1.060 & 10.590 & 10.213 & 10.057 & 0.025 & -0.051 & 9.946 & 9.940 & 9.908 & 9.950 & -0.109 & -0.052 \\
W238 & 0.130 & 0.020 & 0.573 & 0.797 & 0.685 & 11.539 & 11.274 & 11.078 & 0.038 & 0.062 & 10.941 & 10.803 & 10.883 & 10.732 & 0.064 & 0.145 \\
W239 & 0.182 & 0.019 & 0.802 & 0.756 & 0.779 & 10.396 & 10.274 & 10.164 & -0.136 & -0.042 & 10.087 & 10.072 & 9.899 & 9.294 & -0.070 & 0.598 \\
W243 & 0.209 & 0.024 & 0.921 & 0.980 & 0.950 & 11.985 & 11.632 & 11.395 & 0.038 & 0.051 & 11.107 & 11.108 & 11.221 & 11.335 & -0.104 & -0.123 \\
W251 & 0.211 & 0.022 & 0.930 & 0.878 & 0.904 & 11.399 & 11.044 & 10.884 & 0.055 & -0.017 & 10.785 & 10.839 & 10.715 & \_ & \_ & \_ \\
W259 & 0.254 & 0.023 & 1.119 & 0.919 & 1.019 & 9.614 & 9.315 & 9.193 & -0.039 & -0.077 & 9.104 & 9.124 & 9.033 & 9.152 & -0.131 & -0.128 \\
W267 & 0.171 & 0.021 & 0.754 & 0.837 & 0.796 & 11.653 & 11.462 & 11.366 & -0.073 & -0.059 & 11.237 & 11.237 & 11.190 & 11.360 & -0.086 & -0.177 \\
W269 & 0.288 & 0.026 & 1.269 & 1.069 & 1.169 & 11.303 & 10.854 & 10.647 & 0.061 & -0.021 & 10.432 & 10.366 & 10.255 & \_ & \_ & \_ \\
W273 & 0.194 & 0.023 & 0.855 & 0.947 & 0.901 & 12.256 & 11.901 & 11.624 & 0.056 & 0.101 & 11.615 & 11.554 & 11.222 & \_ & \_ & \_ \\
W275 & 0.207 & 0.020 & 0.912 & 0.801 & 0.857 & 10.810 & 10.651 & 10.530 & -0.125 & -0.046 & 10.530 & 10.447 & 10.631 & 10.769 & -0.010 & -0.146 \\
W276 & 0.163 & 0.021 & 0.718 & 0.862 & 0.790 & 11.973 & 11.573 & 11.392 & 0.138 & 0.027 & 11.307 & 11.277 & 11.204 & 11.075 & -0.056 & 0.122 \\
\hline
\end{tabular}
\label{IR2MASS}
\end{table*}
}
\tiny{
\addtocounter{table}{-1}
\begin{table*}[tbph]
\caption{continued}
\centering
\begin{tabular}{@{\ }l@{\ \ \ }l@{\ \ \ }l@{\ \ \ }l@{\ \ \ }l@{\ \ \ }l@{\ \ \ }l@{\ \ \ }l@{\ \ \ }l@{\ \ \ }l@{\ \ \ }l@{\ \ \ }l@{\ \ \ }l@{\ \ \ }l@{\ \ \ }l@{\ \ \ }l@{\ \ \ }l@{\ }}
\hline
W281 & 0.150 & 0.023 & 0.661 & 0.927 & 0.794 & 12.087 & 11.928 & 11.773 & -0.104 & 0.000 & 11.796 & 11.747 & 11.778 & \_ & \_ & \_ \\
W289 & 0.182 & 0.021 & 0.802 & 0.850 & 0.826 & 11.186 & 10.996 & 10.895 & -0.084 & -0.060 & 10.611 & 10.732 & 10.650 & 10.476 & -0.211 & 0.166 \\
W292 & 0.167 & 0.018 & 0.736 & 0.732 & 0.734 & 12.483 & 12.280 & 12.137 & -0.040 & 0.000 & 12.036 & 12.047 & 11.642 & \_ & \_ & \_ \\
W299 & 0.146 & 0.020 & 0.644 & 0.801 & 0.722 & 12.445 & 12.080 & 11.664 & 0.126 & 0.275 & 12.040 & 12.117 & \_ & \_ & -0.155 & \_ \\
W300 & 0.200 & 0.021 & 0.881 & 0.842 & 0.861 & 11.185 & 10.973 & 10.582 & -0.074 & 0.223 & 10.845 & 10.822 & 10.685 & \_ & \_ & \_ \\
W305 & 0.312 & 0.023 & 1.375 & 0.943 & 1.159 & 10.488 & 10.050 & 9.795 & 0.054 & 0.029 & 9.641 & 9.638 & 9.598 & \_ & \_ & \_ \\
W306 & 0.186 & 0.021 & 0.820 & 0.862 & 0.841 & 10.863 & 10.593 & 10.539 & -0.009 & -0.110 & 10.435 & 10.485 & 10.537 & \_ & \_ & \_ \\
W307 & 0.216 & 0.027 & 0.952 & 1.081 & 1.017 & 11.733 & 11.314 & 10.341 & 0.082 & 0.774 & 10.954 & 10.762 & \_ & \_ & \_ & \_ \\
W313 & 0.168 & 0.019 & 0.740 & 0.789 & 0.765 & 11.594 & 11.383 & 11.247 & -0.043 & -0.013 & 11.140 & 11.104 & 10.910 & 10.709 & -0.047 & 0.194 \\
W323 & 0.182 & 0.022 & 0.802 & 0.886 & 0.844 & 11.906 & 11.655 & 11.534 & -0.029 & -0.044 & 11.447 & 11.462 & 11.683 & \_ & \_ & \_ \\
W336 & 0.185 & 0.020 & 0.815 & 0.825 & 0.820 & 11.804 & 11.612 & 11.455 & -0.080 & -0.003 & 11.391 & 11.232 & 10.900 & \_ & \_ & \_ \\
W343 & 0.267 & 0.022 & 1.177 & 0.911 & 1.044 & 9.476 & 9.119 & 8.900 & 0.011 & 0.015 & 8.742 & 8.744 & 8.669 & 8.831 & -0.115 & -0.172 \\
W344 & \_ & 0.014 & \_ & 0.581 & 0.581 & 11.907 & 11.527 & 11.382 & 0.187 & 0.032 & \_ & 11.266 & \_ & \_ & \_ & \_ \\
W347 & 0.184 & 0.024 & 0.809 & 0.972 & 0.890 & 12.512 & 12.199 & 12.077 & 0.018 & -0.052 & 11.949 & 12.015 & 11.625 & \_ & \_ & \_ \\
W351 & 0.254 & 0.020 & 1.119 & 0.809 & 0.964 & 9.994 & 9.783 & 9.670 & -0.109 & -0.075 & 9.667 & 9.608 & 9.578 & 9.368 & -0.046 & 0.201 \\
W364 & 0.164 & 0.019 & 0.723 & 0.768 & 0.746 & 11.734 & 11.418 & 11.131 & 0.069 & 0.141 & 10.717 & 10.948 & 10.476 & \_ & \_ & \_ \\
W368 & 0.125 & 0.014 & 0.551 & 0.553 & 0.552 & 12.091 & 11.875 & 11.694 & 0.033 & 0.073 & 11.704 & 11.735 & 11.790 & \_ & \_ & \_ \\
W371 & 0.175 & 0.021 & 0.771 & 0.833 & 0.802 & 11.671 & 11.398 & 11.163 & 0.007 & 0.078 & 10.907 & 10.845 & 10.829 & \_ & \_ & \_ \\
W388 & 0.165 & 0.026 & 0.727 & 1.053 & 0.890 & 12.186 & 11.973 & 11.812 & -0.082 & -0.013 & 11.621 & 11.518 & 11.271 & \_ & \_ & \_ \\
W389 & 0.070 & 0.015 & 0.309 & 0.626 & 0.467 & 12.982 & 12.812 & 12.631 & 0.015 & 0.090 & 12.439 & \_ & \_ & \_ & \_ & \_ \\
W400 & 0.182 & 0.018 & 0.802 & 0.732 & 0.767 & 11.001 & 10.689 & 10.394 & 0.058 & 0.145 & 10.166 & 10.018 & 9.825 & 9.378 & 0.065 & 0.440 \\
W409 & 0.152 & 0.015 & 0.670 & 0.622 & 0.646 & 11.585 & 11.403 & 9.908 & -0.032 & 1.369 & 11.405 & 11.254 & \_ & \_ & 0.081 & \_ \\
W444 & 0.221 & 0.021 & 0.974 & 0.854 & 0.914 & 10.300 & 9.958 & 9.744 & 0.039 & 0.035 & 9.647 & 9.668 & 9.458 & 9.088 & -0.120 & 0.361 \\
W445 & 0.084 & 0.017 & 0.370 & 0.675 & 0.523 & 12.840 & 12.728 & 12.596 & -0.061 & 0.030 & 12.744 & 12.690 & \_ & \_ & -0.003 & \_ \\
W455 & \_ & 0.022 & \_ & 0.882 & 0.882 & 10.815 & 10.667 & 10.540 & -0.145 & -0.045 & 10.528 & 10.518 & 10.348 & 10.252 & -0.086 & 0.088 \\
W469 & \_ & 0.020 & \_ & 0.817 & 0.817 & 9.576 & 9.449 & 9.310 & -0.144 & -0.021 & 9.283 & 9.122 & 8.740 & \_ & \_ & \_ \\
W472 & 0.177 & 0.020 & 0.780 & 0.829 & 0.805 & 11.610 & 11.507 & 11.397 & -0.164 & -0.047 & 11.441 & 11.366 & 11.406 & \_ & \_ & \_ \\
W473 & 0.140 & 0.017 & 0.617 & 0.703 & 0.660 & 11.508 & 11.416 & 11.333 & -0.127 & -0.046 & 11.272 & 11.243 & 10.913 & 10.160 & -0.043 & 0.747 \\
W484 & 0.182 & 0.018 & 0.802 & 0.740 & 0.771 & 11.237 & 11.144 & 11.052 & -0.163 & -0.059 & 10.982 & 11.008 & 11.000 & \_ & \_ & \_ \\
W490 & 0.151 & 0.019 & 0.666 & 0.785 & 0.725 & 11.896 & 11.798 & 11.733 & -0.142 & -0.077 & 11.707 & 11.626 & 11.221 & \_ & \_ & \_ \\
W495 & 0.115 & 0.019 & 0.507 & 0.781 & 0.644 & 12.734 & 12.746 & 12.555 & -0.226 & 0.065 & 12.667 & 12.521 & \_ & \_ & 0.076 & \_ \\
W496 & 0.113 & 0.020 & 0.498 & 0.805 & 0.651 & 12.523 & 12.346 & 12.173 & -0.039 & 0.046 & 12.074 & 12.103 & \_ & \_ & -0.100 & \_ \\
W504 & 0.147 & 0.020 & 0.648 & 0.821 & 0.735 & 11.677 & 11.603 & 11.510 & -0.170 & -0.051 & 11.504 & 11.443 & 11.206 & \_ & \_ & \_ \\
W515 & 0.150 & 0.018 & 0.661 & 0.732 & 0.696 & 11.496 & 11.181 & 11.036 & 0.084 & 0.009 & 11.036 & 10.955 & 11.021 & \_ & \_ & \_ \\
W519 & 0.094 & 0.019 & 0.414 & 0.756 & 0.585 & 12.083 & 11.772 & 11.653 & 0.117 & 0.005 & 11.588 & 11.525 & 11.647 & \_ & \_ & \_ \\
W520 & 0.162 & 0.020 & 0.713 & 0.825 & 0.769 & 10.498 & 10.358 & 10.264 & -0.115 & -0.056 & 10.231 & 10.144 & 10.054 & 9.744 & 0.004 & 0.303 \\
W536 & 0.200 & 0.018 & 0.881 & 0.732 & 0.807 & 10.471 & 10.357 & 10.260 & -0.154 & -0.061 & 10.248 & 10.234 & 10.130 & 9.639 & -0.074 & 0.483 \\
W541 & 0.216 & 0.020 & 0.952 & 0.797 & 0.874 & 11.579 & 11.351 & 11.194 & -0.062 & -0.014 & 11.181 & 11.183 & 11.095 & 11.032 & -0.097 & 0.055 \\
W550 & 0.174 & 0.019 & 0.767 & 0.781 & 0.774 & 12.135 & 11.931 & 11.789 & -0.053 & -0.009 & 11.782 & 11.754 & 11.340 & \_ & \_ & \_ \\
\hline
\end{tabular}
\label{IR2MASS}
\end{table*}
}
\tiny{
\addtocounter{table}{-1}
\begin{table*}[tbph]
\caption{continued}
\centering
\begin{tabular}{@{\ }l@{\ \ \ }l@{\ \ \ }l@{\ \ \ }l@{\ \ \ }l@{\ \ \ }l@{\ \ \ }l@{\ \ \ }l@{\ \ \ }l@{\ \ \ }l@{\ \ \ }l@{\ \ \ }l@{\ \ \ }l@{\ \ \ }l@{\ \ \ }l@{\ \ \ }l@{\ \ \ }l@{\ }}
\hline
W567 & 0.168 & 0.018 & 0.740 & 0.732 & 0.736 & 11.047 & 10.943 & 10.839 & -0.140 & -0.040 & 10.886 & 10.955 & 10.796 & 10.912 & -0.149 & -0.123 \\
W568 & \_ & 0.014 & \_ & 0.577 & 0.577 & 12.015 & 11.757 & 11.387 & 0.067 & 0.257 & 11.543 & 11.459 & 11.272 & \_ & 0.021 & \_ \\
W570 & \_ & 0.015 & \_ & 0.606 & 0.606 & 11.164 & 10.893 & 10.814 & 0.070 & -0.039 & 10.810 & 10.778 & 10.629 & 10.674 & -0.034 & -0.051 \\
W582 & \_ & 0.014 & \_ & 0.549 & 0.549 & 10.837 & 10.490 & 10.334 & 0.165 & 0.049 & 10.369 & 10.314 & 10.334 & 10.210 & -0.005 & 0.119 \\
W587 & 0.149 & 0.017 & 0.657 & 0.675 & 0.666 & 10.909 & 10.834 & 10.763 & -0.146 & -0.059 & 10.729 & 10.855 & 10.849 & \_ & -0.198 & \_ \\
W588 & 0.170 & 0.012 & 0.749 & 0.480 & 0.614 & 11.321 & 11.160 & 11.028 & -0.043 & 0.012 & 11.030 & 11.159 & 10.812 & 10.958 & -0.196 & -0.152 \\
W590 & 0.070 & 0.014 & 0.309 & 0.565 & 0.437 & 11.396 & 11.261 & 11.190 & -0.010 & -0.014 & 11.148 & 11.168 & 11.444 & \_ & -0.067 & \_ \\
W591 & 0.146 & 0.017 & 0.644 & 0.703 & 0.673 & 10.679 & 10.599 & 10.494 & -0.143 & -0.026 & 10.489 & 10.451 & 10.550 & \_ & -0.035 & \_ \\
W596 & \_ & 0.012 & \_ & 0.480 & 0.480 & 11.505 & 11.233 & 10.723 & 0.113 & 0.416 & 10.985 & 11.087 & 11.076 & \_ & -0.154 & \_ \\
W597 & 0.179 & 0.018 & 0.789 & 0.740 & 0.764 & 11.241 & 11.134 & 11.071 & -0.146 & -0.086 & 11.012 & 11.056 & 11.032 & \_ & -0.127 & \_ \\
W607 & 0.140 & 0.016 & 0.617 & 0.646 & 0.632 & 11.158 & 10.865 & 10.713 & 0.083 & 0.029 & 10.644 & 10.656 & 10.516 & 10.669 & -0.081 & -0.159 \\
W625 & 0.112 & 0.023 & 0.494 & 0.947 & 0.720 & 12.489 & 12.223 & 12.119 & 0.027 & -0.037 & 12.084 & 11.977 & 11.987 & \_ & \_ & \_ \\
W626 & \_ & 0.017 & \_ & 0.699 & 0.699 & 12.045 & 11.656 & 11.544 & 0.157 & -0.025 & 11.409 & 11.446 & 11.498 & \_ & \_ & \_ \\
W627 & 0.200 & 0.019 & 0.881 & 0.789 & 0.835 & 12.504 & 11.349 & 10.574 & 0.878 & 0.612 & \_ & \_ & \_ & \_ & \_ & \_ \\
W632 & 0.080 & 0.018 & 0.353 & 0.740 & 0.546 & 12.816 & 12.620 & 12.480 & 0.015 & 0.033 & 12.366 & 12.393 & \_ & \_ & -0.086 & \_ \\
W633 & \_ & 0.016 & \_ & 0.650 & 0.650 & 12.049 & 11.712 & 10.695 & 0.121 & 0.890 & 11.541 & 11.520 & 11.339 & 11.785 & -0.050 & -0.452 \\
W639 & 0.130 & 0.017 & 0.573 & 0.699 & 0.636 & 11.637 & 11.440 & 11.357 & -0.014 & -0.041 & 11.246 & 11.199 & 10.850 & \_ & -0.022 & \_ \\
\hline                                 
WFI017 & 0.172 & 0.018 & 0.757 & 0.728 & 0.742 & 12.581 & 11.882 & 11.083 & 0.453 & 0.654 & 9.708 & 9.108 & 8.455 & 7.022 & 0.519 & 1.426 \\
W031 & 0.177 & 0.025 & 0.781 & 1.016 & 0.898 & 12.729 & 12.502 & 12.420 & -0.071 & -0.093 & 12.438 & 12.279 & 12.069 & 11.5* & 0.062 & 0.561* \\
W080 & 0.552 & 0.032 & 2.432 & 1.301 & 1.866 & 10.103 & 9.596 & 9.304 & -0.112 & -0.072 & 9.062 & 9.015 & 8.936 & 8.900 & -0.155 & 0.019 \\
W235 & 0.240 & 0.022 & 1.058 & 0.898 & 0.978 & 8.563 & 8.121 & 7.714 & 0.118 & 0.216 & 7.180 & 6.844 & 6.525 & 6.226 & 0.230 & 0.290 \\
W301 & 0.217 & 0.021 & 0.956 & 0.854 & 0.905 & 10.631 & 10.402 & 10.304 & -0.071 & -0.079 & 10.248 & 10.244 & 10.145 & 10.390 & -0.094 & -0.253 \\
W483 & 0.172 & 0.018 & 0.758 & 0.736 & 0.747 & 9.875 & 9.762 & 9.608 & -0.135 & 0.008 & 9.453 & 9.293 & 9.146 & 8.956 & 0.079 & 0.183 \\
W500 & 0.151 & 0.017 & 0.666 & 0.703 & 0.684 & 10.055 & 9.904 & 9.785 & -0.076 & -0.015 & 9.646 & 9.447 & 9.332 & 9.072 & 0.125 & 0.254 \\
W503 & 0.232 & 0.017 & \_ & 0.687 & 0.687 & 8.726 & 8.430 & 8.105 & 0.034 & 0.171 & 7.044 & 6.794 & 6.594 & 6.305 & 0.164 & 0.282 \\
W601 & \_ & 0.017 & \_ & 0.691 & 0.691 & 9.703 & 9.523 & 9.398 & -0.049 & -0.010 & 9.298 & 9.285 & 9.261 & 9.134 & -0.062 & 0.121 \\
\hline                                 
W245 & \_ & \_ & \_ & \_ & 0.791 & 11.199 & 10.475 & 9.776 & 0.462 & 0.545 & 8.425 & 7.823 & 7.482 & 7.073 & 0.516 & 0.402 \\
W494 & \_ & \_ & \_ & \_ & 0.791 & 11.770 & 10.690 & 9.543 & 0.818 & 0.993 & 7.736 & 7.146 & 6.428 & 5.465 & 0.504 & 0.956 \\
\hline
\end{tabular}
\label{IR2MASS}
\begin{flushleft}
* As the magnitude is not present, we give to the star the  limiting magnitude of the archive.
\end{flushleft}

\end{table*}
}

\end{landscape}
\end{flushleft}


%
\begin{longtable}{lllllllllllll}
\caption{\label{preuvePMS} 
Indications for each star about its nature:
in col. 2, about the binarity, in col. 3 from IR data of 2MASS (Y:med is for a mediul infrared excess, 
Y: strong for a strong infrared excess),
in col. 4 from IR data of SPITZER (class I, class II, MS for Main Sequence, 
Y:med is for a medium infrared excess without obvious classification in classI/II or MS),
in col. 5, the age interpolated from MS HR diagrams,
in col. 6, according to MS age interpolated too old for this region.
In col. 7, indications on the nature of the stars from previous studies \citep{walker1961,ref2,hillenbrand1993,herbig2001,evans2005,indeb07}
are given. In col. 8, the mass of the star is provided. In col. 9 and 10, the membership probabilities from \citet[][Tu86]{ref1} 
and from \citet[][Bel99]{ref2} are given.
In col. 11, the deduced status (PMS or MS) for each star in function of previous information is given.
In col. 12, the age interpolated from PMS tracks, is given, and in the last column, the age (in Myears) finally selected 
for each star in function of its nature is provided. The error in this age ranges from 0.02 to 25 Myear depending on the value of the
age. '\_' stands for no available information or data.}
\\
\hline
\hline
Star & Bin? & 2MASS & class & age MS & old & other & mass & Tu86 & Bel99 & new status & age PMS & age kept \\
\hline
\endfirsthead
\caption{continued.}\\
\hline
\hline
Star & Bin? & 2MASS & class & age MS & old & other & mass & Tu86 & Bel99 & new status & age PMS & age kept \\
\hline
\hline  
\endhead
\hline
\endfoot
WFI014a & N & N & \_ & 27 & no &  & 8.9 & \_ & \_ & MS? & 0.04 & 27 \\
WFI019 & N & N & \_ & 37 & no &  & 7.8 & \_ & 0 & MS  & 0.03 & 37 \\
WFI125 & N & N & \_ & 27 & no &  & 8.9 & \_ & \_ & MS? & 0.01 & 27 \\
WFI152 & N & N & \_ & 646 & yes &  & 2.3 & \_ & 26 & PMS? & 2.2 & 2.2 \\
W002 & N & N & \_ & \_ & \_ &  & \_ & 29 & 8 & MS & \_ & \_ \\
W018 & N & N & \_ & 33 & no &  & 1.7 & 0 & 11 & MS & 8.6 & 33 \\
W024 & N & N & \_ & 717 & yes &  & 2.4 & \_ & 5 & PMS? & 2 & 2 \\
W025 & yes & N & \_ & 1 & no &  & 9.4 & 94 & 53 & MS & \_ & 1 \\
W026 & N & Y:med & Y:med & 448 & yes &  & 2.7 & 78 & 62 & PMS & 1 & 1 \\
W035 & N & N & \_ & 28 & no &  & 2.3 & 90 & 51 & PMS? & \_ & 28$>$ \\
W036 & N & N & \_ & 510 & yes & PMS$^{1}$ & 1.8 & 71 & 49 & PMS? & 0.9 & 0.9 \\
W041 & N & N & \_ & \_ & \_ &  & \_ & 0 & 42 & MS & \_ & \_ \\
W064 & N & N & \_ & 55 & yes &  & 6.5 & 0 & 11 & MS & 0.02 & 55 \\
W090 & N & N & \_ & 62 & yes &  & 4.3 & 93 & 77 & PMS & 1 & 1 \\
W125 & yes & N & MS & \_ & \_ &  & \_ & 94 & 79 & MS & \_ & \_ \\
W161 & yes & N & MS & 3 & no & MS$^{2}$ & 23 & 94 & 70 & MS & \_ & 3 \\
W166 & N & N & MS & \_ & \_ &  & \_ & 94 & 95 & MS & \_ & \_ \\
W175 & yes & N & MS & \_ & \_ & MS$^{2}$ & \_ & 94 & 95 & MS & \_ & \_ \\
W188 & yes & N & MS & 0.5 & no & MS$^{2}$ & 15.7 & 87 & 68 & MS & \_ & 0.5 \\
W194 & N & N & \_ & \_ & \_ &  & \_ & 73 & 67 & MS & \_ & \_ \\
W201 & N & N & \_ & 64 & yes &  & 6.2 & 0 & 5 & MS? & 0.03 & 64? \\
W202 & N & N & \_ & 122 & yes & PMS$^{2}$ & 3.4 & 0 & 0 & PMS & 1.1 & 1.1 \\
W203 & N & \_ & \_ & 473 & yes &  & 2.4 & 93 & 90 & PMS & 2.3 & 2.3 \\
W223 & N & N & MS & 0.9 & no &  & 9.8 & 94 & 89 & MS & \_ & 0.9 \\
W228 & N & N & \_ & 1.4 & no &  & 7.4 & 91 & 86 & MS & \_ & 1.4 \\
W231 & N & N & MS & 0.9 & no &  & 9.6 & 93 & 78 & MS & \_ & 0.9 \\
W238 & N & N & \_ & 459 & yes &  & 2.5 & 94 & 56 & PMS & 0.9 & 0.9 \\
W239 & yes & N & Y:med & 2.5 & no &  & 6.8 & 94 & 67 & MS & \_ & 2.5 \\
W243 & yes & N & MS & 4.9 & no &  & 4.4 & 93 & 93 & MS & \_ & 4.9 \\
W251 & N & N & \_ & 2.5 & no &  & 6.8 & 0 & \_ & MS & \_ & 2.5 \\
W259 & N & N & MS & 0.5 & no &  & 14.1 & 94 & 87 & MS & \_ & 0.5 \\
W267 & yes & N & MS & 82 & yes & PMS$^{2}$ & 5 & 92 & 87 & PMS & 0.3 & 0.3 \\
W269 & N & N & \_ & 1.2 & no &  & 7.7 & \_ & 26 & MS & \_ & 1.2 \\
W273 & N & N & \_ & 333 & yes & PMS$^{2}$ & 2.9 & 94 & 58 & PMS & 0.9 & 0.9 \\
W275 & yes & N & MS & 1.3 & no &  & 7.7 & 0 & 0 & MS & \_ & 1.3 \\
W276 & N & N & \_ & 53 & yes? & PMS$^{2}$ & 3.7 & 89 & 69 & PMS & 1 & 1 \\
W281 & N & N & \_ & 52 & yes &  & 6.7 & 94 & 96 & PMS & 0.02 & 0.02 \\
W289 & N & N & \_ & 5.5 & no &  & 5.9 & 91 & 95 & MS? & \_ & 5.5 \\
W292 & N & N & \_ & 232 & yes &  & 2.7 & 29 & 81 & PMS & 2 & 2 \\
W299 & yes & Y:med & \_ & 33 & no & PMS$^{2}$ & 1.7 & 0 & 55 & PMS & 10 & 10 \\
W300 & N & N & \_ & 5.3 & no &  & 6 & \_ & 16 & MS & \_ & 5.3 \\
W305 & N & N & \_ & 0.4 & no &  & 11.3 & 92 & 84 & MS & \_ & 0.4 \\
W306 & N & N & \_ & 1.3 & no &  & 7.5 & 89 & 96 & MS & 0.4 & 1.3 \\
W307 & N & Y:strong  & \_ & 25 & no &  & 6.3 & \_ & 76 & PMS & 0.03 & 0.03 \\
W313 & yes & N & \_ & 104 & yes & PMS$^{2}$ & 4.3 & 81 & 90 & PMS & 2 & 2 \\
W323 & N & N & \_ & 81 & yes &  & 3.9 & 2 & 58 & PMS? & 1.1 & 1.1? \\
W336 & N & N & \_ & 106 & yes &  & 4.6 & 87 & 88 & PMS & 0.3 & 0.3 \\
W343 & yes & N & MS & \_ & \_ &  & \_ & 93 & 83 & MS & \_ & \_ \\
W344 & N & N & \_ & \_ & \_ &  & \_ & 0 & 0 & MS? & \_ & \_ \\
W347 & N & N & \_ & 442 & yes &  & 2.8 & \_ & 66 & PMS & 0.6 & 0.6 \\
W351 & N & N & \_ & 0.6 & no &  & 10.8 & 94 & 91 & MS & \_ & 0.6 \\
W364 & yes & N & \_ & 193 & yes? &  & 3.2 & 93 & \_ & PMS & 1 & 1 \\
W368 & N & N & \_ & 83 & yes? &  & 5.5 & 27 & 10 & MS? & 0.08 & 83? \\
W371 & N & N & \_ & 3 & no &  & 4.7 & 40 & 72 & MS & \_ & 3 \\
W388 & N & N & \_ & 94 & yes & PMS$^{2}$ & 4.1 & 91 & 90 & PMS & 2 & 2 \\
W389 & N & N & \_ & 587 & yes &  & 2.4 & 0 & 1 & PMS? & 2 & 2? \\
W400 & yes & N & Y:med & 211 & yes & PMS$^{2}$ & 3.5 & 93 & 82 & PMS & 1 & 1 \\
W409 & yes & Y:strong  & \_ & 8 & no &  & 5.1 & 94 & 66 & PMS & \_ & \_ \\
W444 & yes & N & Y:med & 1.2 & no &  & 7.7 & 0 & 70 & PMS & \_ & \_ \\
W445 & N & N & \_ & 37 & no &  & 2.1 & 22 & 58 & PMS? & 7 & 7 \\
W455 & N & N & \_ & 442 & yes & PMS$^{2}$ & 2.8 & 7 & 7 & PMS & 0.3 & 0.3 \\
W469 & yes & N & \_ & \_ & \_ &  & \_ & 94 & 70 & MS & \_ & \_ \\
W472 & yes & N & \_ & 73 & yes &  & 5.8 & 91 & 12 & PMS? & 0.2 & 0.2? \\
W473 & yes & N & Y:med & 92 & yes & accr disk$^{3}$ & 5.1 & 92 & 48 & PMS & 0.1 & 0.1 \\
W484 & N & N & \_ & 122 & yes? &  & 3.4 & 94 & 38 & PMS? & 1.2 & 1.2? \\
W490 & N & N & \_ & 259 & yes &  & 3.3 & 92 & 67 & PMS & 0.5 & 0.5 \\
W495 & N & N & \_ & 38 & yes? &  & 1.9 & 90 & 56 & PMS & 6 & 6 \\
W496 & N & N & \_ & 512 & yes &  & 2.5 & 27 & 34 & PMS? & 0.8 & 0.8? \\
W504 & N & N & \_ & 180 & yes &  & 3.5 & 92 & 25 & PMS? & 1 & 1? \\
W515 & N & N & \_ & 33 & yes? &  & 1.7 & 90 & 68 & PMS & 8 & 8 \\
W519 & N & N & \_ & 128 & yes &  & 4.5 & 0 & 0 & MS? & 0.7 & 128? \\
W520 & N & N & Y:med & 110 & yes &  & 4.8 & 94 & 46 & PMS & 0.2 & 0.2 \\
W536 & yes & N & Y:med & 6 & no &  & 7.6 & 94 & 39 & PMS? & \_ & 6$>$? \\
W541 & N & N & \_ & 12 & yes? &  & 6.2 & 92 & 56 & MS? & 0.2 & 12? \\
W550 & N & N & \_ & 33 & yes? &  & 1.7 & 91 & 38 & PMS? & 8 & 8? \\
W567 & N & N & MS & 16 & no &  & 5.7 & 5 & 17 & MS & 0.03 & 16 \\
W568 & N & Y:med & \_ & \_ & \_ &  & \_ & 29 & 15 & MS & \_ & \_ \\
W570 & N & N & MS & \_ & \_ &  & \_ & 0 & 5 & MS & \_ & \_ \\
W582 & yes & N & \_ & 457 & yes &  & 2.8 & 0 & 7 & PMS? & 0.7 & 0.7? \\
W587 & N & N & \_ & 6 & no &  & 5.7 & \_ & 35 & MS & \_ & 6 \\
W588 & N & N & MS & 637 & yes &  & 2.1 & \_ & 18 & PMS? & 2 & 2? \\
W590 & N & N & \_ & 286 & yes &  & 3.4 & 92 & 34 & PMS? & 0.4 & 0.4? \\
W591 & N & N & \_ & 102 & yes &  & 4.3 & 82 & 11 & PMS? & 0.2 & 0.2? \\
W596 & N & Y:strong & \_ & \_ & \_ &  & \_ & 15 & 16 & PMS & \_ & \_ \\
W597 & N & N & \_ & 7 & no &  & 5.4 & 94 & 38 & MS & \_ & 7 \\
W607 & N & N & MS & 4 & no &  & 4.5 & \_ & 18 & MS & \_ & 4 \\
W625 & N & N & \_ & 271 & yes &  & 2.8 & 88 & 57 & PMS & 0.9 & 0.9 \\
W626 & N & N & \_ & \_ & no &  & \_ & 1 & 23 & MS & \_ & \_ \\
W627 & N & Y:strong  & \_ & 28 & no &  & 2.3 & 92 & 21 & PMS? & 2.8 & 2.8? \\
W632 & N & N & \_ & 498 & yes &  & 2.4 & 0 & 0 & PMS? & 1.6 & 1.6? \\
W633 & N & Y:strong  & MS & \_ & \_ &  & \_ & 0 & 2 & MS? & \_ & \_ \\
W639 & N & N & \_ & 33 & no &  & 1.7 & 93 & 15 & MS? & 8 & 33? \\
\hline                         
WFI017 & N & Y:strong, HBe/Ae & class I  & 239 & yes & accr disk$^{3}$ & 2.3 & \_ & 1 & PMS & 3 & 3 \\
W031 & N & N & class II? & 432 & no? &  & 2.7 & 1 & 21 & PMS? & 1 & 1 \\
W080 & N & N & MS & 0.5 & no &  & 8.9 & 0 & 24 & MS & \_ & 0.5 \\
W235 & N & Y:med & Y:med & 12: & yes &  & 13.9: & 94 & 96 & PMS & 0.02 & 0.02 \\
W301 & N & N & MS & 7 & no &  & 7 & 94 & 92 & MS & \_ & 7 \\
W483 & N & N & Y:med & 79 & yes & PMS$^{4}$ & 5.4 & 93 & 72 & PMS & 0.1 & 0.1 \\
W500 & N & N & Y:med & 85 & yes & PMS$^{4}$ & 5.4 & 94 & 54 & PMS & 0.1 & 0.1 \\
W503 & yes & Y:med & Y:med & 7: & yes &  & 21.7: & 94 & 40 & MS?, bin & 0.02 & 7? \\
W601 & N & N & MS? & \_ & \_ & PMS$^{5}$ & 10.2$^{5}$ & 93 & 43 & PMS? & 0.016$^{5}$ & 0.016$^{5}$ \\
\hline                         
W245 & N & Y:strong, HBe/Ae & class II & \_ & \_ & PMS$^{2}$ & \_ & 0 & 1 & PMS & \_ & \_ \\
W494 & N & Y:strong, HBe/Ae & class II & \_ & \_ & PMS$^{2}$ & \_ & 1 & \_ & PMS & \_ & \_ \\
\end{longtable}
\begin{flushleft}
$^{1}$ indication from \citet{ogura02}\\
$^{2}$ indication from \citet{walker1961}\\
$^{3}$ indication from \citet{indeb07}\\
$^{4}$ indication from \citet{ref1}\\
$^{5}$ indication from \citet{alecian08}.
\end{flushleft}

%
%

%
%

%
%

%
%


\twocolumn
\addtocounter{figure}{+6}
\begin{figure}[htpb]
    \centering
    \resizebox{\hsize}{!}{\includegraphics[angle=-90]{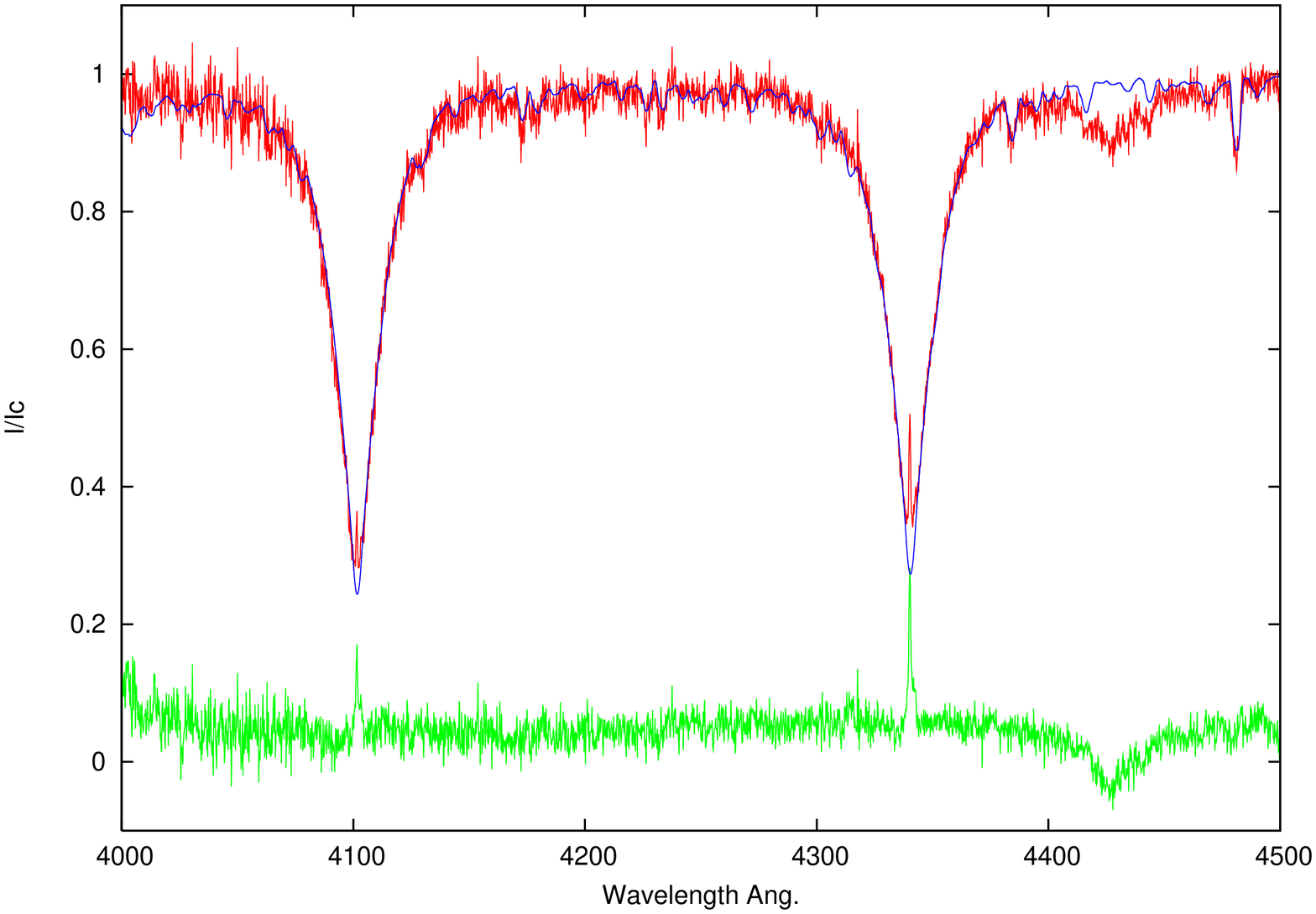}}
    \resizebox{\hsize}{!}{\includegraphics[angle=0]{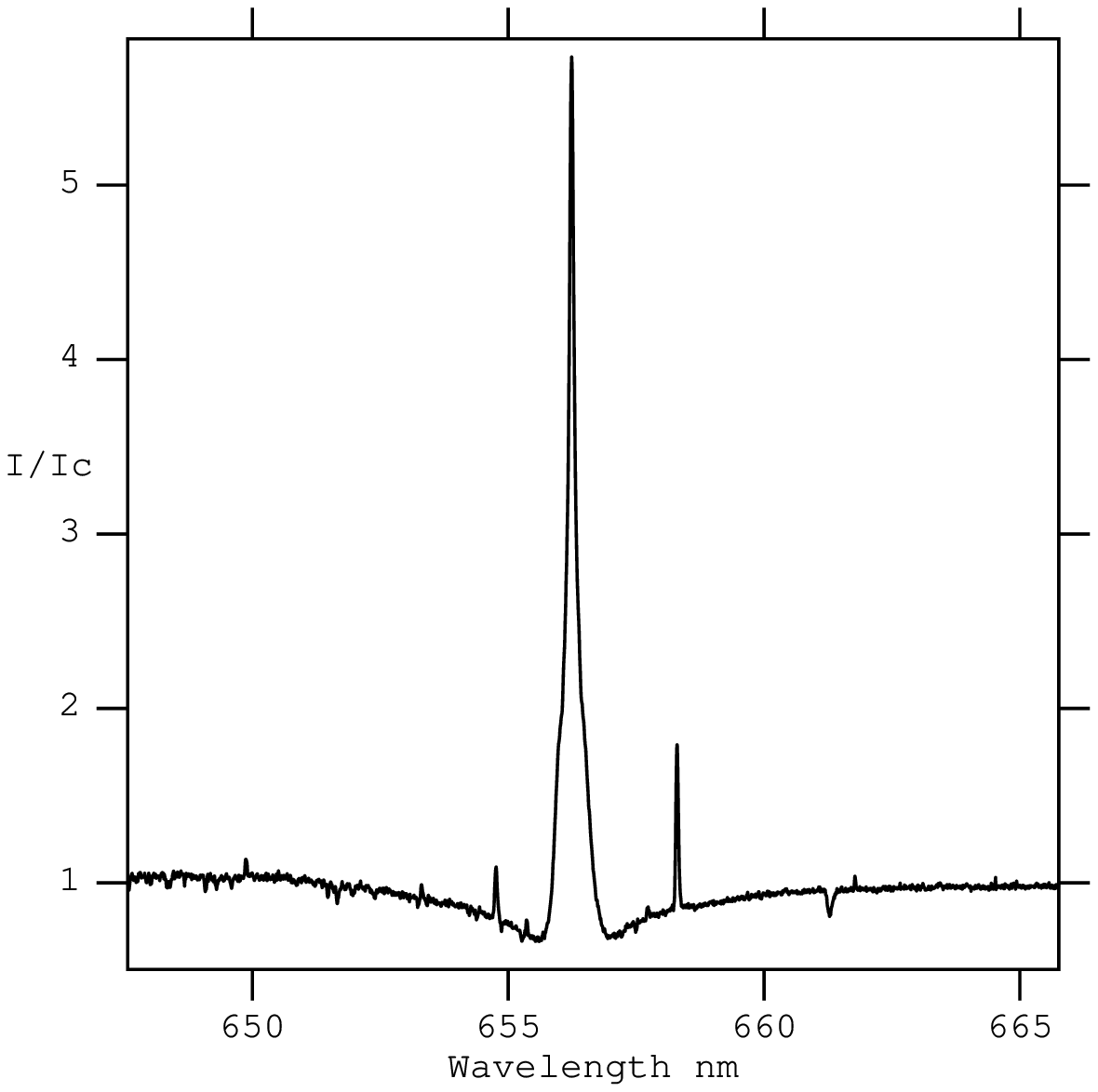}}
    \resizebox{\hsize}{!}{\includegraphics[angle=0]{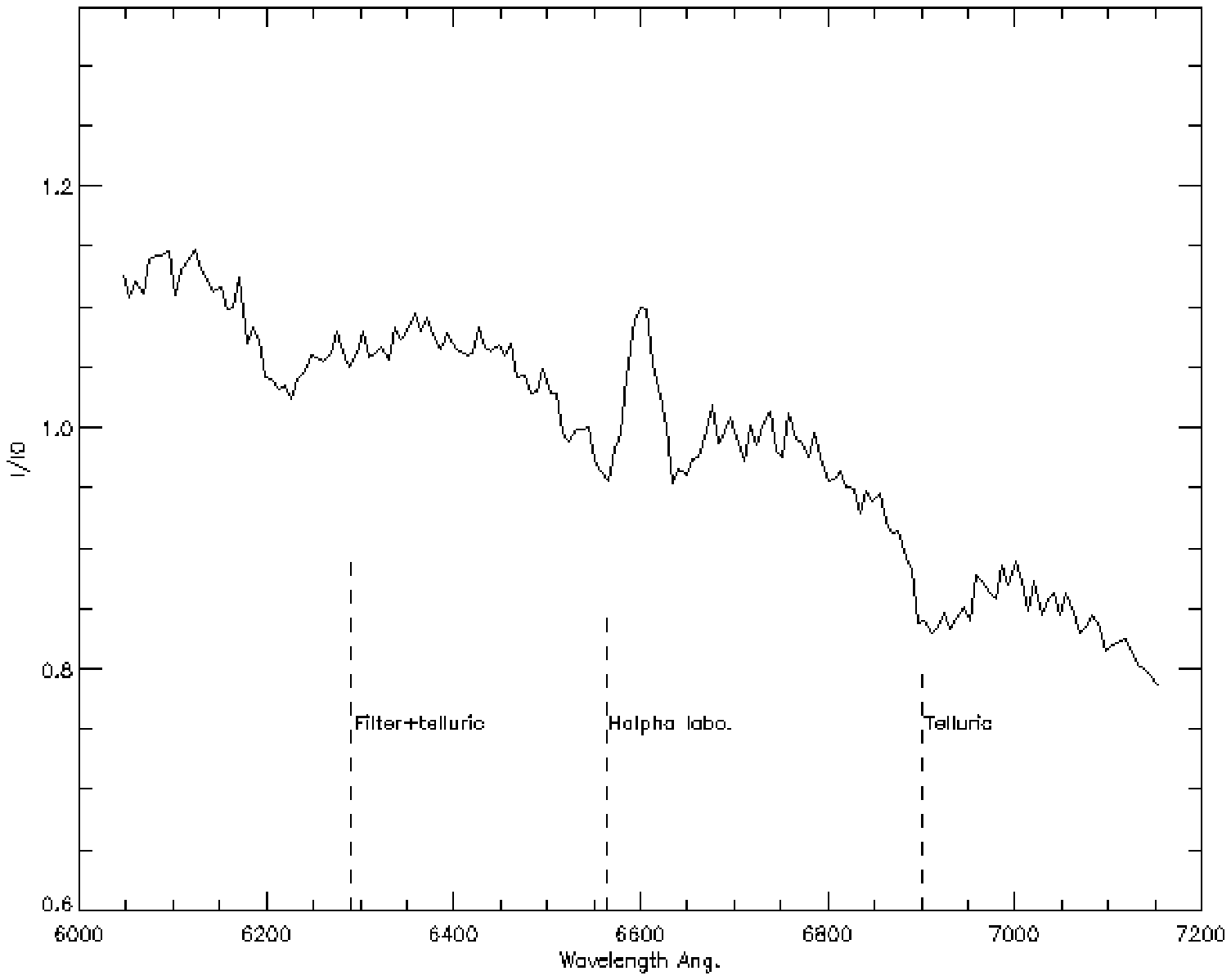}}
    \caption{ELS WFI017. Top panel: fit of the LR2  range of the spectrum for the
fundamental parameters determination. The observed spectrum is in red, the fit
in blue, the residuals in green;  middle panel: H$\alpha$ line from the
VLT-GIRAFFE; bottom panel: H$\alpha$ line from the WFI-spectro.}
    \label{Hastar017}
\end{figure}
\begin{figure}[htpb]
    \centering
    \resizebox{\hsize}{!}{\includegraphics[angle=-90]{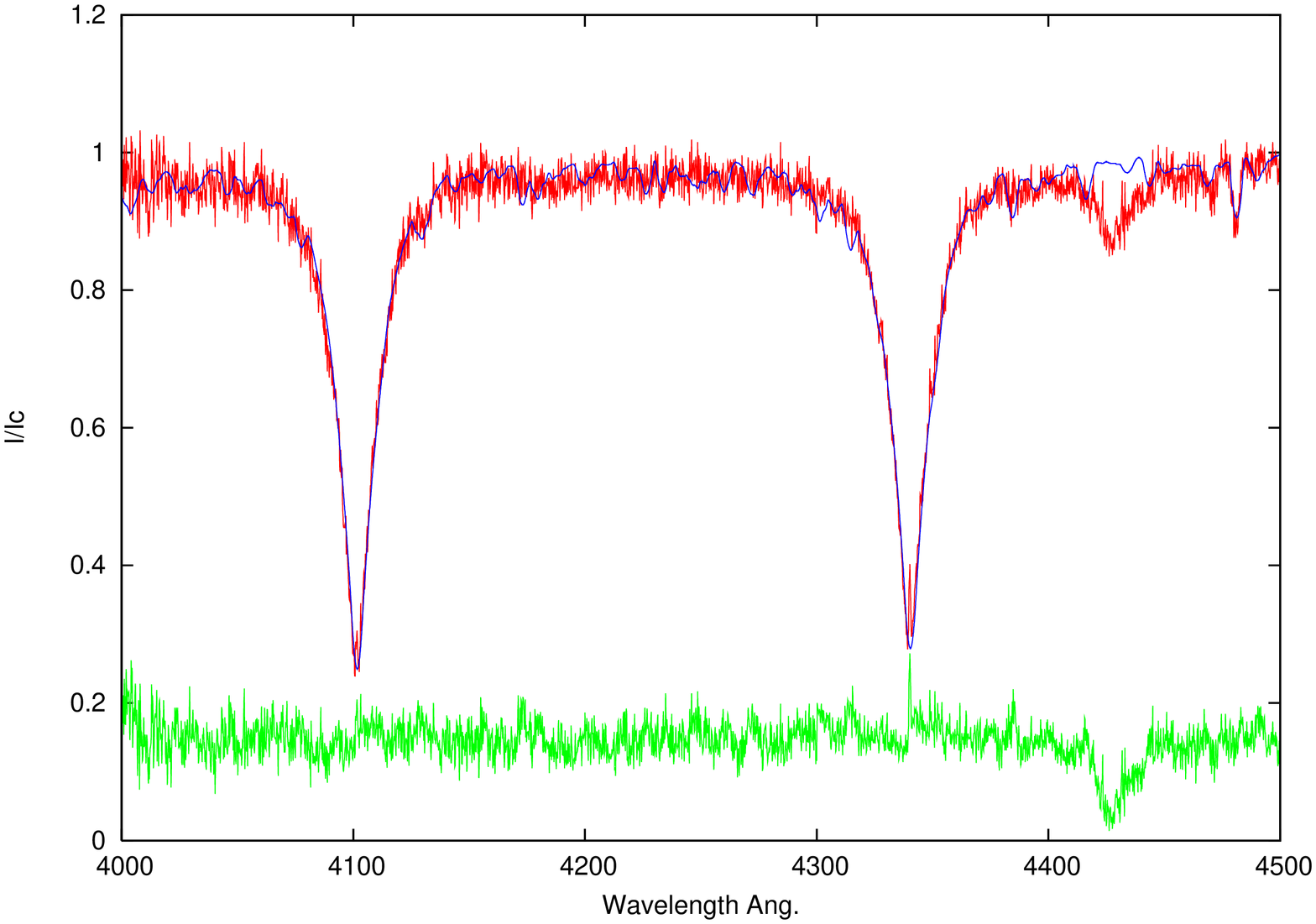}}
    \resizebox{\hsize}{!}{\includegraphics[angle=0]{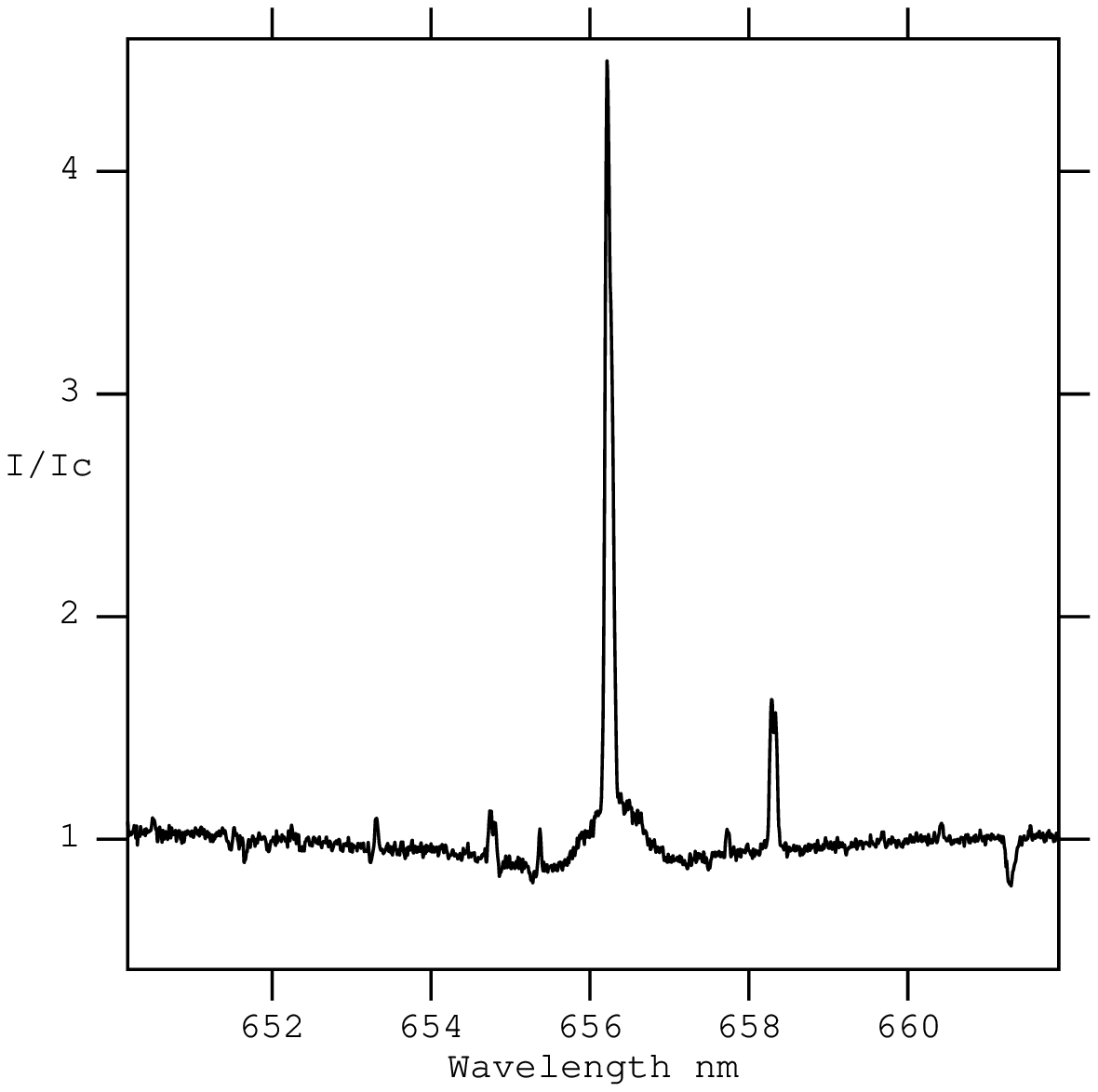}}
    \caption{ELS W031. Top panel: fit of the LR2 range of the spectrum
for the fundamental parameters determination. The observed spectrum is in red,
the fit in blue, the residuals in green; bottom panel: H$\alpha$ line from
the VLT-GIRAFFE.}
    \label{HaW031}
\end{figure}
\begin{figure}[htpb]
    \centering
    \resizebox{\hsize}{!}{\includegraphics[angle=-90]{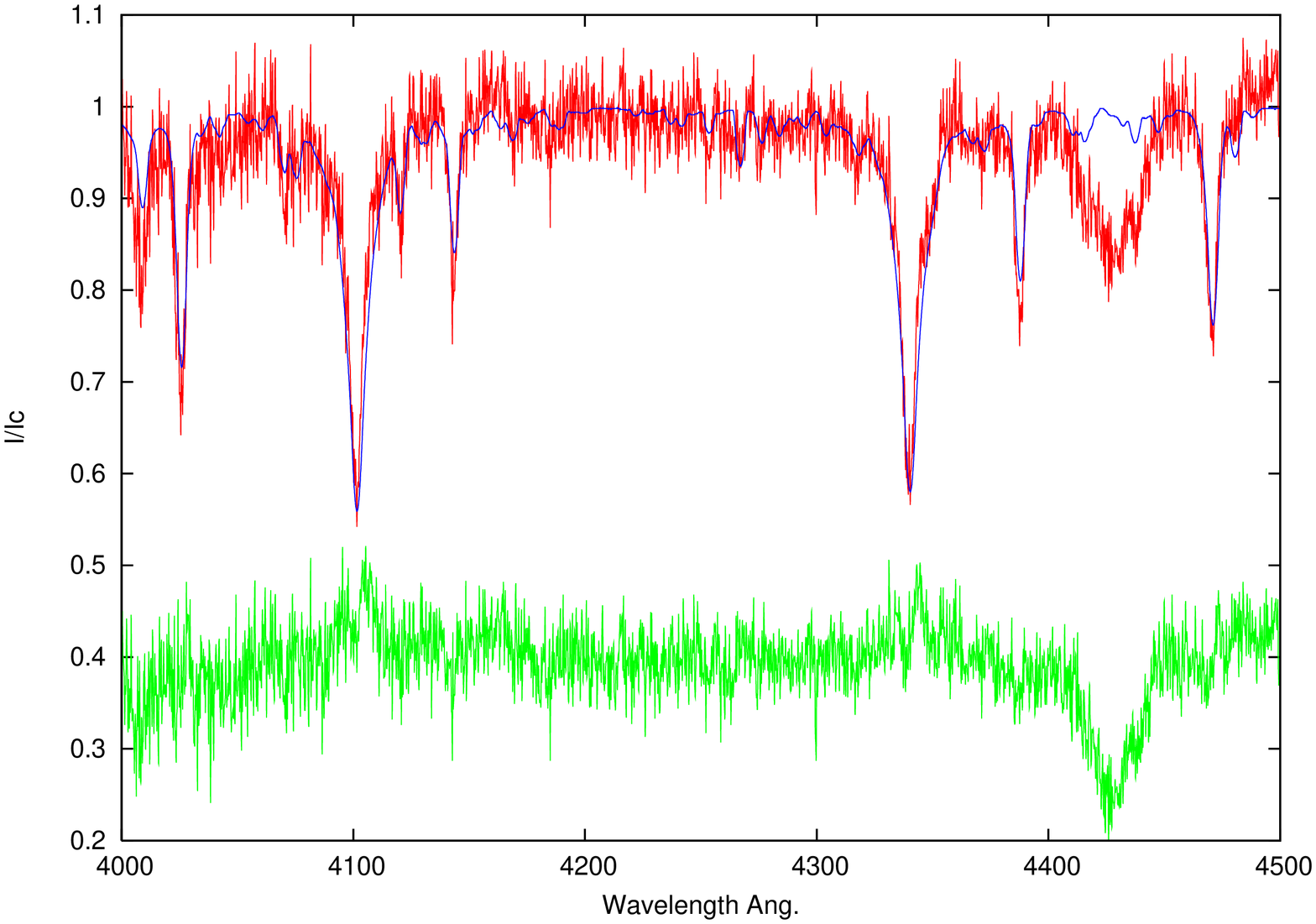}}    
    \resizebox{\hsize}{!}{\includegraphics[angle=0]{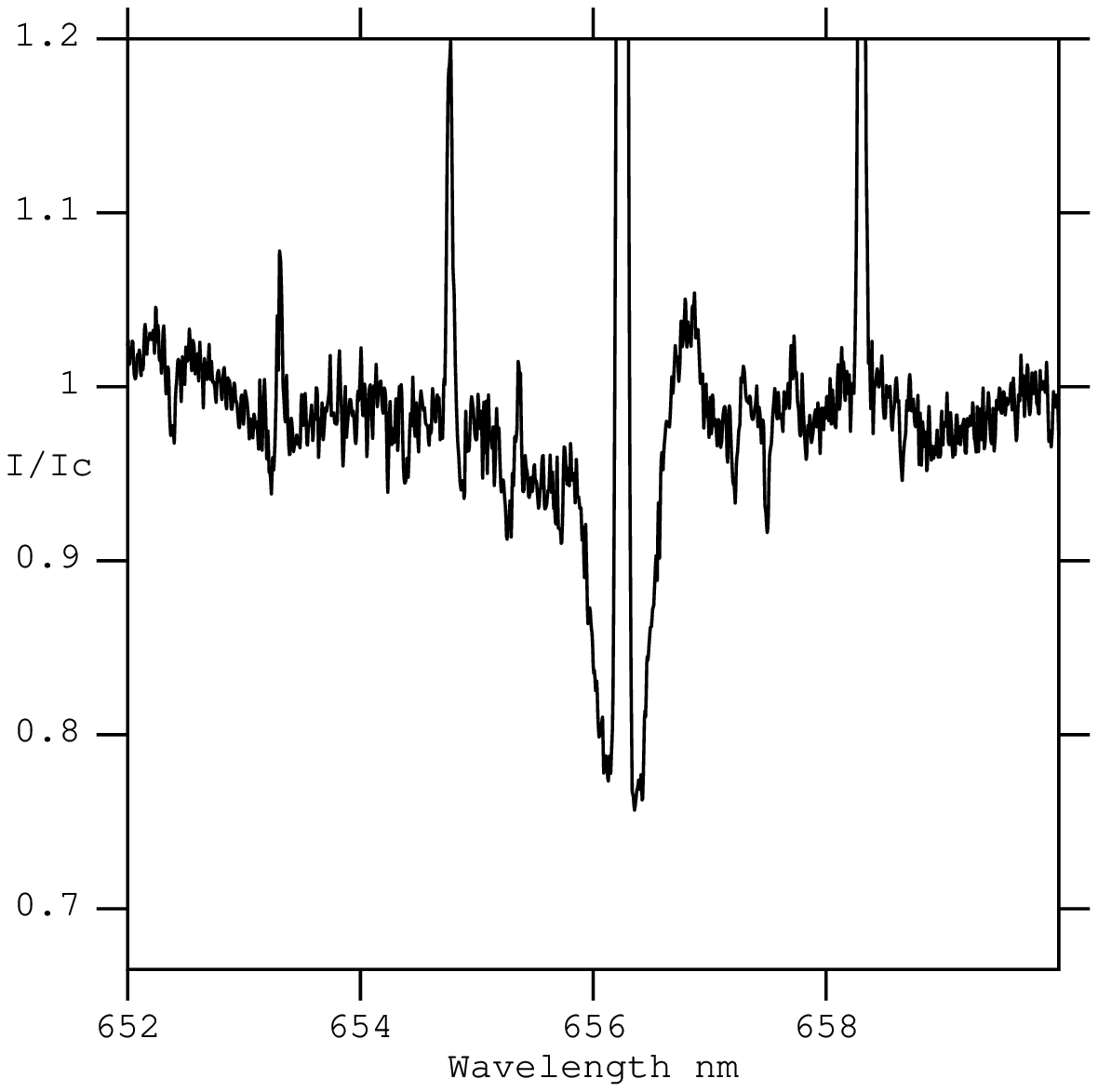}}
    \caption{Same as Fig.~\ref{HaW031} but for ELS W080.}
    \label{HaW080}
\end{figure}

\begin{figure}[htpb]
    \centering
    \resizebox{\hsize}{!}{\includegraphics[angle=-90]{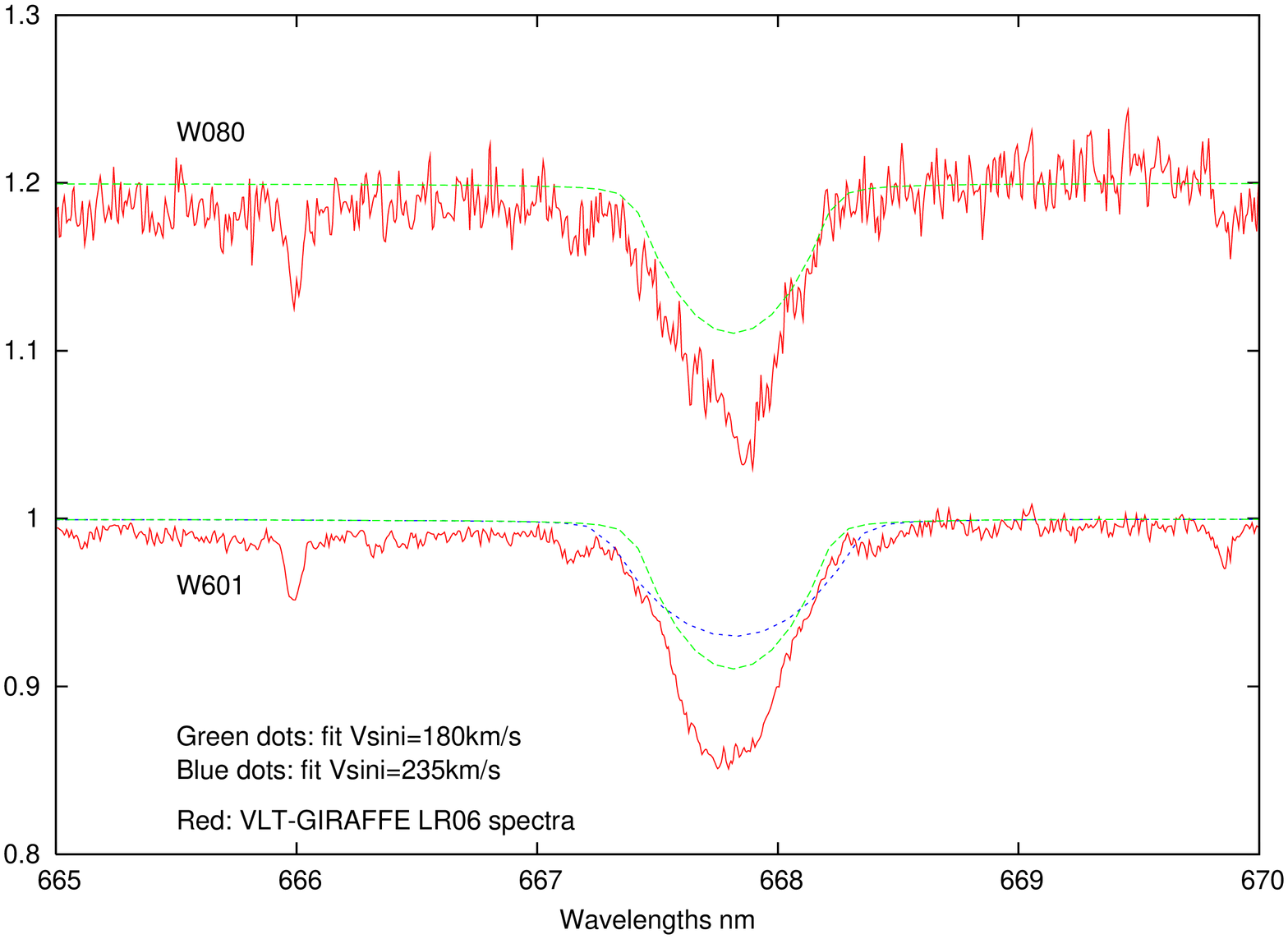}}    
    \caption{NLTE model fitting of the HeI 6678 {\AA}\ from LR06 spectra for the stars W080 (top) and for W601 (bottom).
    The fundamental parameters used for the fits are those determined for the stars; for W601, 2 different \vsini~ 
    were used in function of the studies (\citet{dufton2006} or \citet{alecian08}).}
    \label{W080W601fitsHe}
\end{figure}

\begin{figure}[htpb]
    \centering
    \resizebox{\hsize}{!}{\includegraphics[angle=-90]{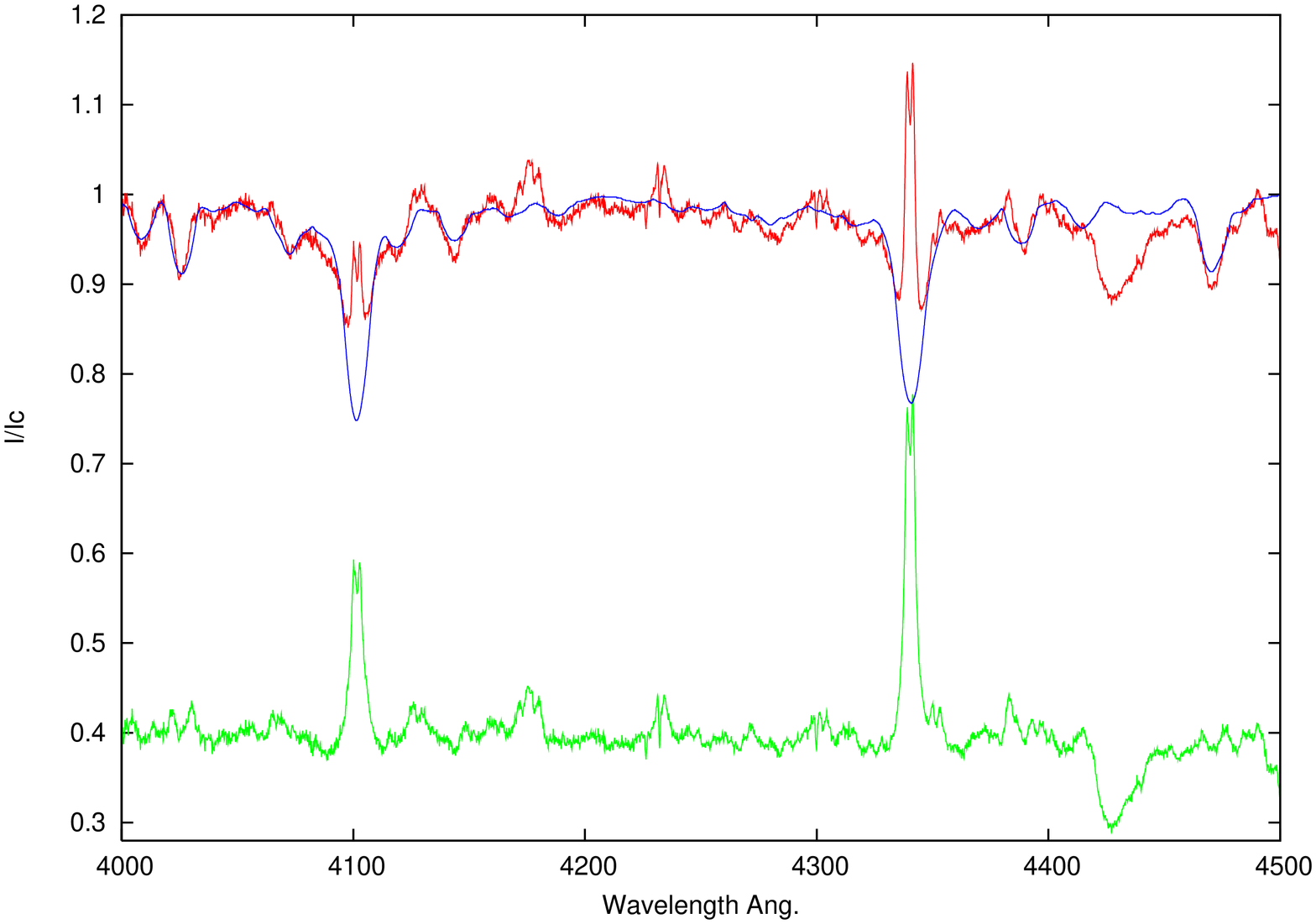}}
    \resizebox{\hsize}{!}{\includegraphics[angle=0]{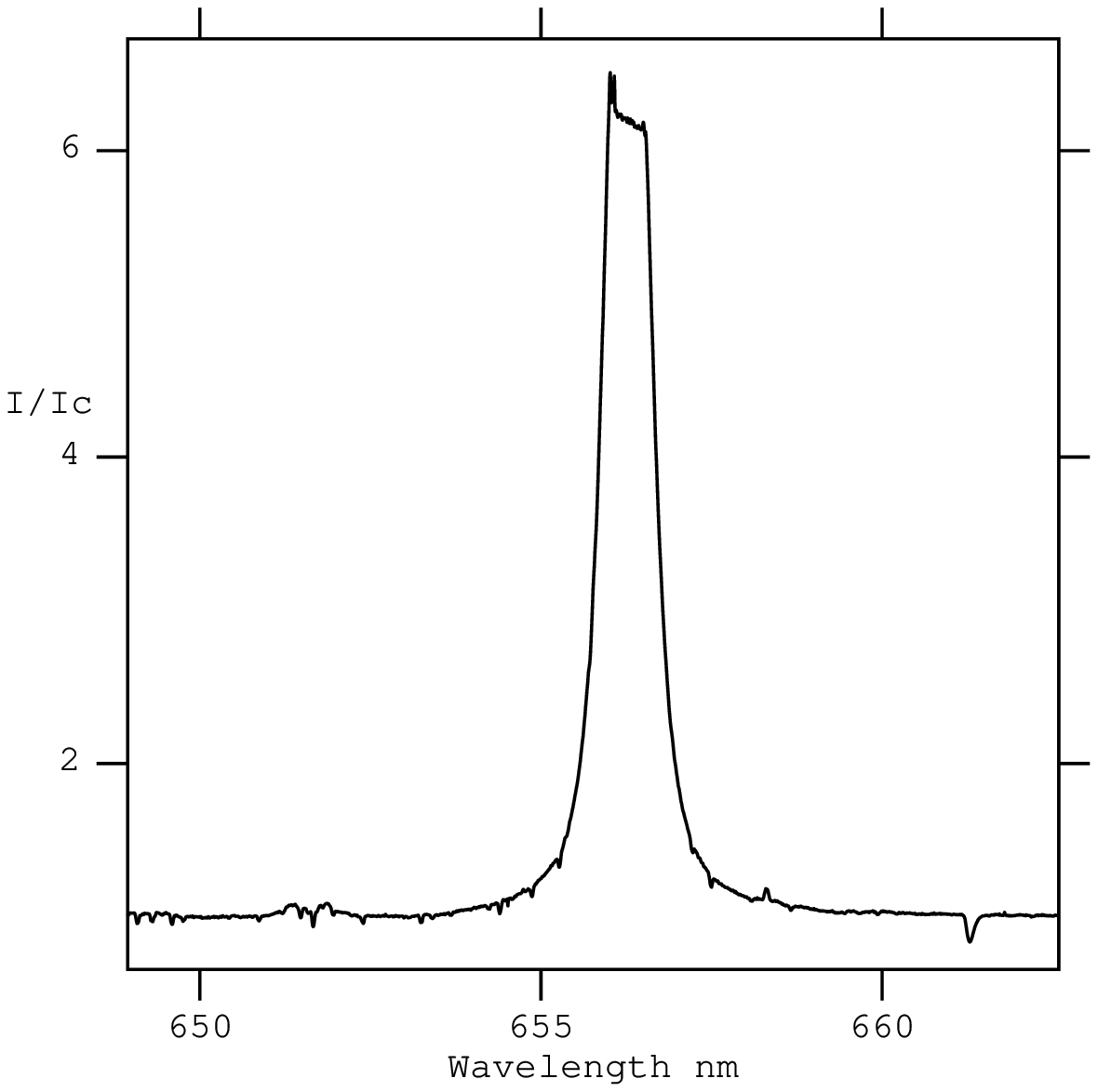}}
    \resizebox{\hsize}{!}{\includegraphics[angle=0]{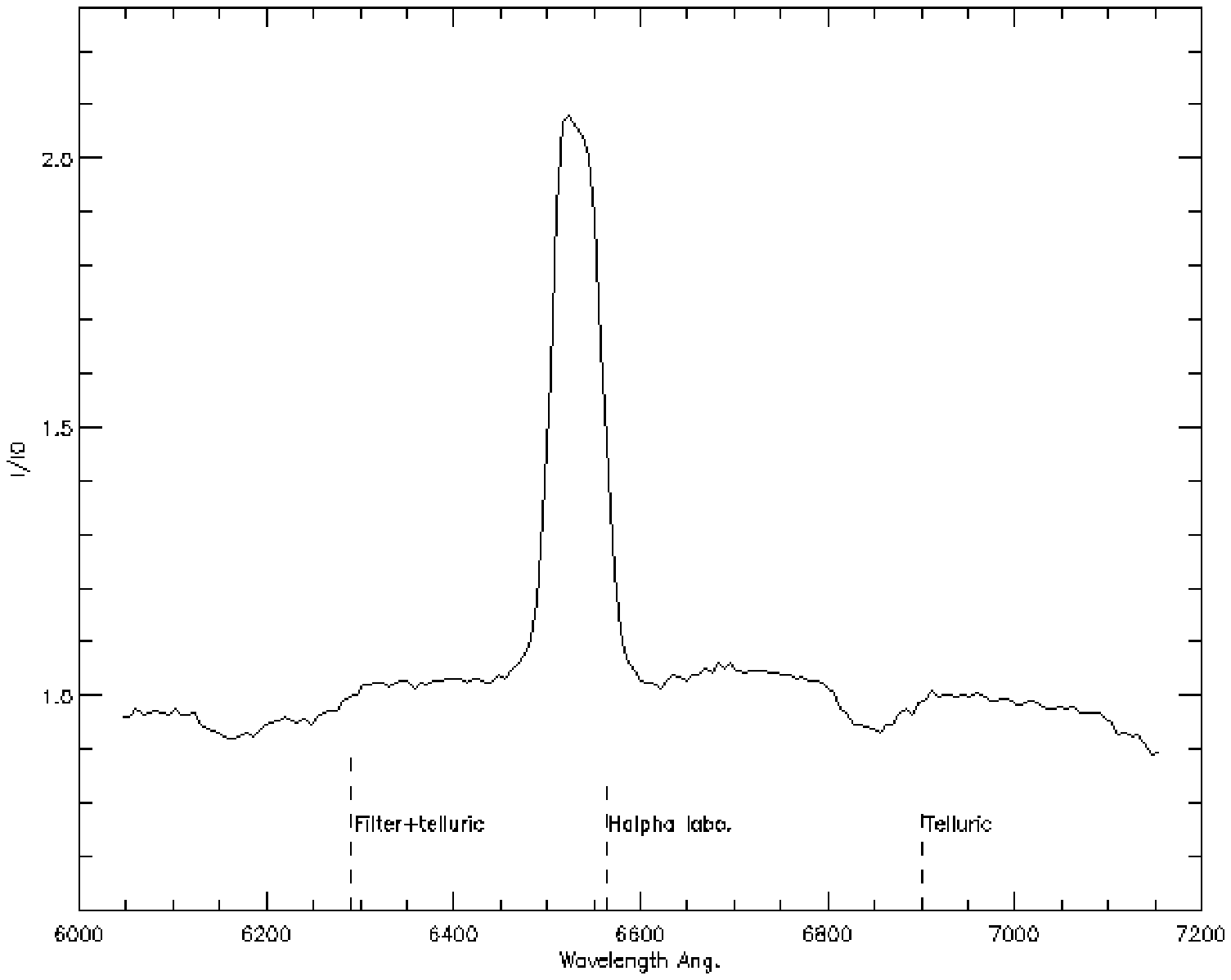}}
    \caption{Same as Fig.~\ref{Hastar017} but for ELS W235.}
    \label{HaW235}
\end{figure}
\begin{figure}[htpb]
    \centering
    \resizebox{\hsize}{!}{\includegraphics[angle=-90]{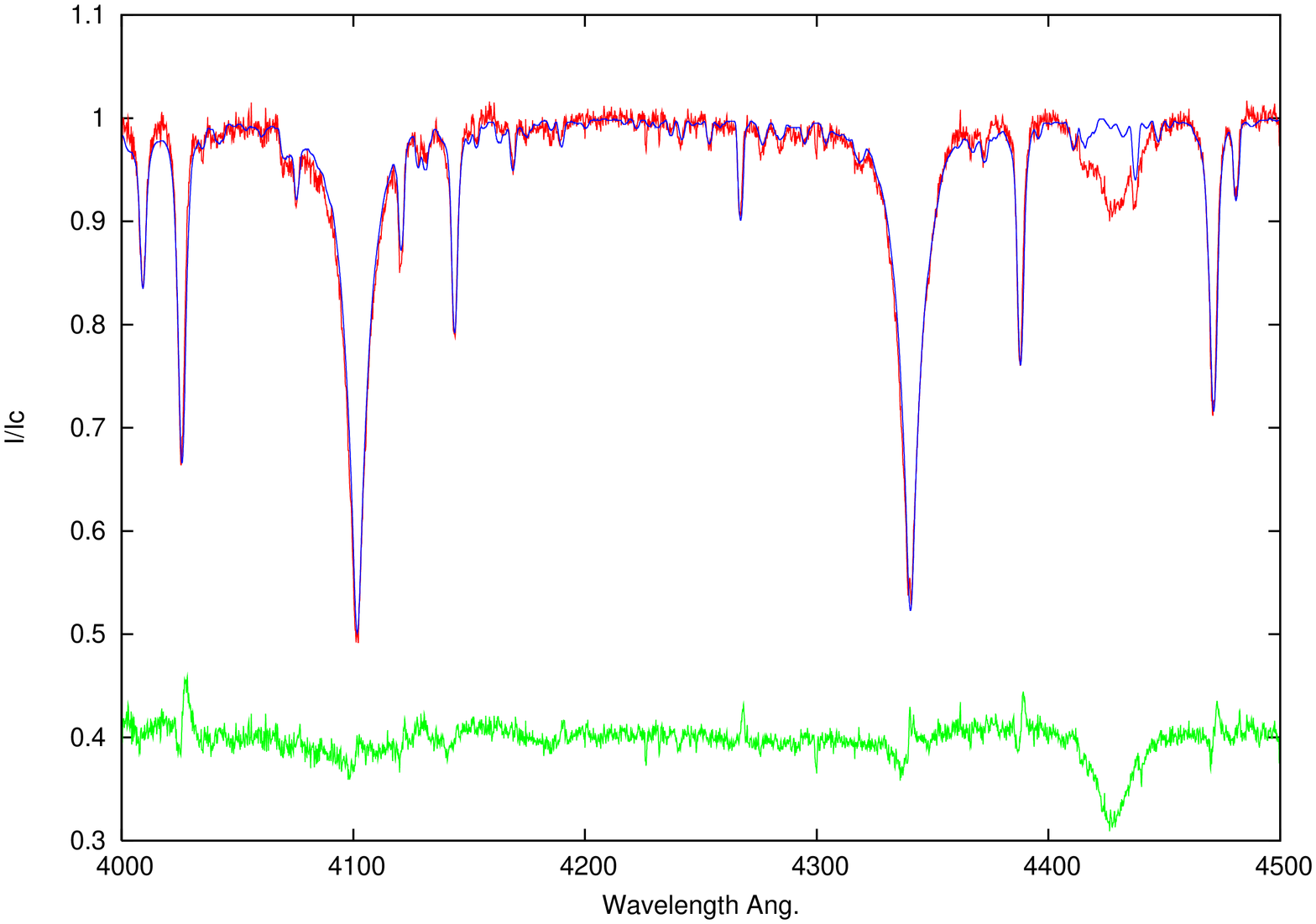}}
    \resizebox{\hsize}{!}{\includegraphics[angle=0]{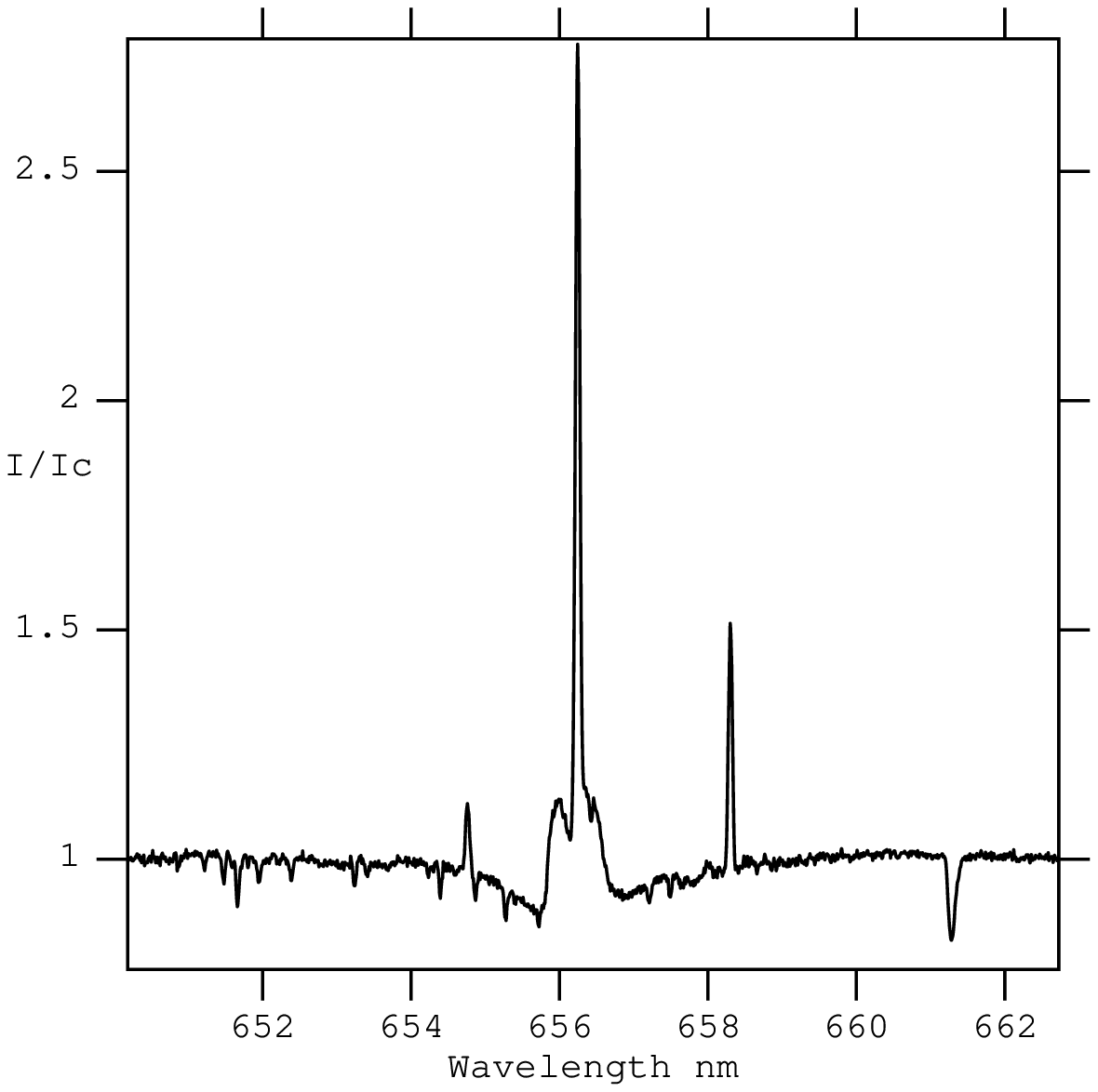}}
    \caption{Same as Fig.~\ref{HaW031} but for ELS W301.}
    \label{HaW301}
\end{figure}
\begin{figure}[htpb]
    \centering
    \resizebox{\hsize}{!}{\includegraphics[angle=-90]{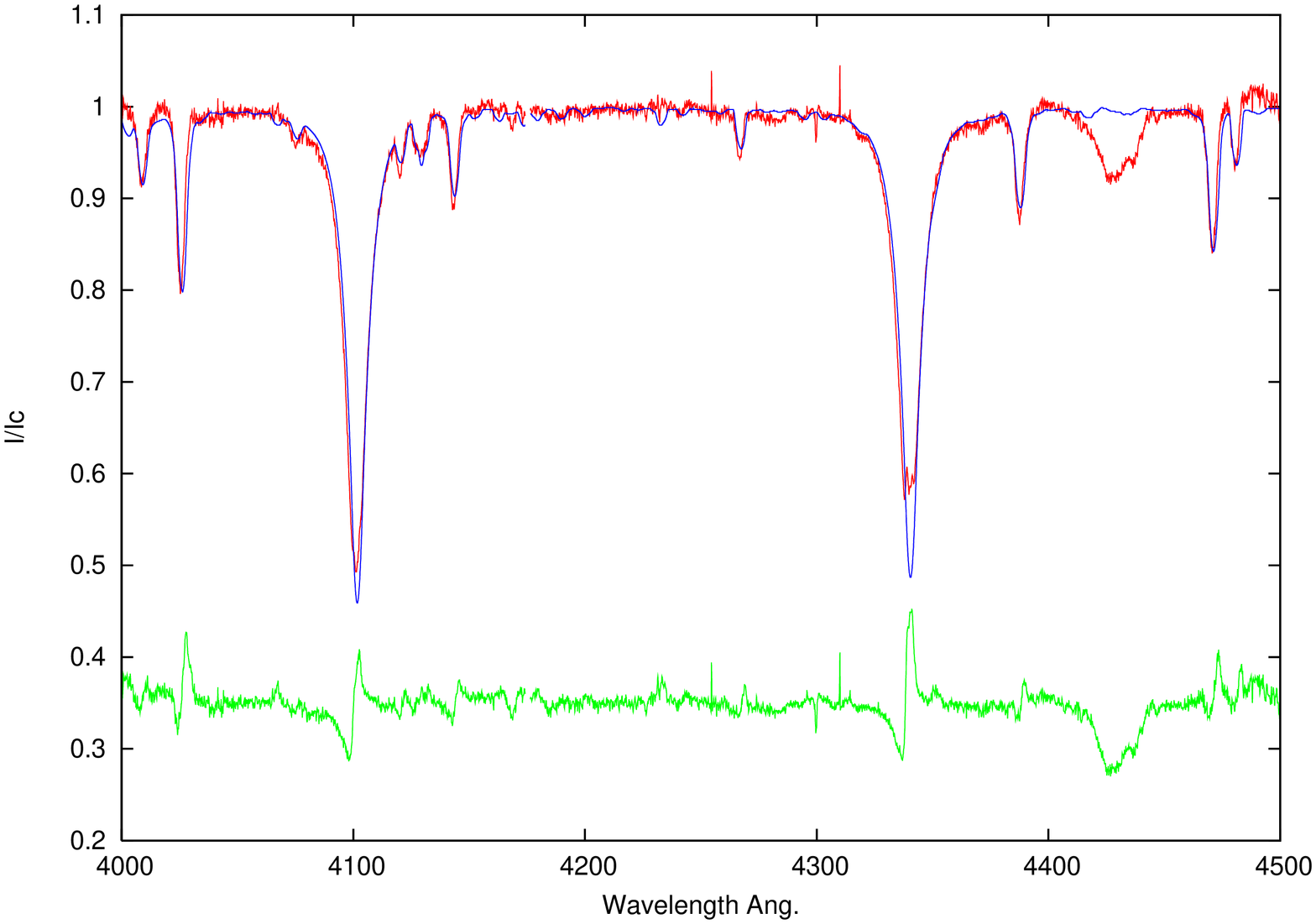}}
    \resizebox{\hsize}{!}{\includegraphics[angle=0]{newHaW483.ps}}
    \resizebox{\hsize}{!}{\includegraphics[angle=0]{wfi_W483.ps}}
    \caption{Same as Fig.~\ref{Hastar017} but for ELS W483.}
    \label{HaW483}
\end{figure}
\begin{figure}[htpb]
    \centering
    \resizebox{\hsize}{!}{\includegraphics[angle=-90]{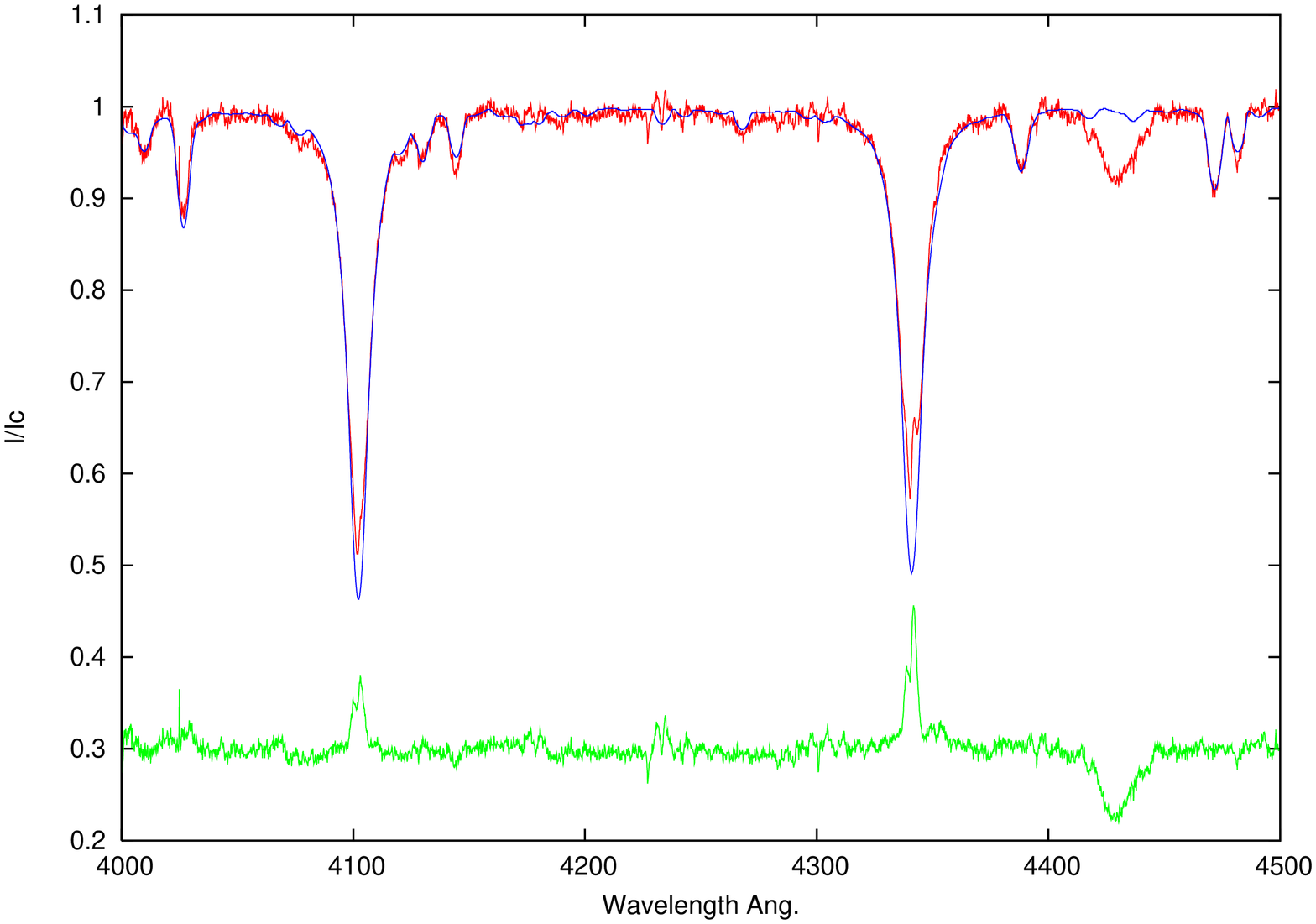}}
    \resizebox{\hsize}{!}{\includegraphics[angle=0]{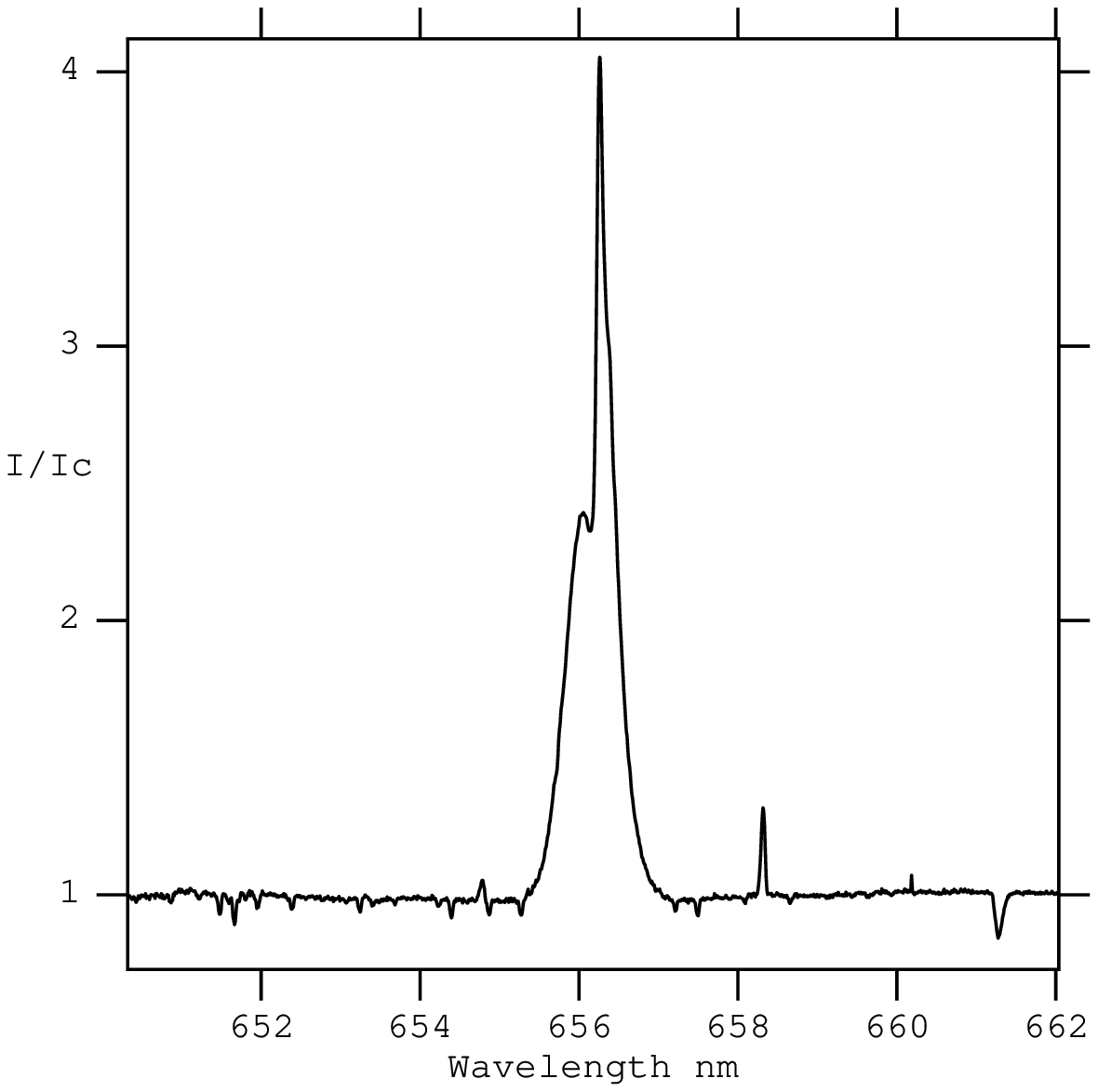}}
    \resizebox{\hsize}{!}{\includegraphics[angle=0]{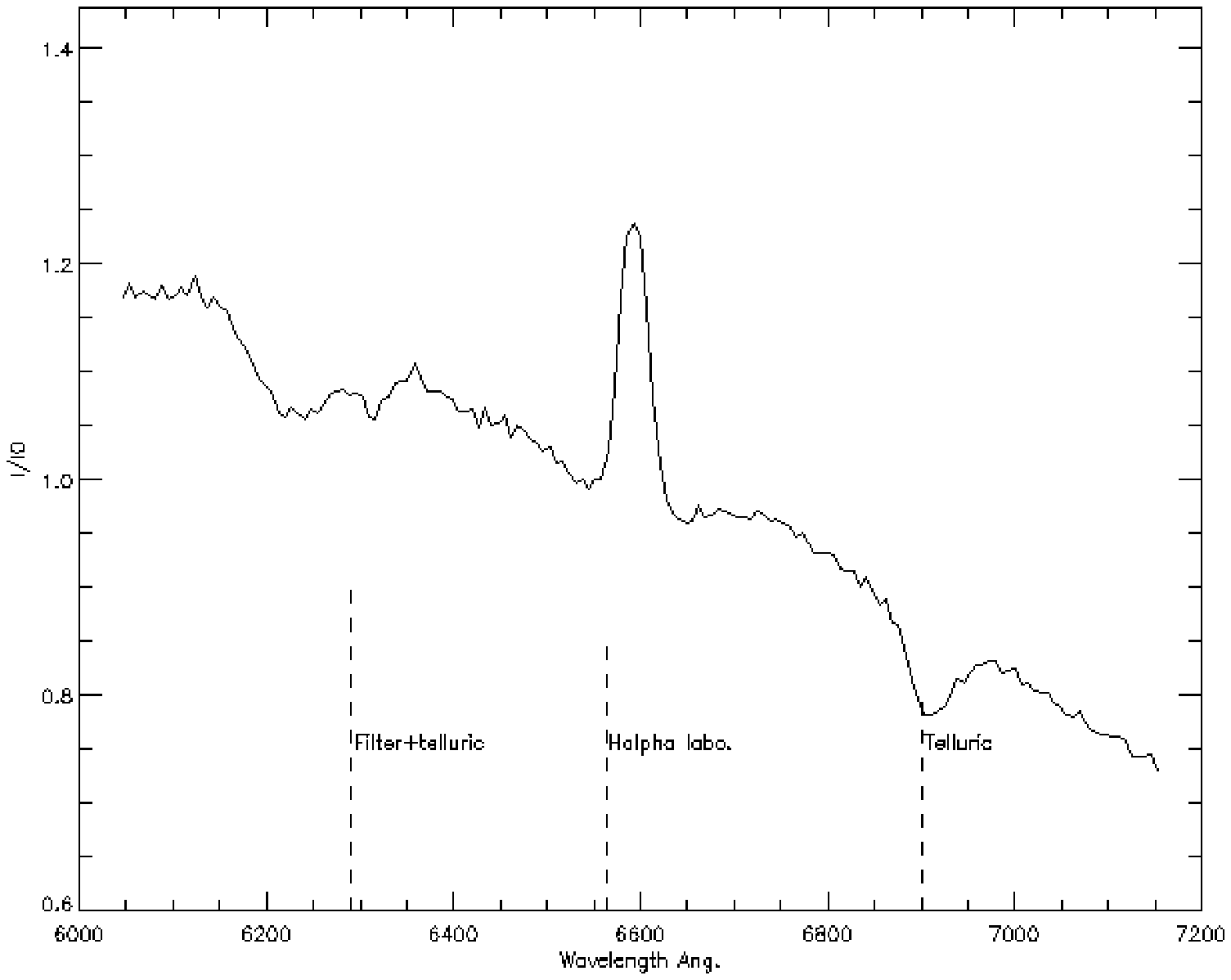}}
    \caption{Same as Fig.~\ref{Hastar017} but for ELS W500.}
    \label{HaW500}
\end{figure}
\begin{figure}[htpb]
    \centering
    \resizebox{\hsize}{!}{\includegraphics[angle=-90]{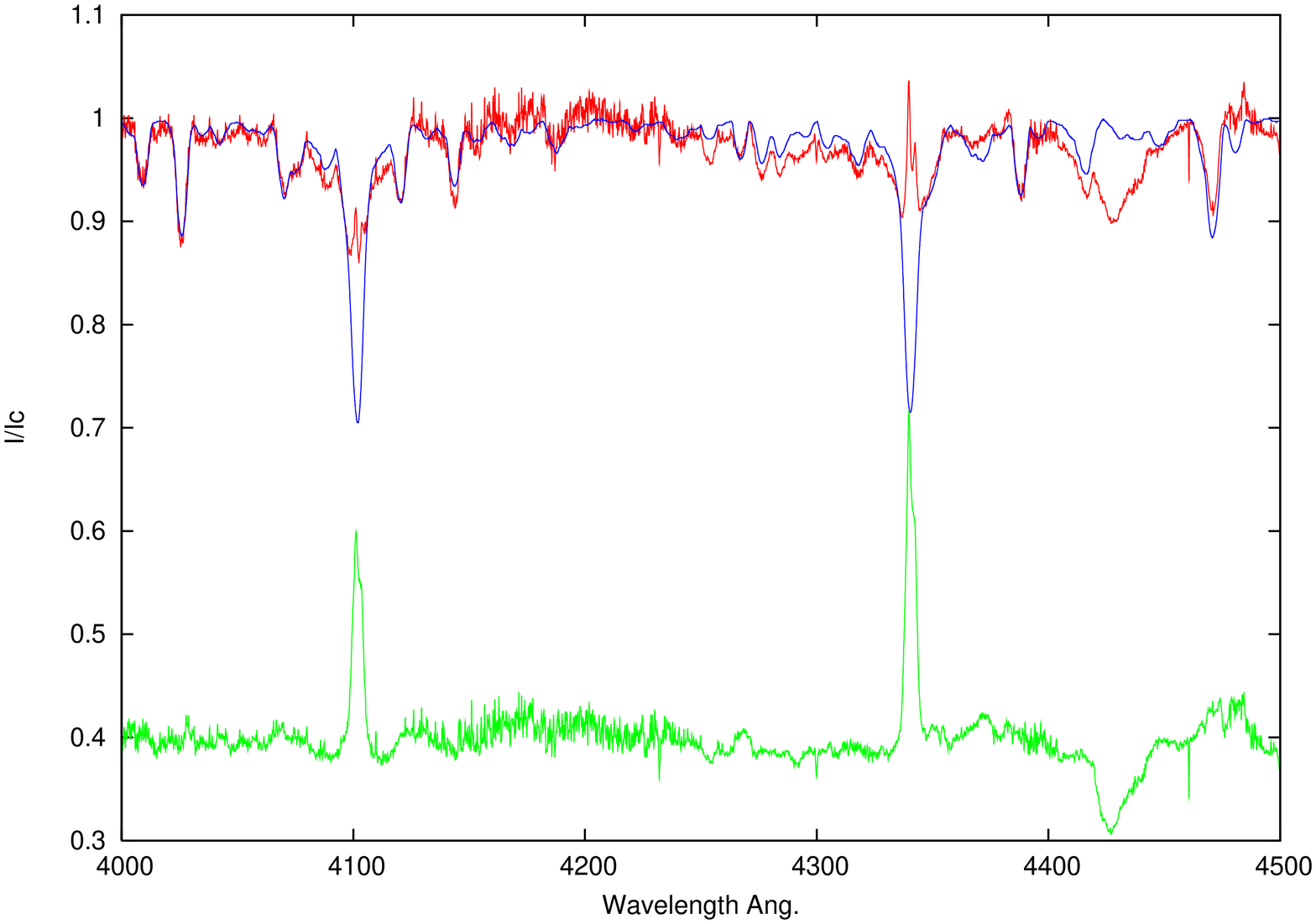}}
    \resizebox{\hsize}{!}{\includegraphics[angle=0]{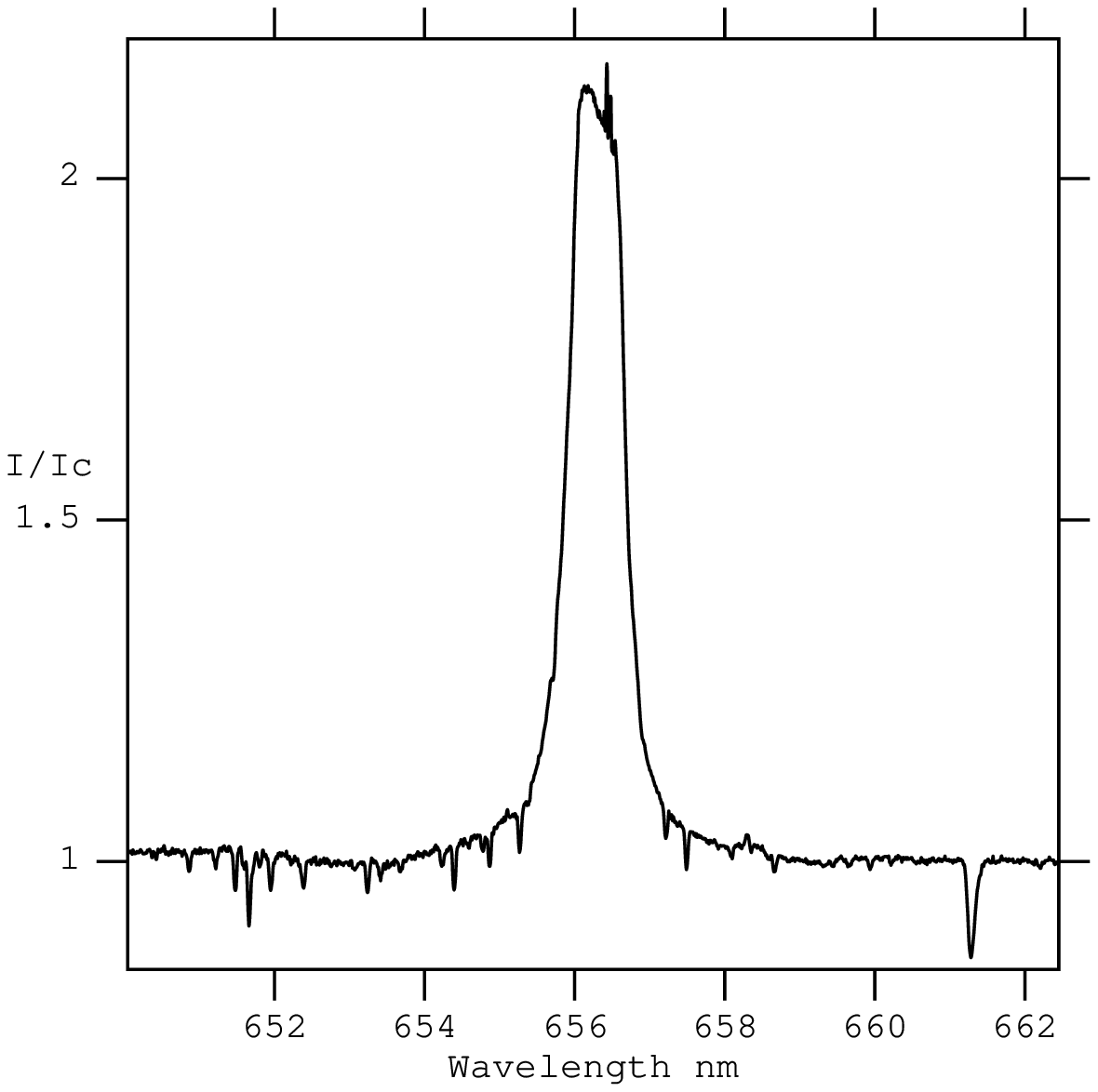}}
    \caption{Same as Fig.~\ref{HaW031} but for ELS W503.}
    \label{HaW503}
\end{figure}
\begin{figure}[htpb]
    \centering
    \resizebox{\hsize}{!}{\includegraphics[angle=0]{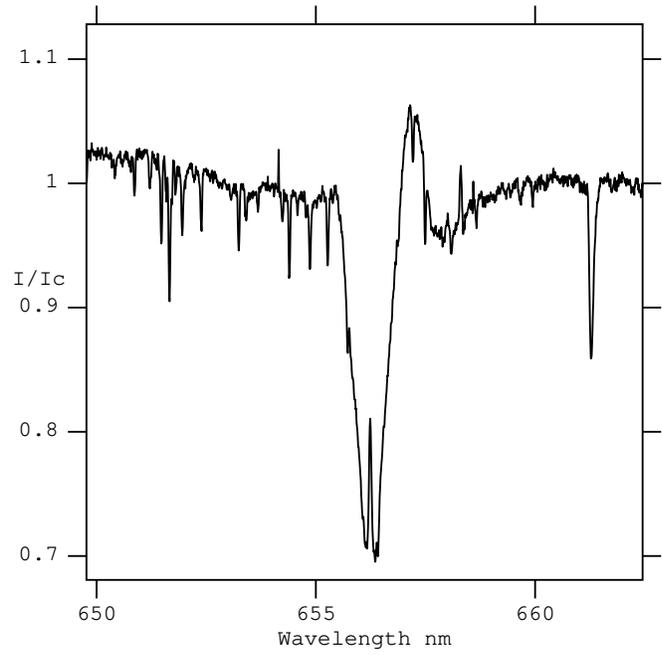}}
    \caption{ELS W601, H$\alpha$ line from the VLT-GIRAFFE.}
    \label{HaW601}
\end{figure}
\onecolumn




\end{document}